\documentclass{article}
\usepackage{jcappub}
\usepackage{epsfig}
\usepackage{graphicx}
\usepackage[latin1]{inputenc}
\usepackage{amsmath}
\usepackage{url}
\def\simlt{\ \raise -2.truept\hbox{\rlap{\hbox{$\sim$}}\raise5.truept   %
\hbox{$<$}\ }}
\def\simgt{\ \raise -2.truept\hbox{\rlap{\hbox{$\sim$}}\raise5.truept   %
\hbox{$>$}\ }}                                                          %

\def\be{\begin{equation}}
\def\ee{\end{equation}}
\def\newline{\hfil\break}

\def\la{\mathrel{\hbox{\rlap{\hbox{\lower4pt\hbox{$\sim$}}}\hbox{$<$}}}}
\def\ga{\mathrel{\hbox{\rlap{\hbox{\lower4pt\hbox{$\sim$}}}\hbox{$>$}}}}

\newcommand{\pd}[3]{\frac{\partial^{#3} #1}{\partial {#2}^{#3}}} 
\newcommand{\td}[3]{\frac{d^{#3} #1}{d {#2}^{#3}}} 
\renewcommand{\v}[1]{\ensuremath{\mathbf{#1}}} 
\newcommand{\gv}[1]{\ensuremath{\mbox{\boldmath$ #1 $}}} 
\renewcommand{\bar}[1]{\ensuremath{\overline{#1}}}

\title{Evolution of Dark Matter Halos and their Radio Emissions}
\author[a,1]{S. Colafrancesco,\note{Corresponding author.}}
\author[a]{P. Marchegiani,}
\author[a]{G. Beck}

\affiliation[a]{School of Physics, University of the Witwatersrand, Private Bag 3, WITS-2050, Johannesburg, South Africa}

\emailAdd{sergio.colafrancesco@wits.ac.za}
\emailAdd{paolo.marchegiani@wits.ac.za}
\emailAdd{geoff.m.beck@gmail.com}

\abstract{
Radio synchrotron emission is expected as a natural by-product of the self-annihilation of super-symmetric dark matter particles.  In this work we discuss the general properties of the radio emission expected in a wide range of dark matter halos, from local dwarf spheroidal galaxies to large and distant galaxy clusters with the aim to determine the neutralino dark matter detection prospects of the Square Kilometre Array (SKA). The analysis of the SKA detection of dark matter(DM)-induced radio emission is presented for structures spanning a wide range of masses and redshifts, and we also analyze the limits that the SKA can set on the thermally averaged neutralino annihilation cross-section in the event of non-detection. To this aim, we construct a model of the redshift evolution of the radio emissions of dark matter halos and apply it to generate predicted fluxes from a range of neutralino masses and annihilation channels for the dark matter halos surrounding dwarf galaxies, galaxies and galaxy clusters. Using the available SKA performance predictions and its ability to determine an independent measure of the magnetic field in cosmic structures, we explore both the detailed detection prospects and the upper-bounds that might be placed on the neutralino annihilation cross-section in the event of non-detection. We find that the SKA can access a neutralino parameter space far larger than that of any preceding indirect-detection experiment, also improving on the realistic CTA detection prospects, with the possibility of setting cross-section upper-bounds up to four orders of magnitude below the thermal relic density bound. Additionally, we find that neutralino radio emissions carry redshift-independent signatures of the dominant annihilation channel and of neutralino mass, offering therefore a means to identify such non-thermal emissions within the observing frequency range of the SKA.}

\begin{document}
 \maketitle

\section{Introduction}
\label{sec.intro}

Modern astrophysical evidence, namely gravitational lensing, galaxy rotation curves, galaxy cluster masses, and cosmic microwave-background anisotropy data suggests that the majority of the matter content within the universe is in the form of dark matter and that this plays a vital role in the evolution of structure within the universe. 
The nature of dark matter remains a major hiatus in the understanding of modern physics, with many contending models proposed to account for this missing puzzle-piece \cite{jungman1996, bertone2005, moffat2007, frampton2010, feng2010, petraki2013} and with many methods of both direct and indirect detection proposed \cite{gaitskell2004, carr2006, Colafrancesco2010, bauer2013}. In this study we focus on one of the current favourites in the dark matter candidate zoo: the so-called ``neutralino", being the lightest particle from the minimal super-symmetric extension of the standard model (MSSM). While there is no experimental evidence yet supporting super-symmetry, the model is attractive as super-symmetric effects may manifest unambiguously in collider experiments. Clearly then, direct detection provides the cleanest route to discovery. However, the possibility of indirect detection through astrophysical signatures remains, as the weakly interacting nature of the neutralino allows for mutual pair annihilation to produce standard model particles which may then be produced as cosmic rays or instead subsequently emit radiation. Additionally, the increasing sensitivity of astronomical measurements and the limitations of current ground-based direct detection experiments combine to make indirect detection methods increasingly attractive and competitive when it comes to setting limits on the nature of dark matter. While many authors have previously studied the high-energy products of neutralino pair annihilation \cite{pieri2011,lamanna2013, funk2013, conrad2012, morselli2010, bergstrom1994, gondolo1994, gondolo1999, pieri2004}, there has been some recent interest in the possibility of detecting radio emission resulting from the annihilation products of such particles \cite{Colafrancesco2006, siffert2011, linden2011, storm2012, spekkens2013}. However, this interest has been largely confined to single source studies, looking at either the galactic centre or towards specific, dark matter-dominated, sources. 
We therefore present here a detailed study of the radio emissions that result from the mutual pair annihilation of neutralinos within the dark matter halos that surround cosmic structures ranging from dwarf galaxies to galaxy clusters observable at various distances from the local environment (e.g. dwarf spheroidal galaxies) to redshifts of $z \sim 5$ (galaxies and galaxy clusters). In this context, we make a thorough survey to determine how dark matter-induced radio emissions depend upon the mass and annihilation channel of the neutralino, and how they evolve with redshift, with an additional focus on the potential detection capabilities of the upcoming Square Kilometre Array (SKA) telescope. Such a comprehensive study, especially one making special reference to the SKA as a ``dark matter machine", is currently absent from the literature.
In this paper we make an extensive study of the radio emissions from dark matter halos varying the neutralino mass, annihilation channel, as well the mass and redshift of the halo while using various dark matter density profiles such as  Einasto, Burkert and Navarro-Frenk-White profiles. We scan the parameter space of structures hosting dark matter, extending from dwarf galaxies to galaxy clusters with redshifts between $0.01$ and $5$. The relatively large redshifts examined are important as they might offer the chance to detect dark matter in structures that are more primitive, having fewer baryon-related astrophysical mechanisms that produce additional foreground synchrotron emissions which make unambiguous dark matter detection more difficult. The choice to examine the halos of both large and small structures is crucial, as dwarf spheroidal galaxies are well known to be highly dark matter dominated but produce faint emissions, while larger structures, even though not immaculate test-beds for dark matter emissions, provide substantially stronger fluxes. This means that a survey of halos of varying mass is essential to locate the best detection prospects for future telescopes like the SKA. In order to model the radio emission we confine our approach to the synchrotron radiation produced by the secondary electrons (we refer to electrons and positrons collectively as `electrons') resulting from the annihilation of super-symmetric neutralinos that compose the halo. In so doing, we model the diffusion of secondary electrons following \cite{Colafrancesco2006}. In the interest of comprehensiveness we utilise several neutralino masses, varying between 10 and 500 GeV in the annihilation channels $b\overline{b}$, $\tau^+\tau^-$ and $W^+ W^-$, using data sourced from the DarkSUSY package~\cite{darkSUSY}. The model we employ to track the redshift evolution of dark matter halos is based on the work of \cite{Bullock2001,diemand2008, Colafrancesco2006,Colafrancesco2007, eke2001}. For our cosmological framework we consider a flat $\Lambda$CDM universe with $\Omega_m (0) = 0.315$ and $H_0 = 67.3$ km s$^{-1}$ Mpc$^{-1}$ in accordance with the PLANCK results~\cite{ade2013}. Additionally we assume the primordial power spectrum to be scale invariant and normalised to the value $\sigma_8 = 0.897$. We also examine three models of the magnetic field radial dependency within galaxy clusters: \textit{i}) a model following the density of the intra-cluster medium, \textit{ii}) a power-law model in accordance with results in \cite{Bonafede2010,Rave2013} which suggest a flatter magnetic field radial profile than that of the thermal gas, and \textit{iii}) a constant field model. In the case of galaxies and dwarf galaxies we use only the constant magnetic field model.

With this approach we examine the general properties of the neutralino-induced synchrotron radiation over the given parameter space and demonstrate that differing annihilation channels leave distinct signatures upon the observed flux from the halo and that these signatures are both redshift and halo mass independent. Additionally we used these results, in conjunction with the expected sensitivity of the SKA, to determine the bounds on the annihilation cross-section that might be imposed by non-detection with the SKA. These bounds are computed as functions of the neutralino mass and annihilation channel as well as the strength of the magnetic field present within the halo. Finally we discuss the exclusion plane that determines the combinations of halo mass and redshift that would be observable to the SKA for thermally-averaged cross section values of $3 \times 10^{-26}$ cm$^3$ s$^{-1}$ or lower. In so doing we determine the ideal cases for SKA detection of neutralino-induced synchrotron emission and demonstrate that the SKA has the potential to produce upper-bounds on the cross section that are far more stringent than those from existing experiments, as found in \cite{bottino2008,ackerman2010,hooper2011}, suggesting the SKA will have a relevant role in determining the nature of dark matter. Moreover, the difficulty of disentangling the effects of the electron spectrum and the magnetic field itself in synchrotron radiation makes the accuracy of magnetic field models within prospective radio sources highly important. It is therefore of considerable advantage to radio searches for dark matter that the SKA will be also capable of determining magnetic field structures through both polarisation and Faraday Rotation measurements.

A breakdown of this paper is as follows: we first present a discussion of the models and assumptions employed in our work in sections 2, then we present the results of our synchrotron flux calculations for varying halo masses, redshifts and neutralino properties in section 3, following this we present the non-detection constraints derived from SKA sensitivity data and finally we discuss the implications of these results in section 4.

\section{Radio Emission from Dark Matter Halos}

\subsection{Dark Matter Halos}
In the study of dark matter halo structures we will make use of the prominent halo density profiles, i.e. the Navarro-Frenk-White (NFW)~\cite{nfw1996}, Burkert~\cite{Burkert1995}, and Einasto~\cite{einasto1968} models. These are given respectively by
\begin{equation}
\begin{aligned}
\rho_N(r)=\frac{\rho_s}{\frac{r}{r_s}\left(1+\frac{r}{r_s}\right)^{2}} \; ,\\
\rho_{B} (r) = \frac{\rho_s}{\left(1 + \frac{r}{r_s}\right)\left(1+\left(\frac{r}{r_s}\right)^2\right)} \; ,\\
\rho_E(r)=0.25 \rho_s \mbox{e}^{-\frac{2}{\alpha}\left(\left(\frac{r}{r_s} \right)^\alpha - 1\right)} \; ,
\end{aligned}
\label{eq:nfw}
\end{equation}
with $r_s$ being the scale radius of the profile, with $\alpha \approx 0.18$ giving profiles similar to NFW in the outer halo while differing in the inner regions, and $\rho_s$ is the halo characteristic density (the factor of $0.25$ accounts for the difference between $\rho (r_{-2})$ and $\rho_s$ where $r_{-2}$ is the radius at which the logarithmic slope of halo density profile is equal to -2.).
We then define the virial radius $R_{vir}$, of a halo with mass $M_{vir}$, as the radius within which the mean density of the halo is equal to the product of the collapse over-density $\Delta_{c}$ and the critical density $\rho_c$, where 
\begin{align}
\rho_c (z) & = \frac{3 H(z)^2}{8\pi G} \; , \\
M_{vir} & = \frac{4}{3}\pi \Delta_{c} \rho_c R_{vir}^3 \; ,
\end{align}
with $H(z)$ being the Hubble parameter. The density contrast parameter at collapse is given in flat cosmology by the approximate expression
\begin{equation}
\Delta_{c} \approx 18 \pi^2 - 82 x - 39 x^2 \; ,
\end{equation}
and, in the Einstein/de-Sitter universe, by~\cite{bryan1998}
\begin{equation}
\Delta_{c} \approx 18 \pi^2 - 60 x - 32 x^2 \; ,
\end{equation}
with $x = 1.0 - \Omega_m (z)$, where $\Omega_m (z)$ is the matter density parameter at redshift $z$ given by
\begin{equation}
\Omega_m (z)  = \frac{1}{1 + \frac{\Omega_\Lambda (0)}{\Omega_m (0)}(1+z)^{-3}} \; .
\end{equation}
The concentration parameter for the halo can then be defined as
\begin{equation}
\begin{aligned}
c_{vir} = \frac{R_{vir}}{r_{-2}} \; .
\end{aligned}
\end{equation}
Furthermore, we make the identification $r_s = r_{-2}$ for all of the Einasto, Burkert and NFW dark matter density profiles. In the following work we make use of the method discussed in \cite{Bullock2001} to determine the concentration parameter as a function of halo mass
\begin{equation}
c_{vir}= K \frac{1 + z_c}{1 + z} = \frac{c_{vir} (M,z=0)}{1 +z} \; , \label{eq:cvir}
\end{equation}
with $K = 4$ providing a good fit for $\Lambda$CDM models~\cite{eke2001}, and $z_c$ is the collapse redshift of the halo, defined by
\begin{equation}
G(z_c) \sigma (F M_{vir}) = 1.686 \; ,
\end{equation}
with $F$ being a free parameter (we take $F = 0.01$ in accordance with ref.~\cite{Bullock2001}), $G(z)$ being the linear perturbation growth factor and $\sigma (M)$ is the present-day rms density fluctuation of spheres containing the mass $M$~\cite{peebles1980}. We note here that there are more recent results in the virial concentration mass relations~\cite{Munoz2010} indicating larger values of the virial concentration parameter in the case of galaxy cluster-type halos. These in turn result in an amplification of the radio fluxes calculated here for this class of halo (within the redshift $z \le 1$ range), while for higher redshifts (and smaller halos) there is no substantial difference between the two approaches. This implies that the $c_{vir}$ model used here provides slightly conservative estimates on the radio emission from more local DM halos.\\
The dimensionless characteristic density contrast $\frac{\rho_s}{\rho_c}$ can then be defined in terms of $c_{vir}$, following ref.~\cite{ludlow2013},
\begin{equation}
\frac{\rho_s (c_{vir})}{\rho_{c}}=\frac{\Delta_{c}}{3}\frac{c_{vir}^3}{\ln(1+c_{vir})-\frac{c_{vir}}
{1+c_{vir}}} \; .
\end{equation}

In this work we will explore three classes of dark matter halo. Those around galaxy clusters, with masses $M_{vir} > 10^{14} M_\odot$, around galaxies, with  $10^{11} M_\odot \le M_{vir} \le 10^{12} M_\odot$, and finally the halos of dwarf spheroidal galaxies with $M_{vir} < 10^{9} M_\odot$.

\subsection{Halo Substructure}
In order to account for the effect of substructure within a dark matter halo we will divide the total density profile into smooth
and sub-halo components
\begin{equation}
\rho_{tot} (r) = \rho_{sm} (r) + \rho_{sh} (r) \; ,
\end{equation}
where the total profile is
\begin{equation}
\rho_{tot} (r) = \rho_s g(r/r_s) \; ,
\end{equation}
with $g(x)$ being such as to reproduce the relevant density profile (NFW, Burkert or Einasto). Here $\rho_{sm}$ and $\rho_{sh}$ are respectively the smooth and sub-halo contributions to the mass density.

The substructure of the halo is characterised by $f_s$, the fraction of the halo mass concentrated within sub-halos
\begin{equation}
f_s \equiv \frac{M_{sub}^{tot}}{M_{vir}} \; ,
\end{equation}
and by the spatial density of sub-halos, assumed here to be given by same function as the total halo mass density but with a larger scale radius $R_s \sim 7 r_s$ to represent radial biasing of the distribution~\cite{nagai2005}
\begin{equation}
p_{sub} (r) = N g(r/R_s) \; ,
\end{equation}
with normalisation constant $N$ determined by the requirement
\begin{equation}
4\pi \int_0^{R_{vir}} dr \, r^2 p_{sub}(r) = 1 \; .
\end{equation}
An additional characteristic of the halo substructure is the sub-halo mass distribution. In order to account for this we consider the rate of change of the sub-halo number density with respect to sub-halo mass $M_{sub}$~\cite{pieri2011}
\begin{equation}
\td{n_s}{M_{sub}}{} (M_{sub}) = \mathcal{M}_0 \widetilde{M}^{-\mu} \; ,
\end{equation}
with $\mu = 2$ where, for a given sub-halo mass scale $M_*$,
\begin{equation}
\widetilde{M} = \frac{M_{sub}}{M_{*}}
\end{equation}
and $\mathcal{M}_0$ is the normalisation constant, such that the total sub-halo mass contribution
is normalised to $f_s M_{vir}$.
In choosing the limits for mass distribution, we follow the approach of~\cite{pieri2011} in setting $M_{min} = 1\times 10^{-6}M_{\odot}$ and $M_{max} = 0.01 M_{vir}$. Our minimum mass has been chosen to correspond to the WIMP free-streaming scale, although it is possible for there to be contributions from even smaller masses within the wider SUSY frame-work: lower values of  $M_{min}$ could serve to increase the boosting factor which results from the increased density provided by sub-halos.\\
With this in hand one can determine the coefficient $\beta$
\begin{equation}
\beta = \int d M_{s} \, \td{n_s}{M_{s}}{} \int d c_s^{\prime} \, \mathcal{P}(c_s^{\prime}) \mathcal{F}(c_s^{\prime}) \; ,
\end{equation}
such that
\begin{equation}
\mathcal{N}_{\chi}^{sub} (r) \propto p_{sub}(r) \beta \; ,
\end{equation}
is the average sub-halo contribution to the local WIMP-pair density.
The function $\mathcal{P} (c_s)$ is a log-normal distribution in sub-halo concentration parameter
\begin{equation}
\mathcal{P} (c_s) = \frac{1}{\sqrt{2\pi}\sigma_c c_s}\exp{-\left(\frac{\log{c_s} - \log{\bar{c}_s}}{\sqrt{2}\sigma_c}\right)^2} \; ,
\end{equation}
with $\bar{c}_s$ is the average sub-halo concentration from \eqref{eq:cvir} and $\sigma_c = 0.14$~\cite{Bullock2001,pieri2011}. The limits of integration are chosen to ensure the whole Gaussian is captured.
Additionally, we calculate the average concentration parameter of a sub-halo of mass $\widetilde{M}$ as being $\sim 1.5$ times greater than that of an isolated halo of equivalent mass, this is in order to account for the tendency for average sub-halo concentrations to exceed that of the parent halo~\cite{Bullock2001,Colafrancesco2006}. Finally, the function $\mathcal{F}$ is given by
\begin{equation}
\mathcal{F}(x) = 4 \pi (r_s^{\prime})^3\int_0^x d\xi \, \xi^2 \left(\rho^{\prime} (\xi)\right)^2 \; ,
\end{equation}
where $\xi = \frac{r}{r_s^{\prime}}$ and we assume sub-halos have the same form as parent halo, with the sub-halo density profile $\rho^{\prime} (\xi)$ given by
\begin{equation}
\rho^{\prime} (\xi) = \rho_s (c_s) g\left(\xi\right) \; ,
\end{equation}
where $g(r)$ having its previous definition and $r_s^{\prime}$ being the scale-radius determined by the sub-halo mass $M_{sub}$ and concentration $c_s$.

Now we can express the density of WIMP pairs $\mathcal{N}_{\chi} (r)$ as
\begin{equation}
\mathcal{N}_{\chi} (r) = \left(\left(\rho_{tot} (r) - f_s M_{vir} p_{sub} (r)\right)^2 + p_{sub}(r)\beta\right)/2 M_{\chi}^2 \; . \label{eq:pairs}
\end{equation} 
We note that this expression is valid only when considering unresolved substructure and spherically averaged observables~\cite{Colafrancesco2006}.

\subsection{Electron Source Functions From Neutralino Annihilations}
In this paper we assume dark matter to be composed of the lightest neutralino from the minimal supersymmetric extension to the standard model, following the model used by the DarkSUSY package~\cite{darkSUSY}. The source function for the production of a stable particle $i$, produced promptly by neutralino annihilation or ancillary processes is given by
\begin{equation}
Q_i (r,E) = \langle \sigma V\rangle \sum\limits_{f}^{} \td{N^f_i}{E}{} B_f \mathcal{N}_{\chi} (r) \; ,
\end{equation}
where $\langle \sigma V\rangle$ is the thermally-averaged neutralino annihilation cross-section at $0$ K, the index $f$ labels kinematically justified annihilation final states with branching ratios $B_f$ and spectra $\td{N^f_i}{E}{}$,  and $\mathcal{N}_{\chi} (r)$ is the neutralino pair density at a given halo radius $r$.

The chosen particle physics framework will dictate both $\langle \sigma V\rangle$ and the set of branching ratios.
Since the neutralino is a Majorana fermion, light fermion final states are suppressed in favour heavy fermions and Higgs bosons. In keeping with standard procedure in indirect detection studies we will focus on one annihilation channel at a time and assume a branching ratio of $1$ for the channel of interest. In this study we have focussed on the $b\overline{b}$ annihilation channel, but have also considered $W^+W^-$ and $\tau^+\tau^-$, and employed the results of the Pythia~\cite{pythia} DarkSUSY MonteCarlo routines in order to determine the spectra $\td{N^f_i}{E}{}$ for electron production from the decay of neutralino annihilation products. Finally, the neutralino pair density will be calculated in accordance with eq.\eqref{eq:pairs}.

\subsection{Magnetic Field and Thermal Plasma Models}

In order to treat the diffusion of WIMP annihilation products, as well as resulting synchrotron radiation, we must account for the presence of both magnetic fields and a thermal plasma within host halos. 

For the thermal plasma density we assume it follows a radial scaling of the form
\begin{equation}
n(r) = n_0 \left(1 + \left(\frac{r}{r_s}\right)^2\right)^{-q_e} \; ,
\end{equation}
with the central value $n_0$ and the scaling exponent $q_e$ being specified as parameters. Then $\overline{n}$ is simply the volume average of $n(r)$ assuming spherical symmetry.

In the case of the magnetic field, we then assume it is subject to Kolmogorov-type turbulence, in accordance with previous work on the diffusion of secondary electrons from WIMP annihilation~\cite{Colafrancesco2006,Colafrancesco2007}. In this case the power spectrum is thus given by~\cite{sreenivasan1997,Colafrancesco1998}
\begin{equation}
P(k) = P(k_0) \left(\frac{k}{k_0}\right)^{-\delta} \; ,
\end{equation}
where $\delta$ is the spectral index of the turbulence, $k = \frac{1}{d}$ where $d$ is a uniformity scale for the magnetic field. Setting $\delta = \frac{5}{3}$ yields the spectrum of Kolmogorov turbulence. We assume that the smallest length-scale at which the field is uniform $d_0 = 20$ kpc.

We will examine three different magnetic field models here. The first being a constant field, while the second model is one where the magnitude decreases with radius, mimicking the profile of the thermal plasma
\begin{equation}
B (r) = B_0 \left(1 + \left(\frac{r}{r_s}\right)^2\right)^{-q_b} \; ,
\label{eq:mag-therm}
\end{equation} 
where $B_0$ is the central magnetic field strength and $q_b$ is scaling exponent chosen to be $0.5$. 
Our final magnetic field model will be an attempt to obtain a self-consistent description of the spatial shape of the magnetic field within a galaxy cluster. To do this we take into account the results of several experiments, simulations and analytical calculations, according to which a magnetic field can be written as $\vec{B}=\vec{B_0}+\vec{b}$, where $\vec{B_0}$ is the mean magnetic field (i.e. the smooth component), and $\vec{b}$ is representing the fluctuations component. It has been shown (see, e.g., \cite{Verma99} and references therein) that, while the power spectrum of the fluctuating component is expected to have a Kolmogorov shape (i.e. $\delta = \frac{5}{3}$), the smooth component is expected to have a spectrum of the type 
\begin{equation}
\tilde{B_0}(k)\propto k^{-w} \;\;\;\; \mbox{ for } \;\; k_{min}\le k \le k_{max},
\label{smooth_b_spectrum}
\end{equation}
with $w\sim 1/3$. We can derive the radial shape of the magnetic field $B(r)$ by taking the inverse Fourier transform of eq.(\ref{smooth_b_spectrum}), and by assuming $k_{min}=R_{max}^{-1}$, where $R_{max} = R_{vir}$ is the maximum radius of the halo, and $k_{max}=R_{min}^{-1}$, where $R_{min}$ is the minimum radius after which the magnetic field can be considered smooth, while under this radius the fluctuations component is dominant (so, it corresponds to the $d_0$ length introduced before). The normalization of the spectrum in eq.\eqref{smooth_b_spectrum} is linked to the normalization of the magnetic field (as the effect of the Parseval's theorem, see also \cite{Colafrancesco1998}) and to the values of $k_{min}$ and $k_{max}$. For this reason, since we have the possibility to choose the value of the mean magnetic field from other considerations (i.e., the results of hydrodynamical simulations or Faraday Rotation measurements in galaxy clusters) we impose either a central or average value on the field and determine the appropriate normalisation thereafter. Thus we derive the required radial shape of the magnetic field, and then will set the normalization of $B_0$ by imposing that energy density of the magnetic field is less than the thermal energy density of the hot gas within the cluster
\begin{equation}
\frac{\langle B^2 \rangle}{8\pi} \le \overline{n} k_b T \; ,
\label{eq:norm_bfield1}
\end{equation}
where CGS units are used, $k_b$ is the Boltzmann constant, $\langle B^2 \rangle$ represents a volume average and $T$ is the intra-cluster medium temperature, or using the virial result~\cite{Shimizu2003}
\begin{equation}
k_b T = \frac{1}{2} \mu_{mp} \frac{G M_{vir}}{R_{vir}} \; , 
\label{eq:norm_bfield2}
\end{equation}
with $G$ being the Newton's constant and $\mu_{mp}$ being the mean molecular mass in the intra-cluster medium.

For a spectrum as in eq.(\ref{smooth_b_spectrum}), with $0<w<1$, the inverse Fourier transform is proportional to $r^{w-1}$, and so 
we obtain that the radial shape of the magnetic field is given by
\begin{equation}
B(r)=B_0 \left(\frac{r}{R_{min}}\right)^{w-1} \;\;\; \mbox{ for } \; R_{min} \le r \le R_{max},
\label{eq:mag-norm}
\end{equation}
and the normalization is fixed by imposing the condition in eq.\eqref{eq:norm_bfield1} and eq.\eqref{eq:norm_bfield2}. For $r < R_{min}$, the magnetic field is dominated by the fluctuations; so we can consider the smooth component as a constant in this region ($B(r)\sim B_0$), superimposed to which the fluctuations can increase the synchrotron emission~\cite{Colafrancesco2005}. For $r > R_{max}$, the magnetic field is assumed to decrease quickly to zero. The normalization in eq.\eqref{eq:norm_bfield2} results in a condition, for a general value of $0<w<1$, given, where $R_{max}>>R_{min}$, by: 
\begin{equation}
B_0^2 \le \frac{8 \pi}{3} (2w+1) \left( \frac{R_{vir}}{R_{min}} \right)^{2-2w} \overline{n} k_b T .
\end{equation}
We note that for $w=1/3$ we obtain a radial shape $B(r)\propto r^{-2/3}$, which is flatter than the one usually found in the literature for the thermal gas in galaxy clusters; this is in agreement with the results obtained from studies of Farady Rotation measurements in galaxy clusters (see, e.g., \cite{Bonafede2010} for the Coma cluster case). We note also that \cite{Rave2013} obtained in several galaxy clusters an almost flat profile for the magnetic field until large radii ($\sim 7 r_s$); however, we remind that the same authors pointed out that in their paper there is an error in the analysis of sources that may have affected the conclusions relative to the homogeneous component of the magnetic field.

\begin{figure}[htbp]
\centering
\includegraphics[scale=0.6]{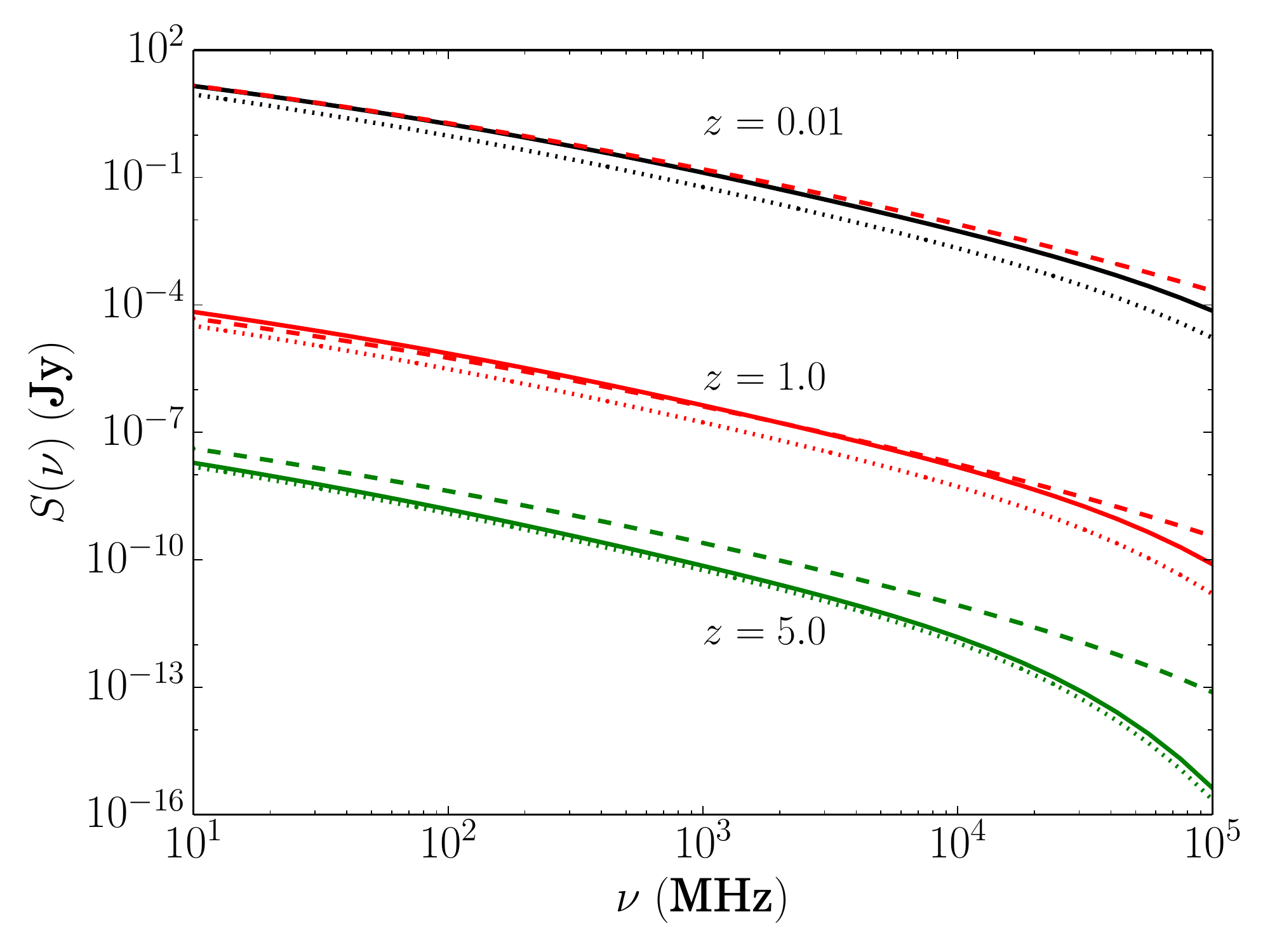}
\caption{The effect of various magnetic field models on the radio flux, calculated according to eq.\eqref{eq:flux}, for a galaxy cluster halo ($M = 10^{15}$ M$_{\odot}$). The solid lines are constant magnetic fields with magnitude 5 $\mu$G, dotted lines have central strength 5 $\mu$G but scale according to the thermal model in eq.\eqref{eq:mag-therm} and the dashed lines represent the model in eq.\eqref{eq:mag-norm} normalised to 0.1 of the thermal energy in the cluster at redshift zero. Black lines correspond to $z = 0.01$, red lines to $z = 1$ and green lines to $z = 5$.}
\label{fig:bfields}
\end{figure}

It is clear from Figure~\ref{fig:bfields}, where we show the variance of the radio flux with magnetic field model, that each of the magnetic field models yields results similar to the constant field case, at most accounting for a difference of less then an order of magnitude at $\nu \approx 100$ Ghz and at redshift $z = 5$ in the case of the power-law model of eq.\eqref{eq:mag-norm}. However, this is largely due to a lack of a good model for the evolution of the B-field normalisation with redshift. Indeed the most notable differences incurred by eq.\eqref{eq:mag-norm} occur only at energies beyond the range of frequencies observable by the SKA and thus are not of importance in the application examined here. Thus it is adequate for the purposes of modelling detection by the SKA to employ constant field models. The reason for the similarity in the predictions of these models is down to their behaviour within the core radius $r_s$, as the high electron density within this region is responsible for most of the observed synchrotron emission and within this region all the field models predict similar average magnetic field strengths, the weak radial dependence of eq. \eqref{eq:mag-norm} ensures that its radial variation in this region is not large enough to impose significant differences although the increased high-frequency flux is caused by it predicting larger values for the field near the centre of the cluster. An additional consideration is the behaviour of synchrotron radiation with increasing B-field strength, where the output power increases until a saturation point, with the rate of power gain being inversely proportional to rate of increase in field strength as is evident in Figure~\ref{fig:sigv_b} (presented in section 3 below).
For each halo examined we will consider two different magnitudes for the constant magnetic field strength. The first has $\overline{B} = 5$ $\mu$G while the second is given by $\overline{B} = 1$ $\mu$G. For the thermal plasma we will assume that values of $n_0$ scale with the mass of the halo, using values of $n_0 = 10^{-3}$ cm$^{-3}$ for large halos ($M_{vir} \ge 10^{14} M_\odot$) and $n_0 = 10^{-6}$ cm$^{-3}$ for smallest halos ($M_{vir} \sim 10^{7} M_\odot$) with intermediate mass halos taking values between these extremes.

\subsection{Diffusion of Secondary Electrons}
\label{sec:diff}

In order to take into account the effects of the magnetic field and thermal plasma on electron diffusion we take average values for the field strength and thermal plasma density, being $\overline{B} \equiv \sqrt{\langle B(r)^2 \rangle}$ and $\overline{n} \equiv \langle n(r) \rangle$, respectively. 
We then define the spatial diffusion coefficient as~\cite{Colafrancesco1998}
\begin{equation}
D(E) = \frac{1}{3}c r_L (E) \frac{\overline{B}^2}{\int^{\infty}_{k_L} dk P(k)} \; ,
\end{equation}
where $r_L$ is the Larmour radius of a relativistic particle with energy $E$ and charge $e$ and $k_L = \frac{1}{r_L}$, and require that
\begin{equation}
\int^{\infty}_{k_0} dk P(k) = \overline{B}^2 \; .
\end{equation}
This leads us to the result that
\begin{equation}
D(E) = D_0 d_0^{\frac{2}{3}} \left(\frac{\overline{B}}{1 \mu\mbox{G}}\right)^{-\frac{1}{3}} \left(\frac{E}{1 \mbox{GeV}}\right)^{\frac{1}{3}}  \; , \label{eq:diff}
\end{equation}
where $D_0 = 3.1\times 10^{28}$ cm$^2$ s$^{-1}$. It is worth noting that the diffusion coefficient is assumed to be lacking radial dependence. While it is possible to implement diffusion without this simplification, we present results here under the assumption we can substitute the averaged value of the magnetic field in the diffusion coefficient as it is evident that the weak radial dependence of the field and the weak dependence of the diffusion coefficient on the field strength imply that our approximation is not unwarranted. 

The diffusion equation for electrons within the halo is then taken to be
\begin{equation}
\begin{aligned}
\pd{}{t}{}\td{n_e}{E}{} = & \; \gv{\nabla} \left( D(E,\v{x})\gv{\nabla}\td{n_e}{E}{}\right) + \pd{}{E}{}\left( b(E,\v{x}) \td{n_e}{E}{}\right)\\& + Q_e(E,\v{x}) \; ,
\end{aligned}
\end{equation}
where $\td{n_e}{E}{}$ is the electron equilibrium spectrum, $D(E,\v{x})$ is the spatial diffusion function, $b(E,\v{x})$ is the energy-loss function and $Q_e(E,\v{x})$ is the electron source function. A detailed analysis of the solution to this equation in the case of electron production via neutralino annihilation can be found in \cite{Colafrancesco2006}. The solution in the case of spherical symmetry and assuming the energy-loss and diffusion functions have no spatial dependence (as in eq.\eqref{eq:diff}) will be of the form
\begin{equation}
\td{n_e}{E}{} (r,E) = \frac{1}{b(E)}  \int_E^{M_\chi} d E^{\prime} \, G(r,E,E^{\prime}) Q_e (r,E^{\prime}) \; ,
\end{equation}
where $G(r,E,E^{\prime})$ is a Green's function which will reduce to unity in the case of larger halos at zero redshift~\cite{Colafrancesco2006}. However, in smaller halos the diffusive effects in $G(r,E,E^{\prime})$ will become significant. Following \cite{Colafrancesco2007} we take $b(E)$ to be
\begin{equation}
\begin{aligned}
b(E) = & b_{IC} E^2 (1+z)^4 + b_{sync} E^2 \overline{B}^2 \\&\; + b_{Coul} \overline{n} (1+z)^3 \left(1 + \frac{1}{75}\log\left(\frac{\gamma}{\overline{n} (1+z)^3}\right)\right) \\&\; + b_{brem} \overline{n} (1+z)^3 \left( \log\left(\frac{\gamma}{\overline{n} (1+z)^3 }\right) + 0.36 \right)
\end{aligned}
\label{eq:loss}
\end{equation}
where $\overline{n}$ is given in cm$^{-3}$ and $b_{IC}$, $b_{synch}$, $b_{col}$, and $b_{brem}$ are the Inverse Compton, synchrotron, Coulomb and Bremsstrahlung energy loss factors, taken to be $0.25$, $0.0254$, $6.13$, and $1.51$ respectively in units of $10^{-16}$ GeV s$^{-1}$. Here $E$ is the energy in GeV and the B-field is in $\mu$G.
\begin{figure}[htbp]
\centering
\includegraphics[scale=0.37]{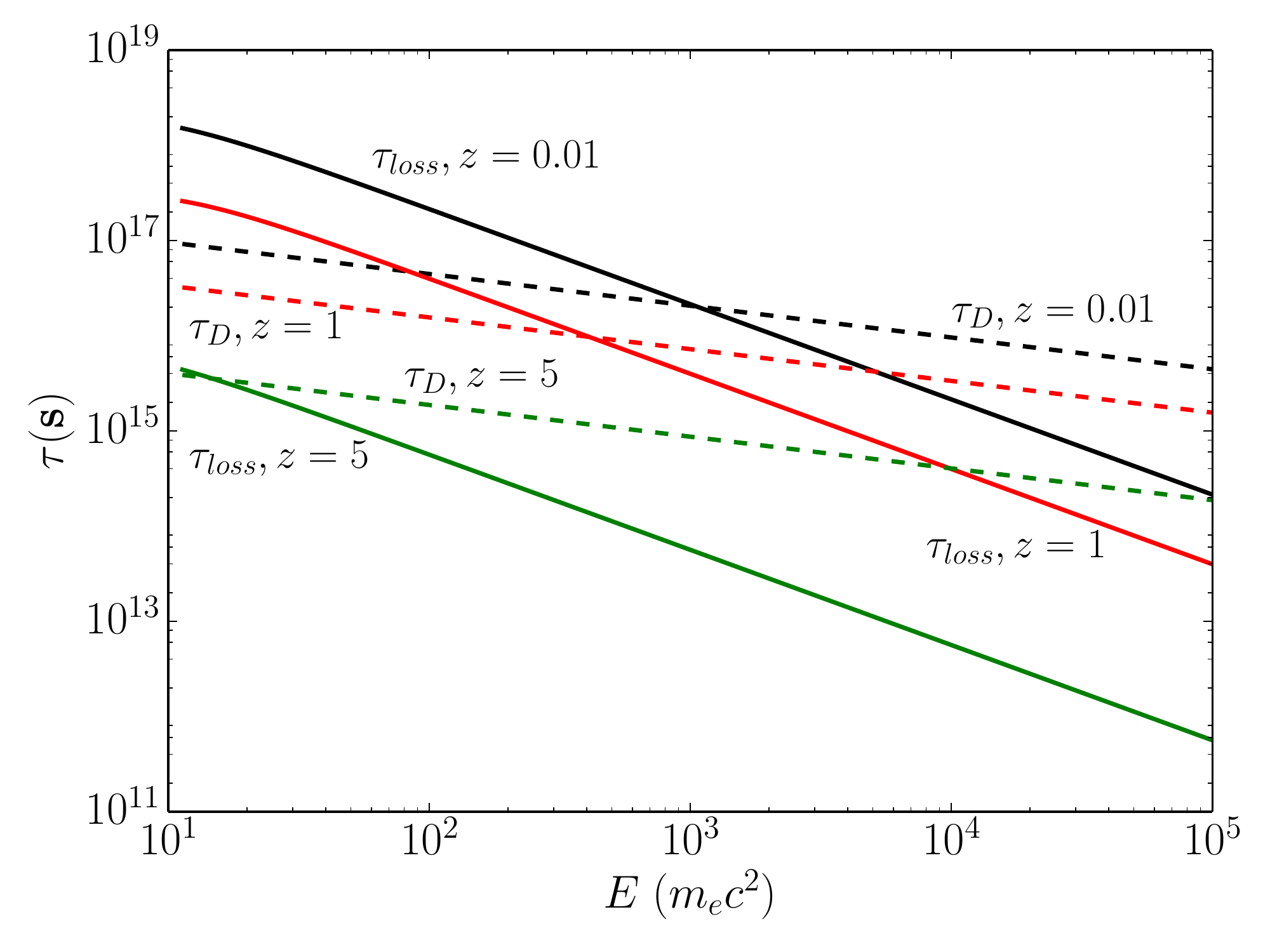}
\includegraphics[scale=0.37]{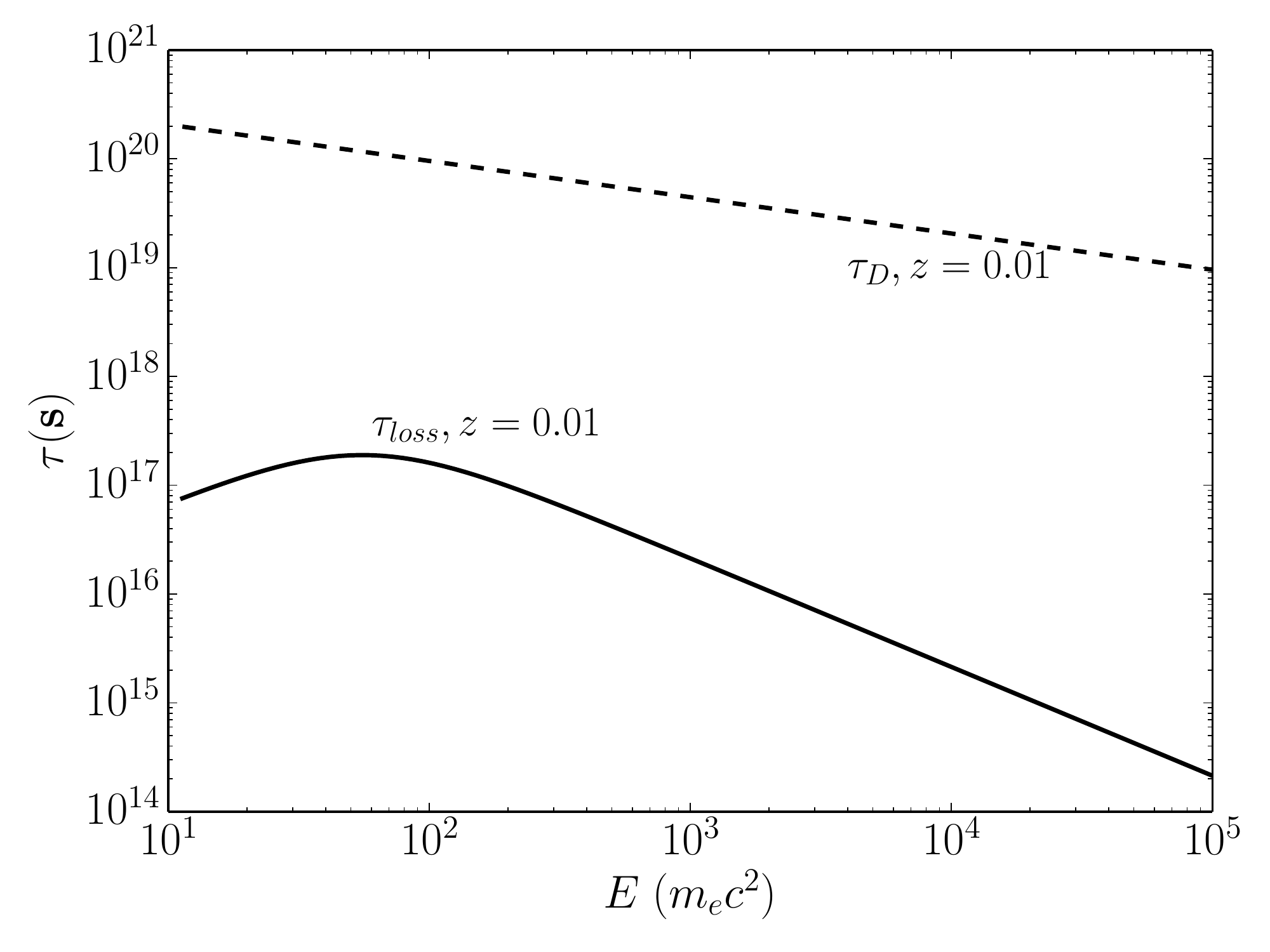}
\caption{Time scales $\tau$ as electron energy $E$ is varied. Left: Halo mass $10^{7}$ M$_{\odot}$, where black lines correspond to $z = 0.01$, red lines to $z = 1.0$ and green lines to $z = 3$. Right: Halo mass $10^{12}$ M$_{\odot}$ at $z = 0.01$. }
\label{fig:loss}
\end{figure}

The time-scales of both diffusive and energy-loss effects are displayed in Figure~\ref{fig:loss}. It is worth noting that as the redshift increases the energy-loss time-scale rapidly decreases, as the inverse-Compton (IC) process comes to dominate energy-loss and to accelerate the loss time-scale to levels that make diffusive effects irrelevant, as these only scale with the dimension of the halo, which shows far weaker redshift dependence than the IC terms. The right-hand panel of Fig.~\ref{fig:loss} shows how drastically insignificant diffusion time-scales become for larger halos. This conclusion very nicely corroborates the assertions made in \cite{Colafrancesco2006} on this matter.

\subsection{Synchrotron Emission}
The average power of the synchrotron radiation at observed frequency $\nu$ emitted by an electron with energy $E$ in a magnetic field with amplitude $B$ is given by~\cite{longair1994}
\begin{equation}
P_{synch} (\nu,E,r,z) = \int_0^\pi d\theta \, \frac{\sin{\theta}^2}{2}2\pi \sqrt{3} r_e m_e c \nu_g F_{synch}\left(\frac{\kappa}{\sin{\theta}}\right) \; ,
\label{eq:power}
\end{equation}
where $m_e$ is the electron mass, $\nu_g = \frac{e B}{2\pi m_e c}$ is the non-relativistic gyro-frequency, $r_e = \frac{e^2}{m_e c^2}$ is the classical electron radius, and the quantities $\kappa$ and $F_{synch}$ are defined as
\begin{equation}
\kappa = \frac{2\nu (1+z)}{3\nu_0 \gamma^2}\left(1 +\left(\frac{\gamma \nu_p}{\nu (1+z)}\right)^2\right)^{\frac{3}{2}} \; ,
\end{equation}
and
\begin{equation}
F_{synch}(x) = x \int_x^{\infty} dy \, K_{5/3}(y) \simeq 1.25 x^{\frac{1}{3}} \mbox{e}^{-x} \left(648 + x^2\right)^{\frac{1}{12}} \; .
\end{equation}

The local synchrotron emissivity can then be found as a function of the electron and positron equilibrium distributions as well as the synchrotron power
\begin{equation}
j_{synch} (\nu,r,z) = \int_{m_e}^{M_\chi} dE \, \left(\td{n_{e^-}}{E}{} + \td{n_{e^+}}{E}{}\right) P_{synch} (\nu,E,r,z) \; .
\label{eq:emm}
\end{equation}
Once one has determined this quantity, it can be used to find all of the observables relevant to our study, namely the flux densities and surface brightnesses of the dark matter halo. The flux density spectrum within a radius $r$ is found via 
\begin{equation}
S_{synch} (\nu,z) = \int_0^r d^3r^{\prime} \, \frac{j_{synch}(\nu,r^{\prime},z)}{4 \pi D_L^2} \; ,
\label{eq:flux} 
\end{equation}
where $D_L$ is the luminosity distance to the halo. Then the azimuthally averaged surface brightness is given by
\begin{equation}
I_{synch} (\nu,\Theta,\Delta\Omega,z) = \int_{\Delta\Omega} d\Omega \, \int_{l.o.s} dl \, \frac{j_{synch}(\nu,l,z)}{4 \pi} \; , 
\label{eq:sb}
\end{equation}
where the integrals are performed over a cone given by the solid angle $\Delta\Omega$ and an axis along the line of sight (l.o.s) which makes an angle $\Theta$ with the direction of the halo centre.

\section{Results}

Throughout this work the SKA sensitivity is drawn from \cite{ska2012} and represents the SKA MID and LOW frequency ranges for the SKA phase 1.

\begin{figure}[htbp]
\centering
\includegraphics[scale=0.37]{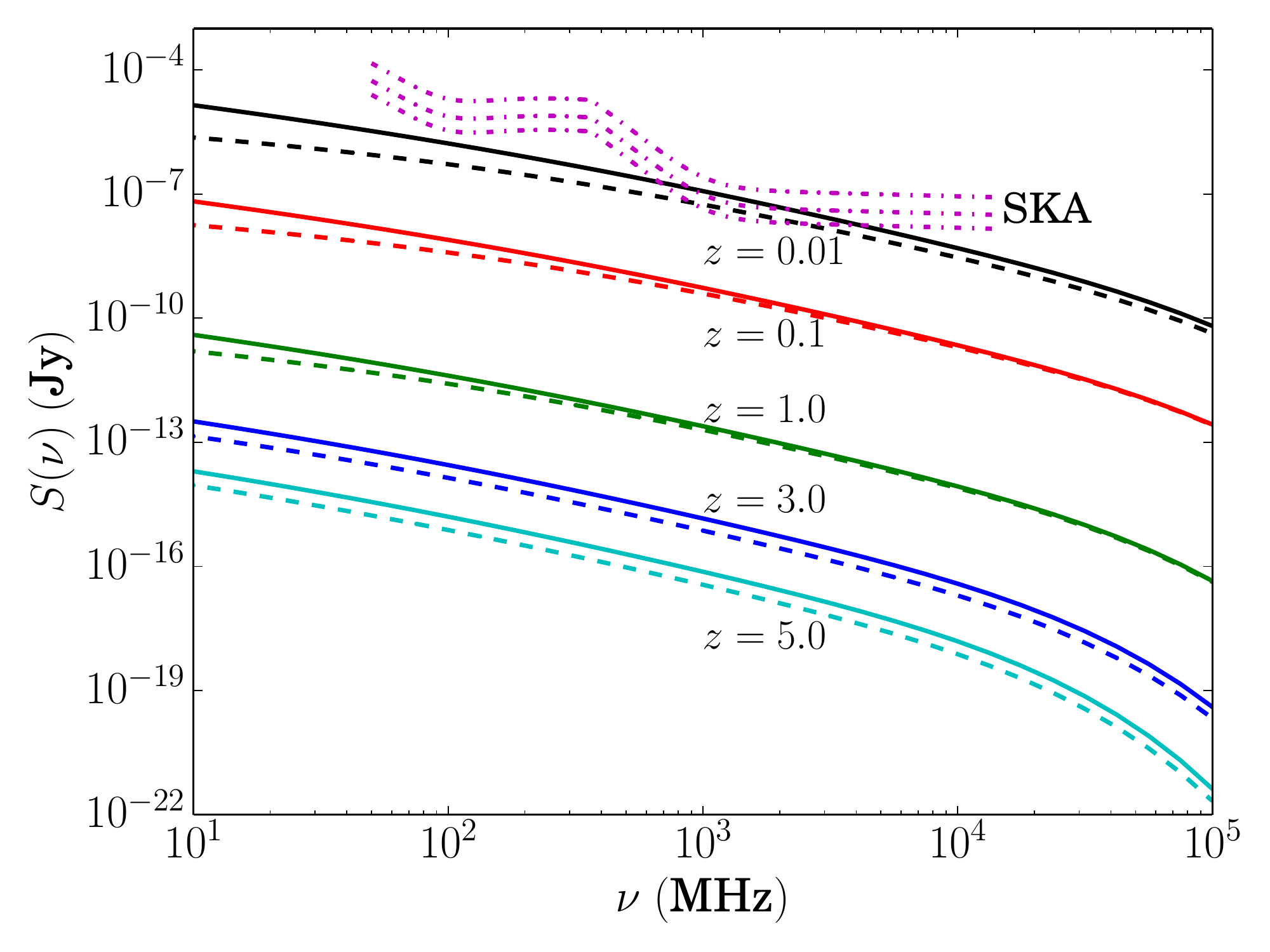}
\includegraphics[scale=0.37]{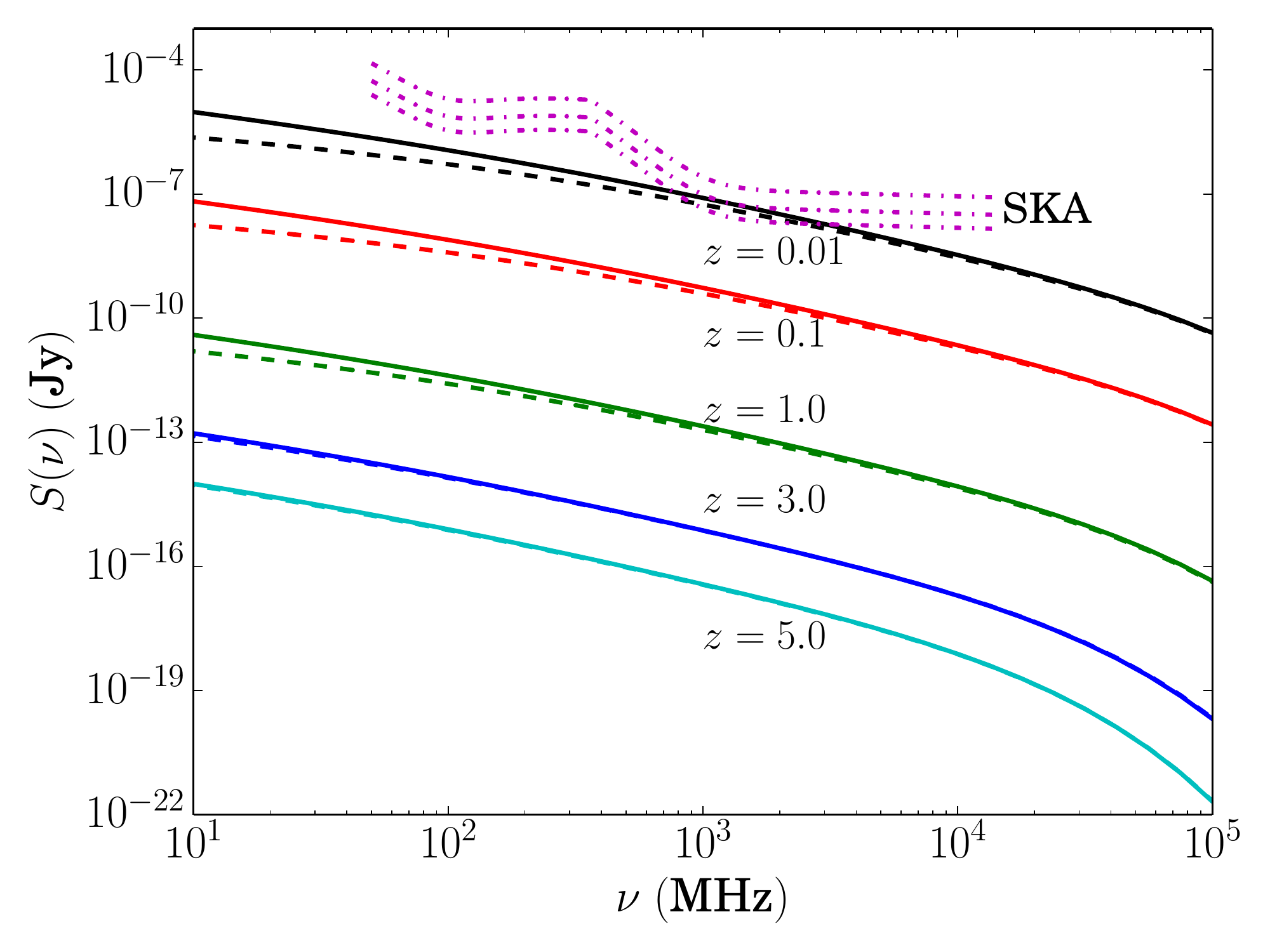}
\caption{Flux densities for dwarf spheroidal galaxies ($M = 10^{7}$M$_{\odot}$), the halo profile is NFW and $\langle B \rangle = 5$ $\mu$G. WIMP mass is $60$ GeV and the composition is $b\overline{b}$. Black lines represent $z = 0.01$, Red signifies $z = 0.1$, Green $z = 1$, Blue $z = 3$, and Cyan $z = 5$. Solid lines are without diffusion, dotted are with diffusion. The pink dash-dotted lines are SKA 1$\sigma$ sensitivity limits for 30, 240, and 1000 hour integration times, respectively. The first (left) plot is the flux within an arcmin of the halo centre and the second (right) is within the virial radius.}
\label{fig:bb60_m7_b5_nfw}
\end{figure}

\begin{figure}[htbp]
\centering
\includegraphics[scale=0.37]{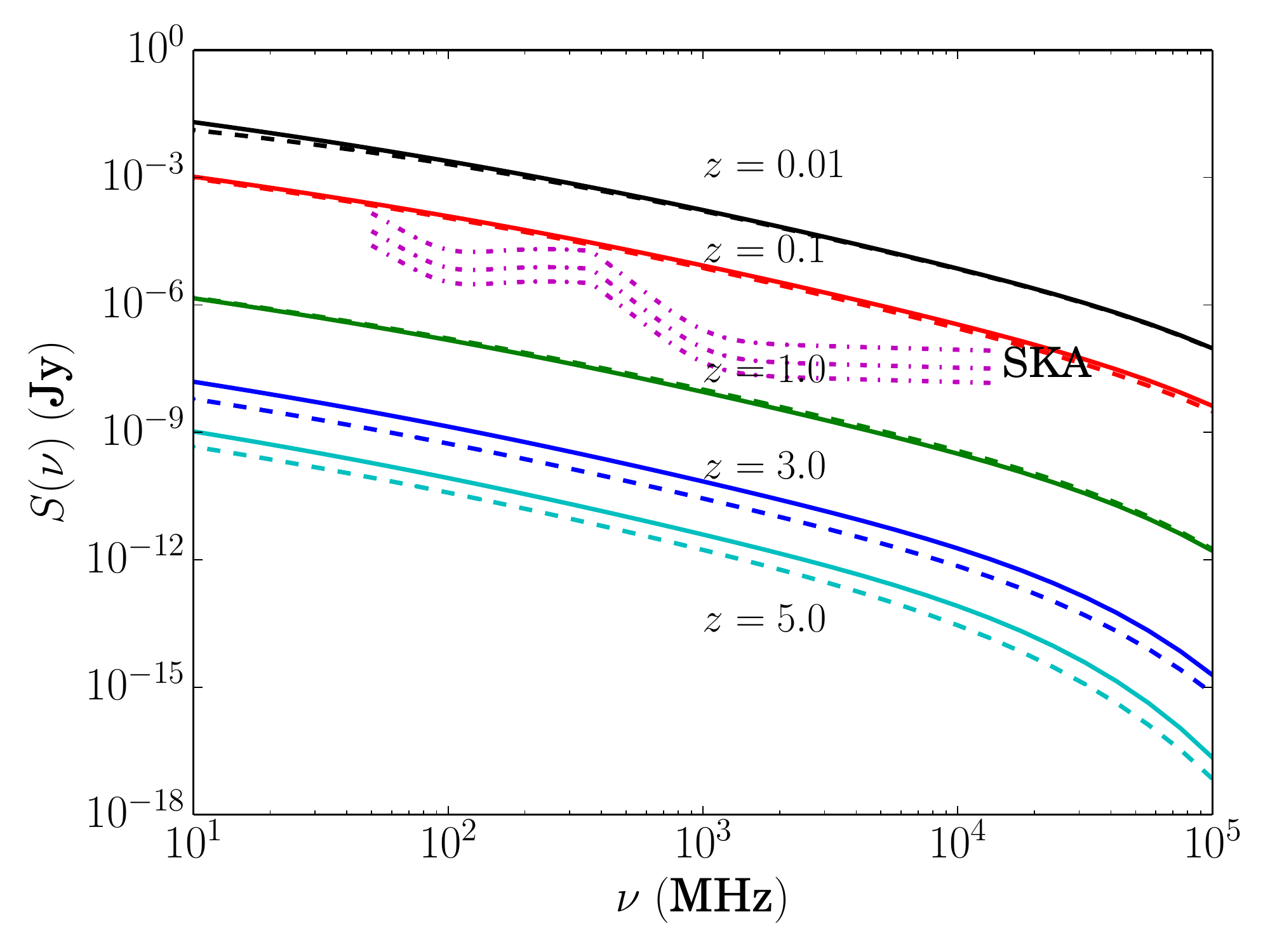}
\includegraphics[scale=0.37]{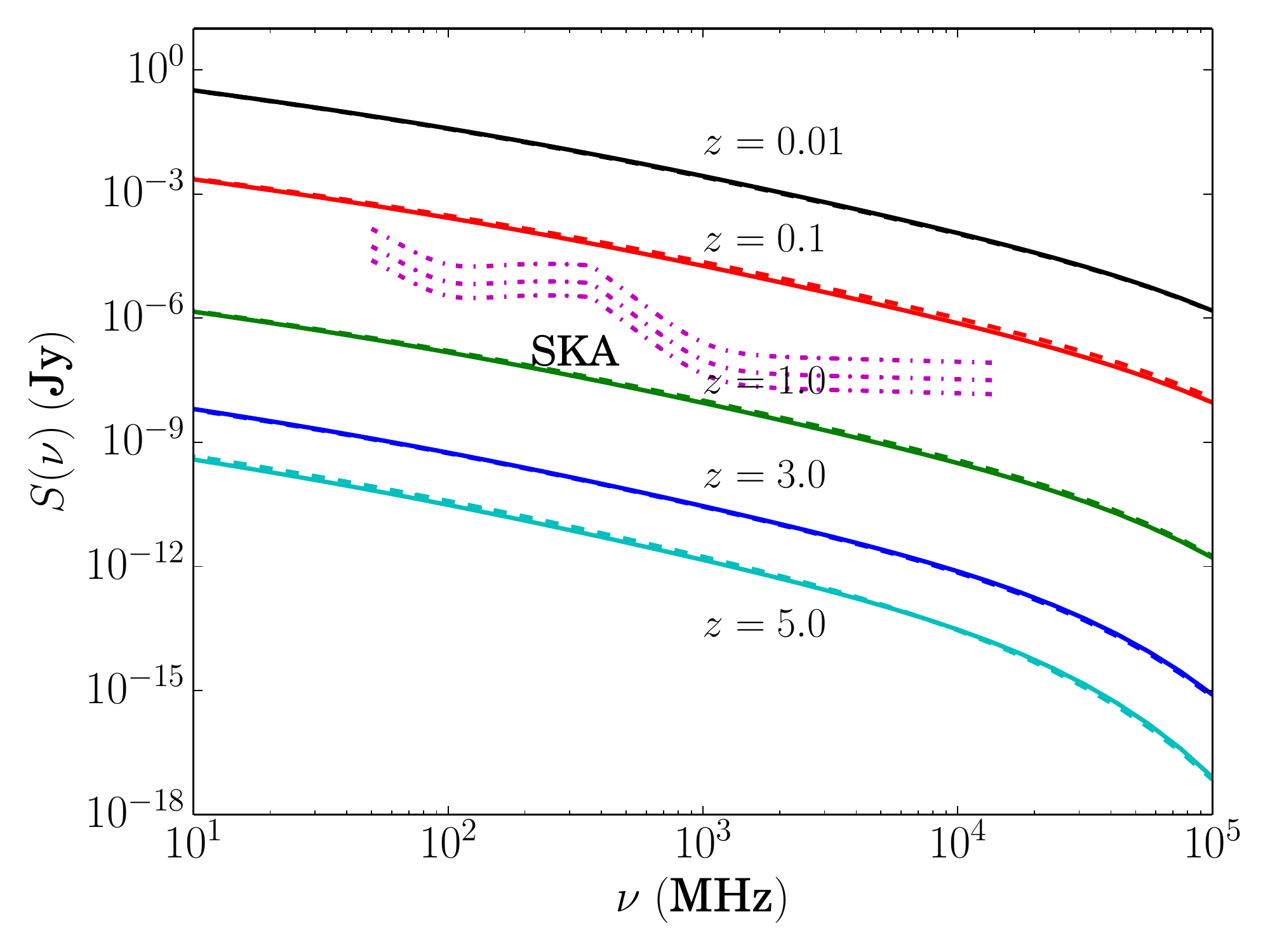}
\caption{Flux densities for galaxies ($M = 10^{12}$M$_{\odot}$), the halo profile is NFW and $\langle B \rangle = 5$ $\mu$G. WIMP mass is $60$ GeV and the composition is $b\overline{b}$. See the caption of Figure~\ref{fig:bb60_m7_b5_nfw} for legend.}
\label{fig:bb60_m12_b5_nfw}
\end{figure}

\begin{figure}[htbp]
\centering
\includegraphics[scale=0.37]{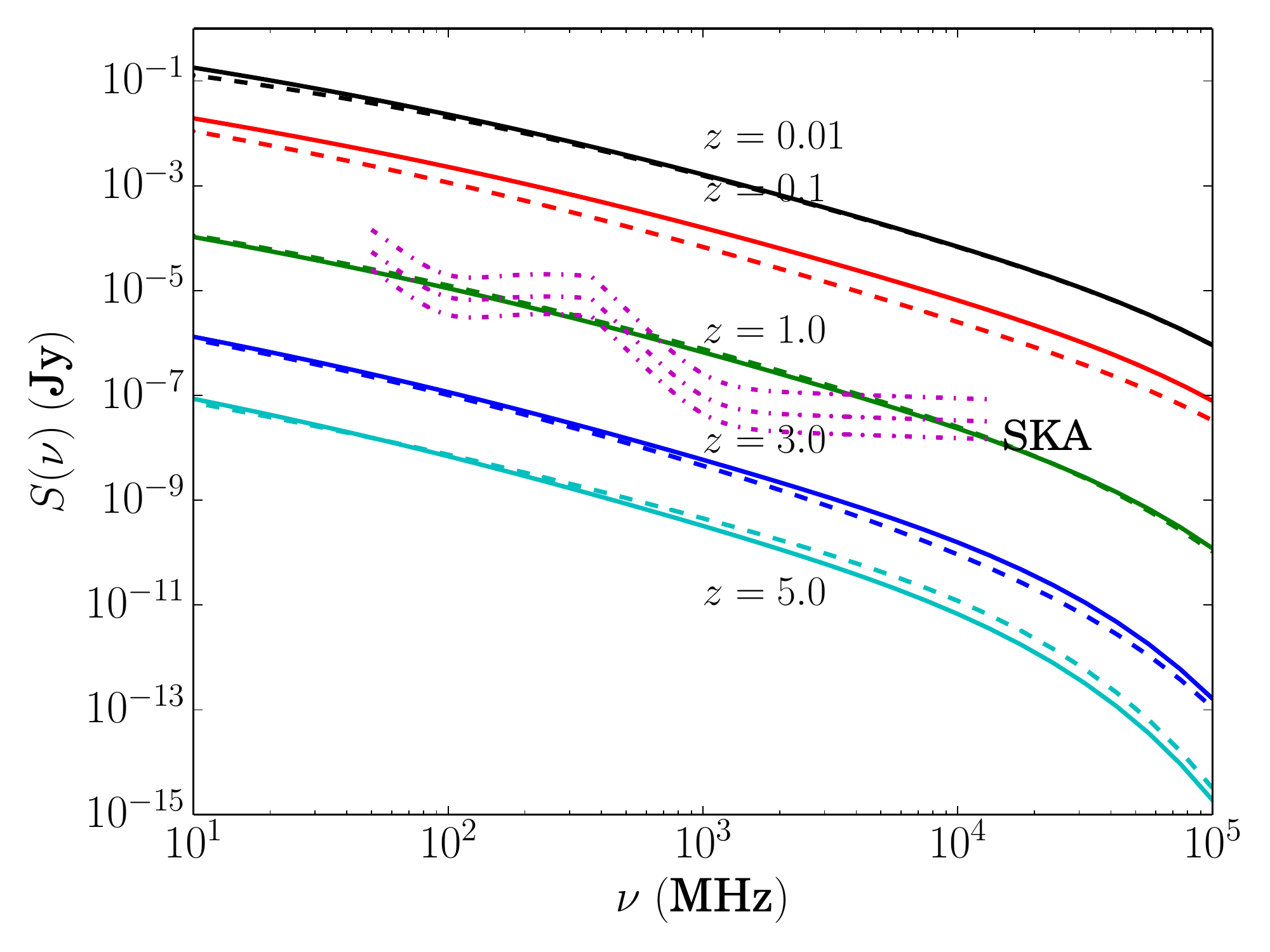}
\includegraphics[scale=0.37]{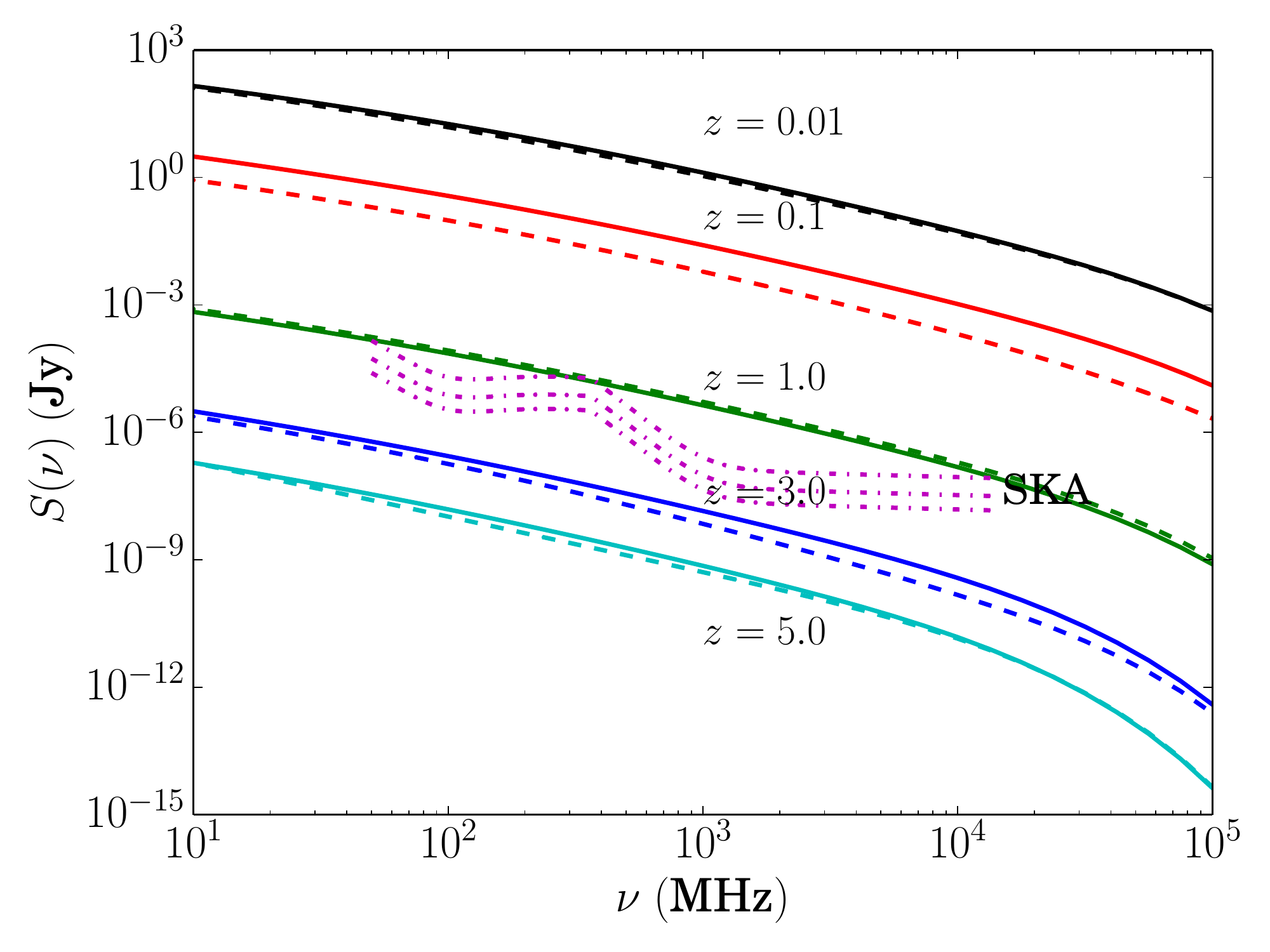}
\caption{Flux densities for galaxy clusters ($M = 10^{15}$M$_{\odot}$), the halo profile is NFW and $\langle B \rangle = 5$ $\mu$G. WIMP mass is $60$ GeV and the composition is $b\overline{b}$. See the caption of Figure~\ref{fig:bb60_m7_b5_nfw} for legend.}
\label{fig:bb60_m15_b5_nfw}
\end{figure}

\begin{figure}[htbp]
\centering
\includegraphics[scale=0.37]{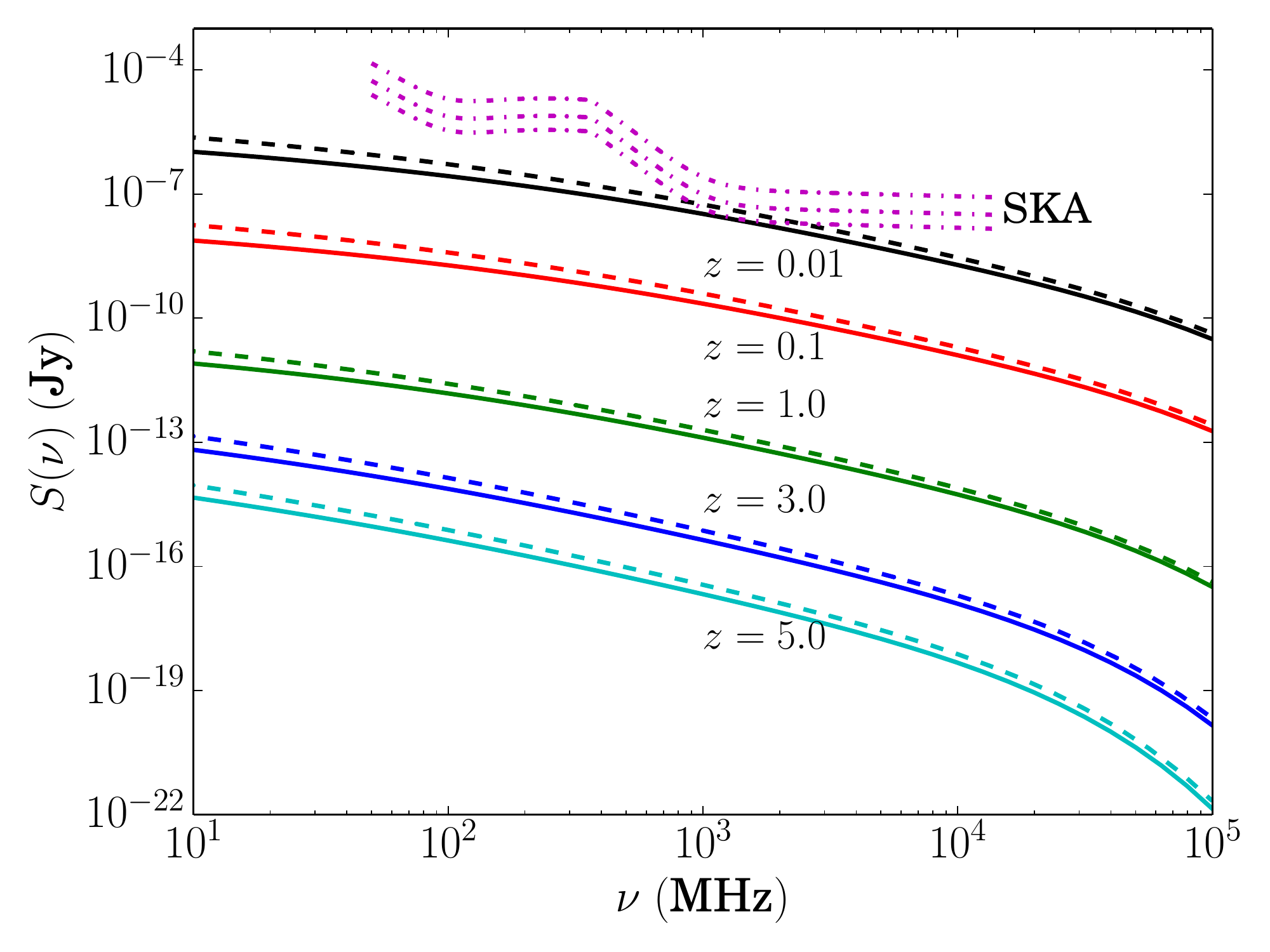}
\includegraphics[scale=0.37]{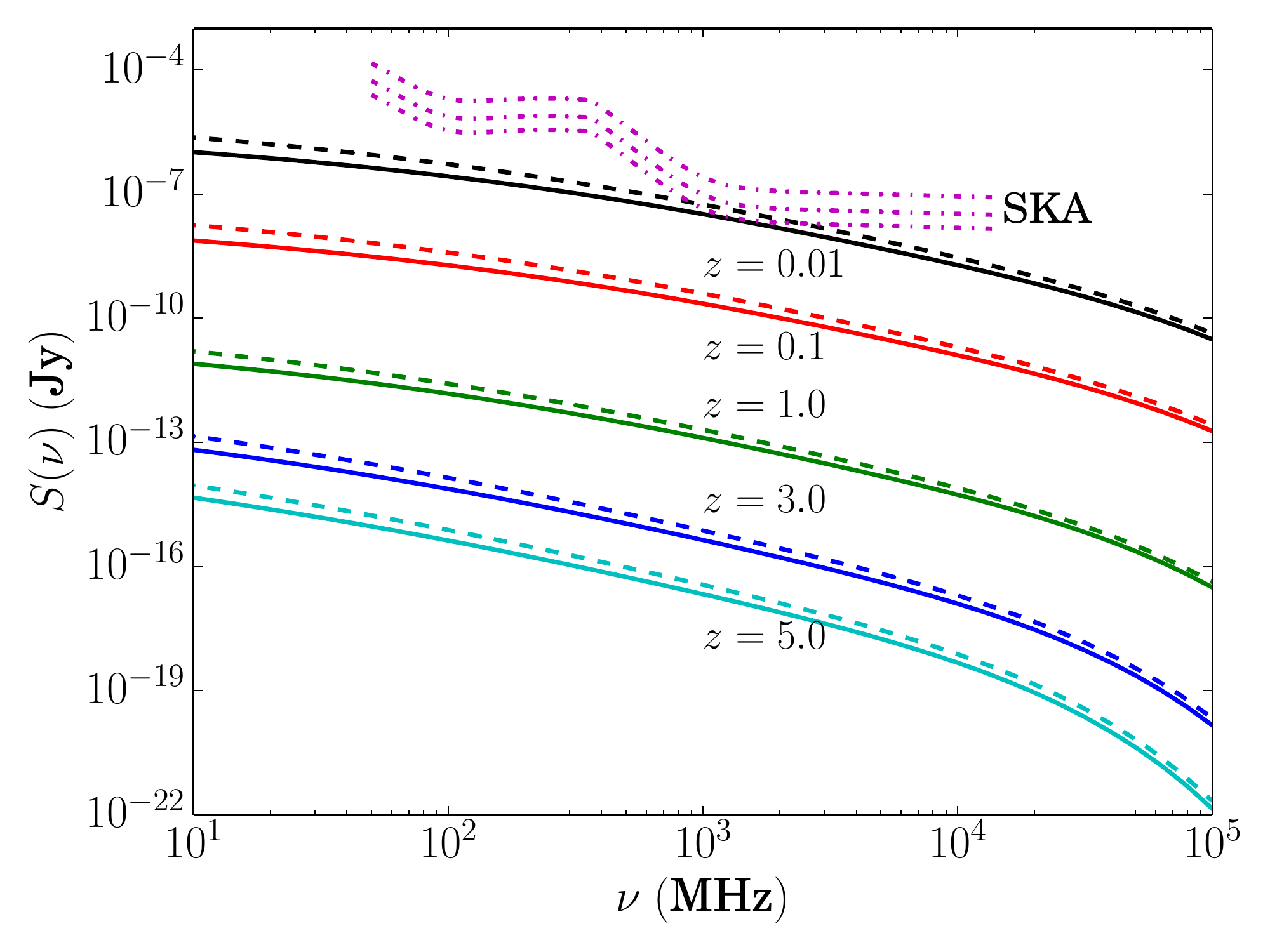}
\caption{Flux densities for dwarf spheroidal galaxies ($M = 10^{7}$M$_{\odot}$), comparing halo profiles NFW (solid lines) and Burkert (dashed lines) with $\langle B \rangle = 5$ $\mu$G. WIMP mass is $60$ GeV and the composition is $b\overline{b}$. See the caption of Figure~\ref{fig:bb60_m7_b5_nfw} for legend (all plots include diffusion).}
\label{fig:bb60_m7_b5_burkert}
\end{figure}

\begin{figure}[htbp]
\centering
\includegraphics[scale=0.37]{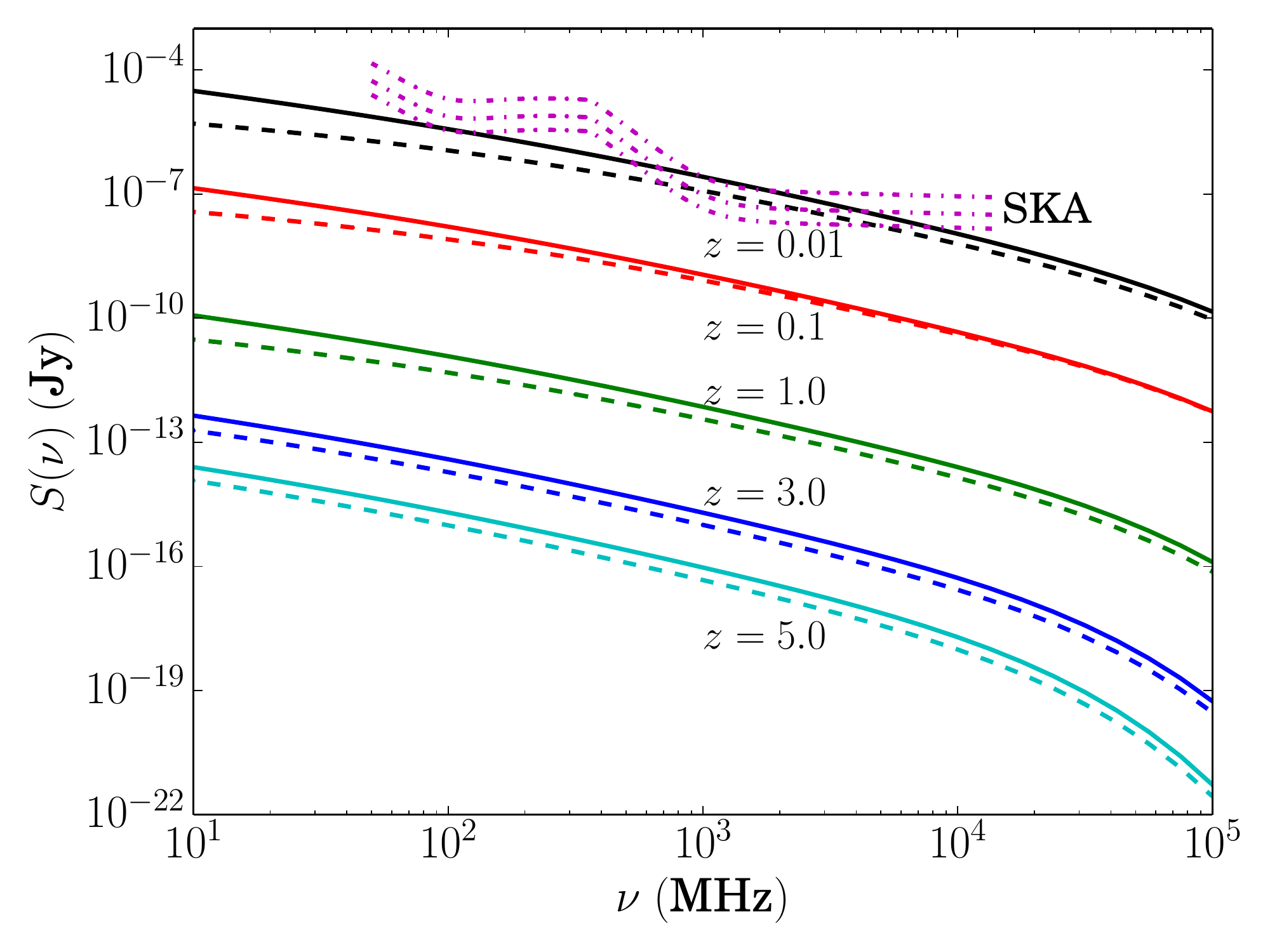}
\includegraphics[scale=0.37]{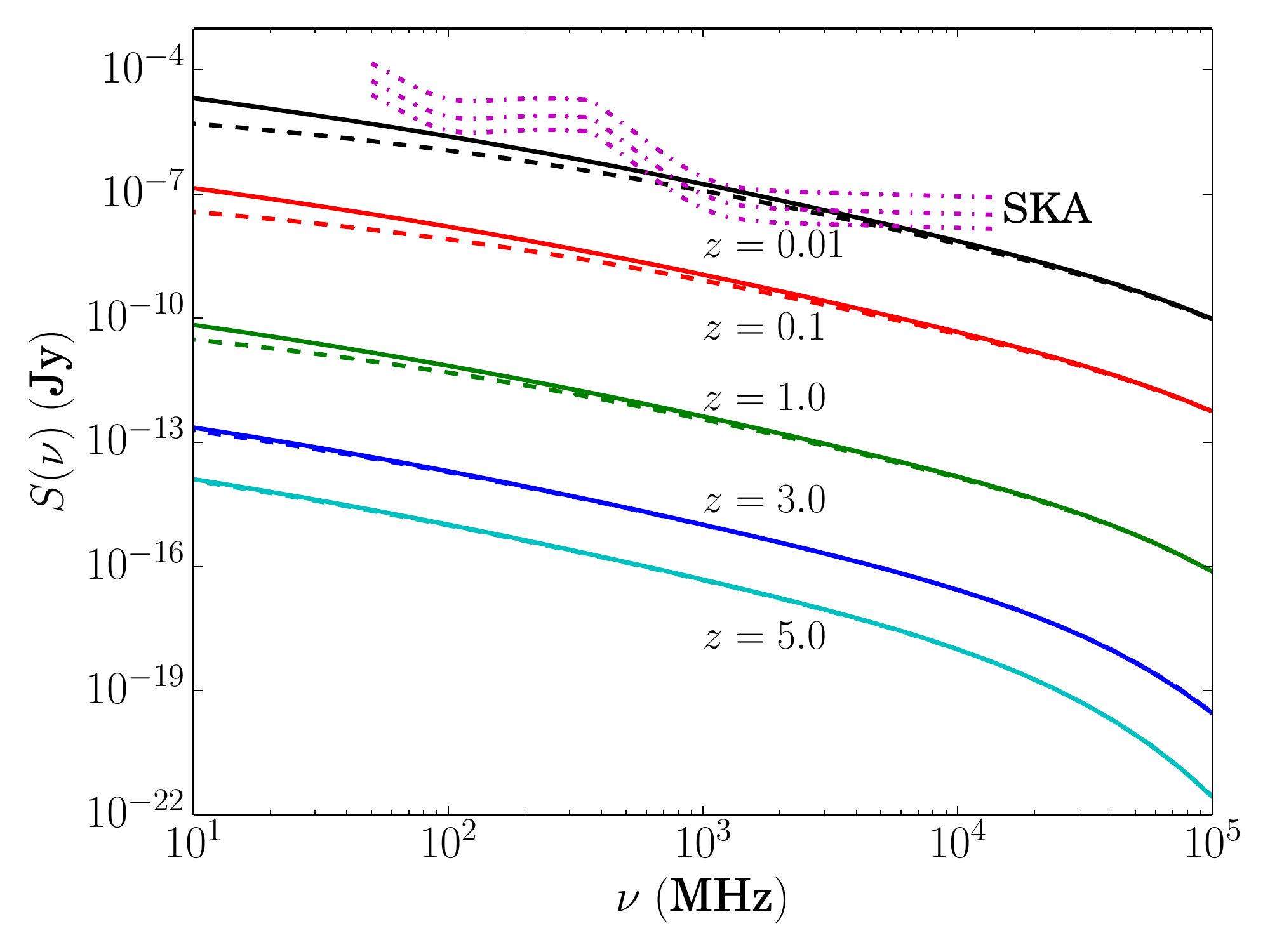}
\caption{Flux densities for dwarf spheroidal galaxies ($M = 10^{7}$M$_{\odot}$), the halo profile is Einasto ($\alpha = 0.18$) and $\langle B \rangle = 5$ $\mu$G. WIMP mass is $60$ GeV and the composition is $b\overline{b}$. See the caption of Figure~\ref{fig:bb60_m7_b5_nfw} for legend.}
\label{fig:bb60_m7_b5_ein}
\end{figure}

\begin{figure}[htbp]
\centering
\includegraphics[scale=0.37]{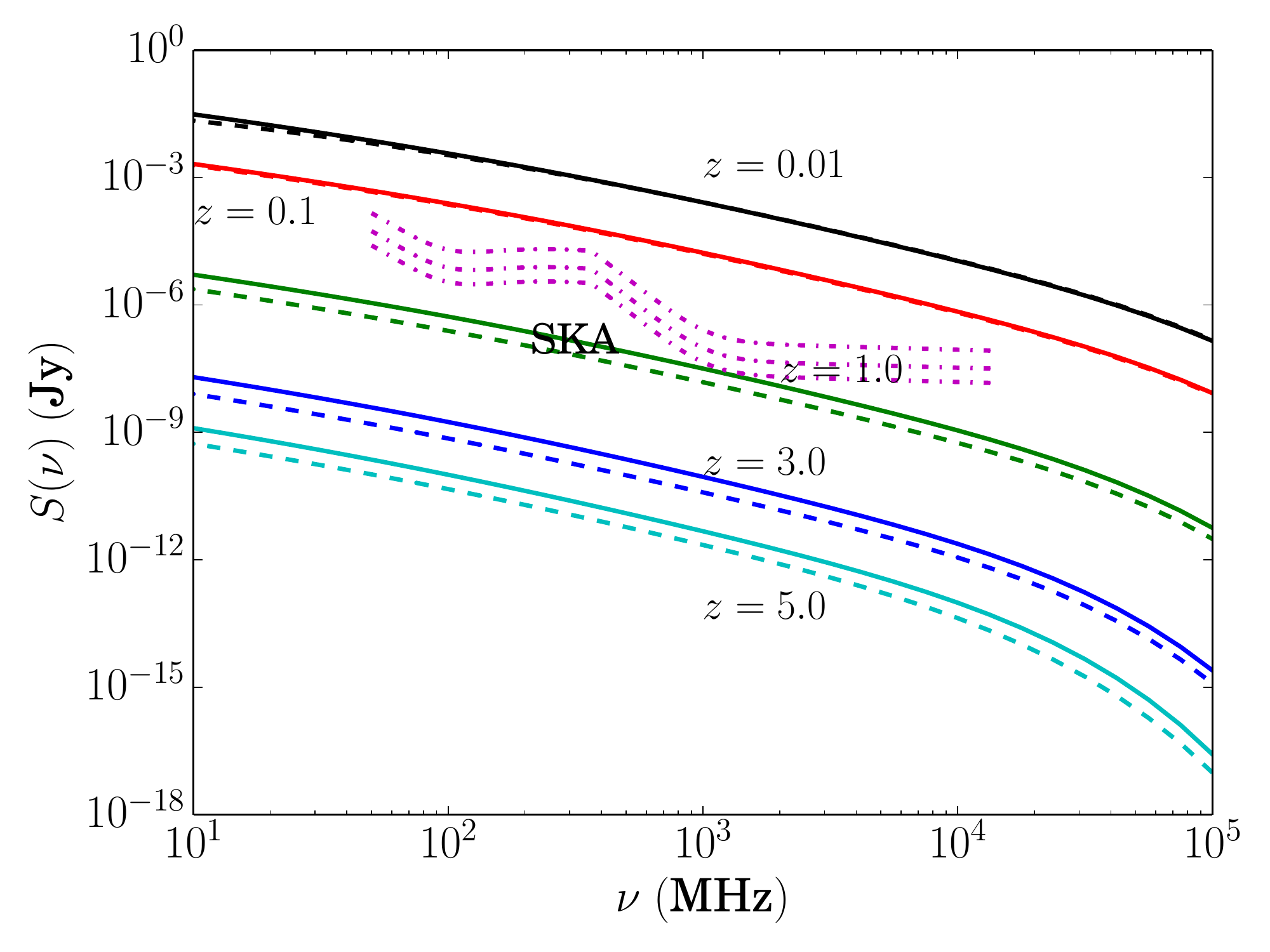}
\includegraphics[scale=0.37]{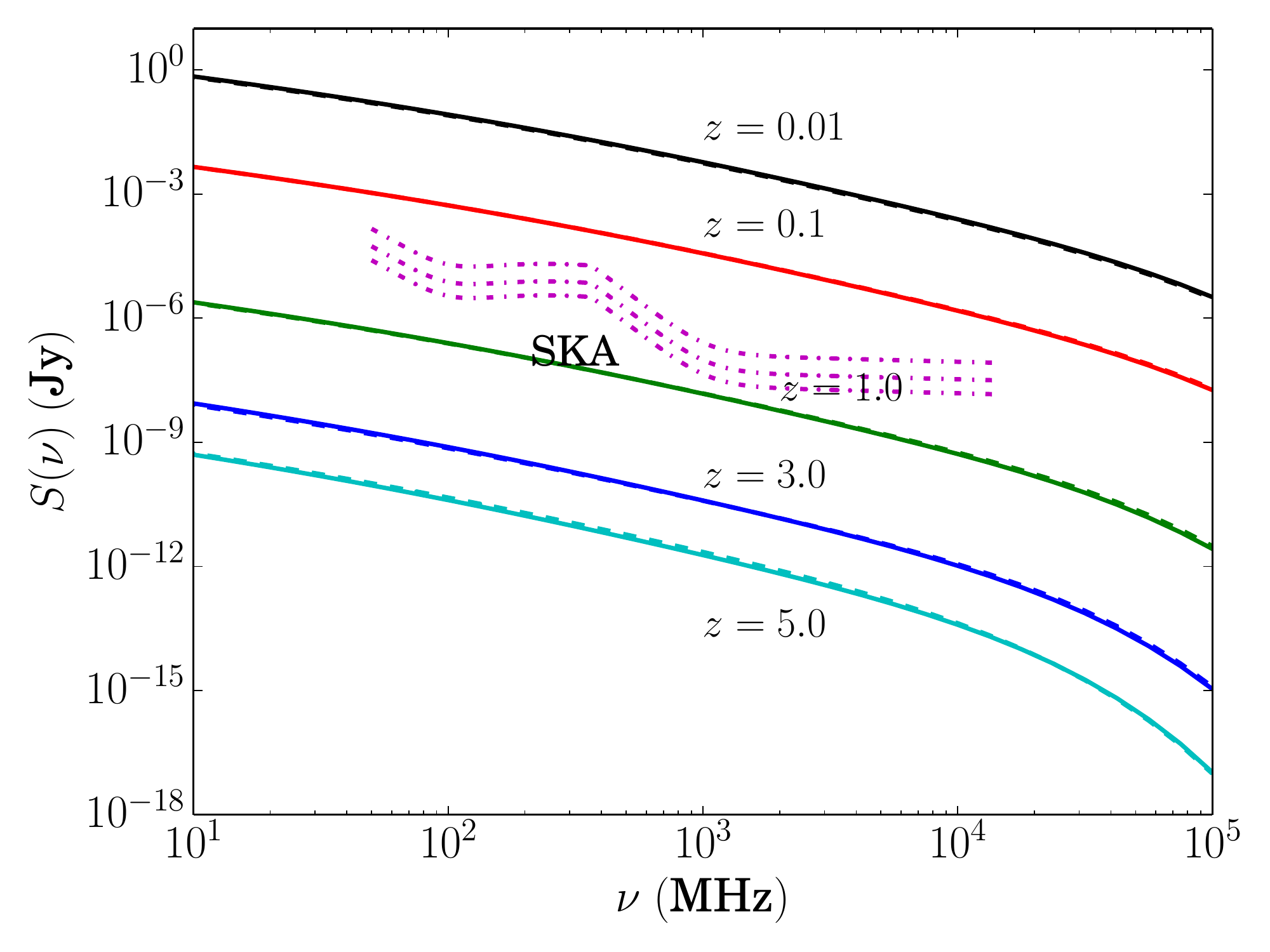}
\caption{Flux densities for galaxies ($M = 10^{12}$M$_{\odot}$), the halo profile is Einasto ($\alpha = 0.18$) and $\langle B \rangle = 5$ $\mu$G. WIMP mass is $60$ GeV and the composition is $b\overline{b}$. See the caption of Figure~\ref{fig:bb60_m7_b5_nfw} for legend.}
\label{fig:bb60_m12_b5_ein}
\end{figure}

\begin{figure}[htbp]
\centering
\includegraphics[scale=0.37]{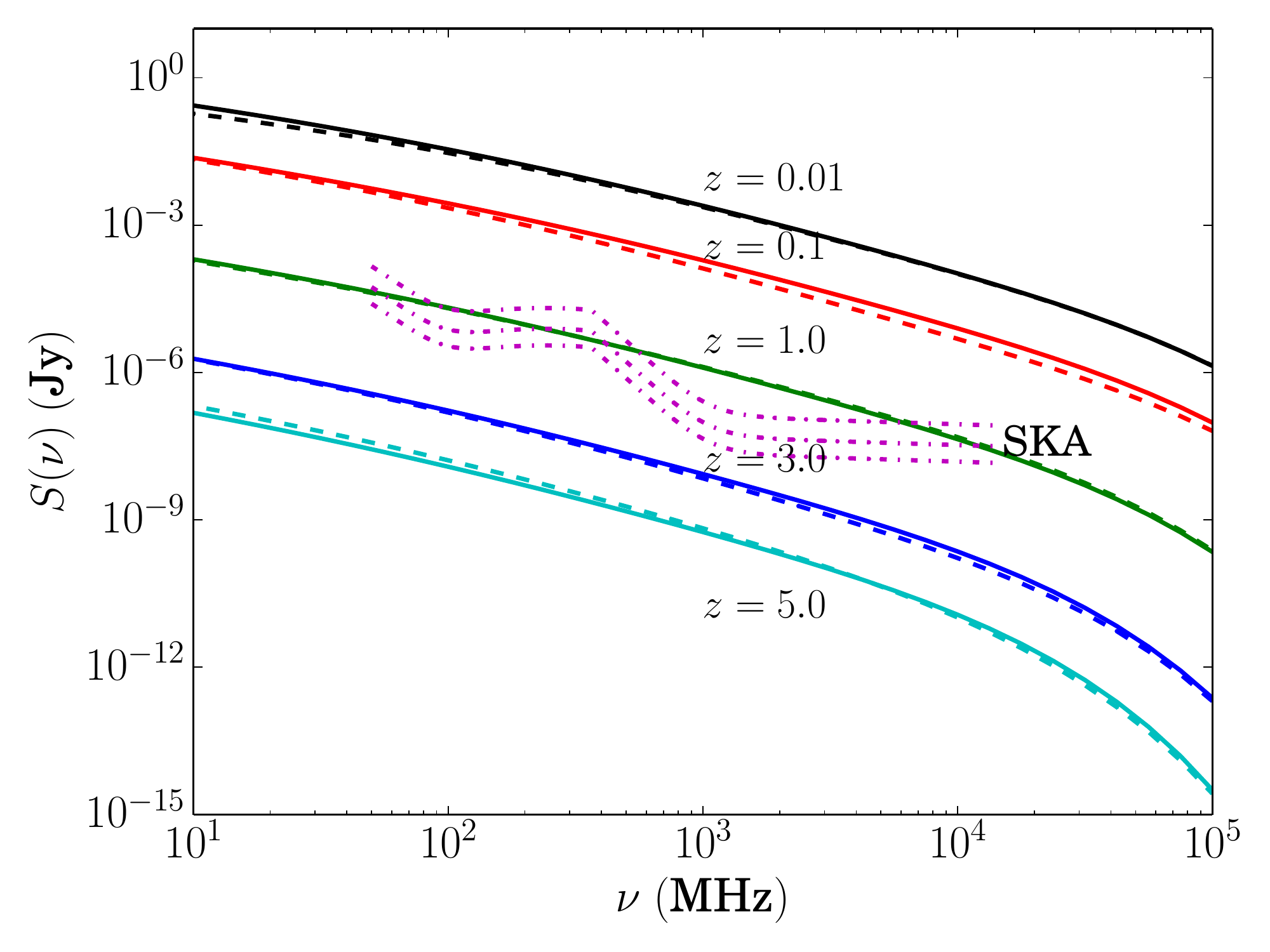}
\includegraphics[scale=0.37]{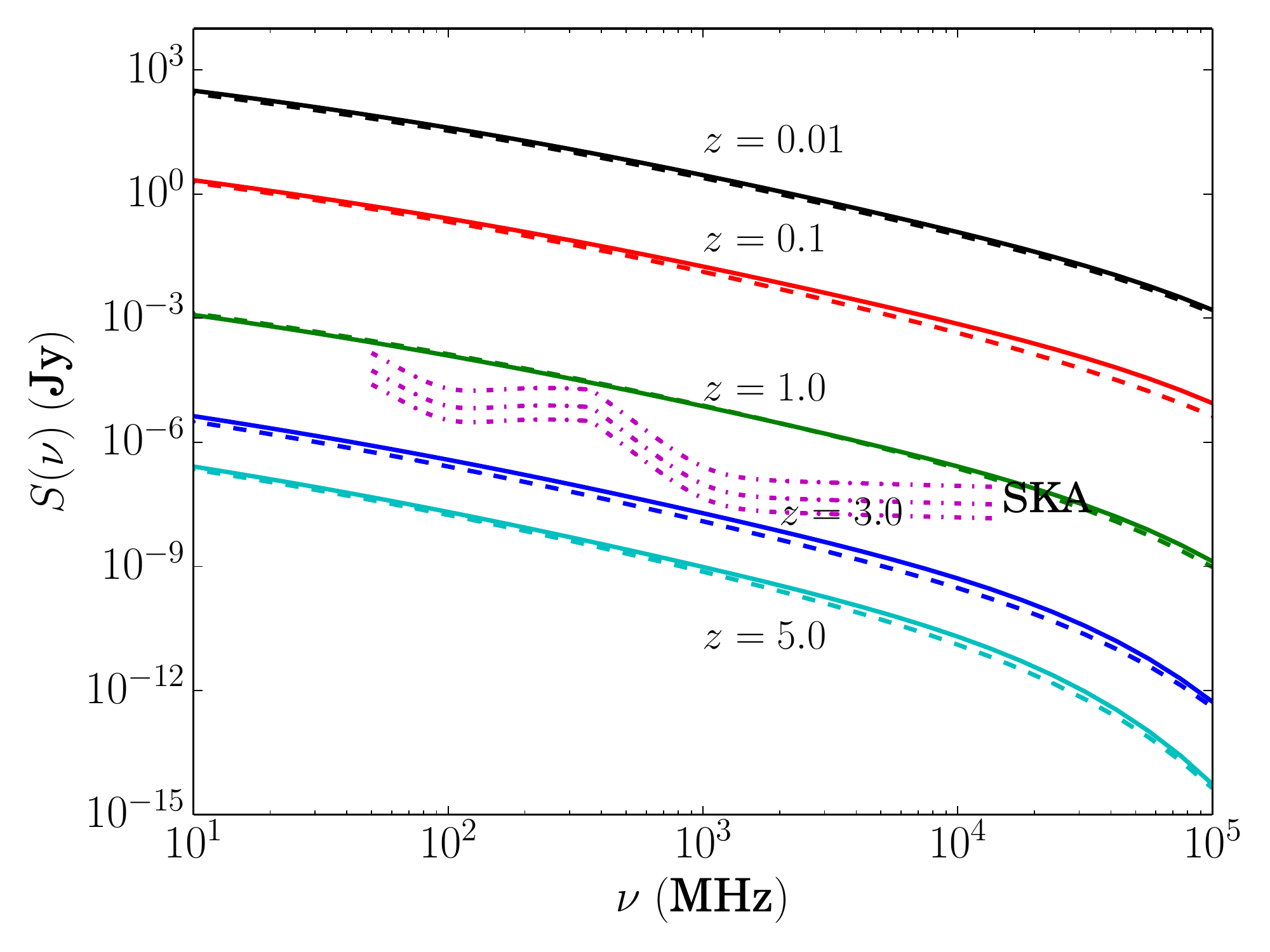}
\caption{Flux densities for galaxy clusters ($M = 10^{15}$M$_{\odot}$), the halo profile is Einasto ($\alpha = 0.18$) and $\langle B \rangle = 5$ $\mu$G. WIMP mass is $60$ GeV and the composition is $b\overline{b}$. See the caption of Figure~\ref{fig:bb60_m7_b5_nfw} for legend.}
\label{fig:bb60_m15_b5_ein}
\end{figure}

\begin{figure}[htbp]
\centering
\includegraphics[scale=0.37]{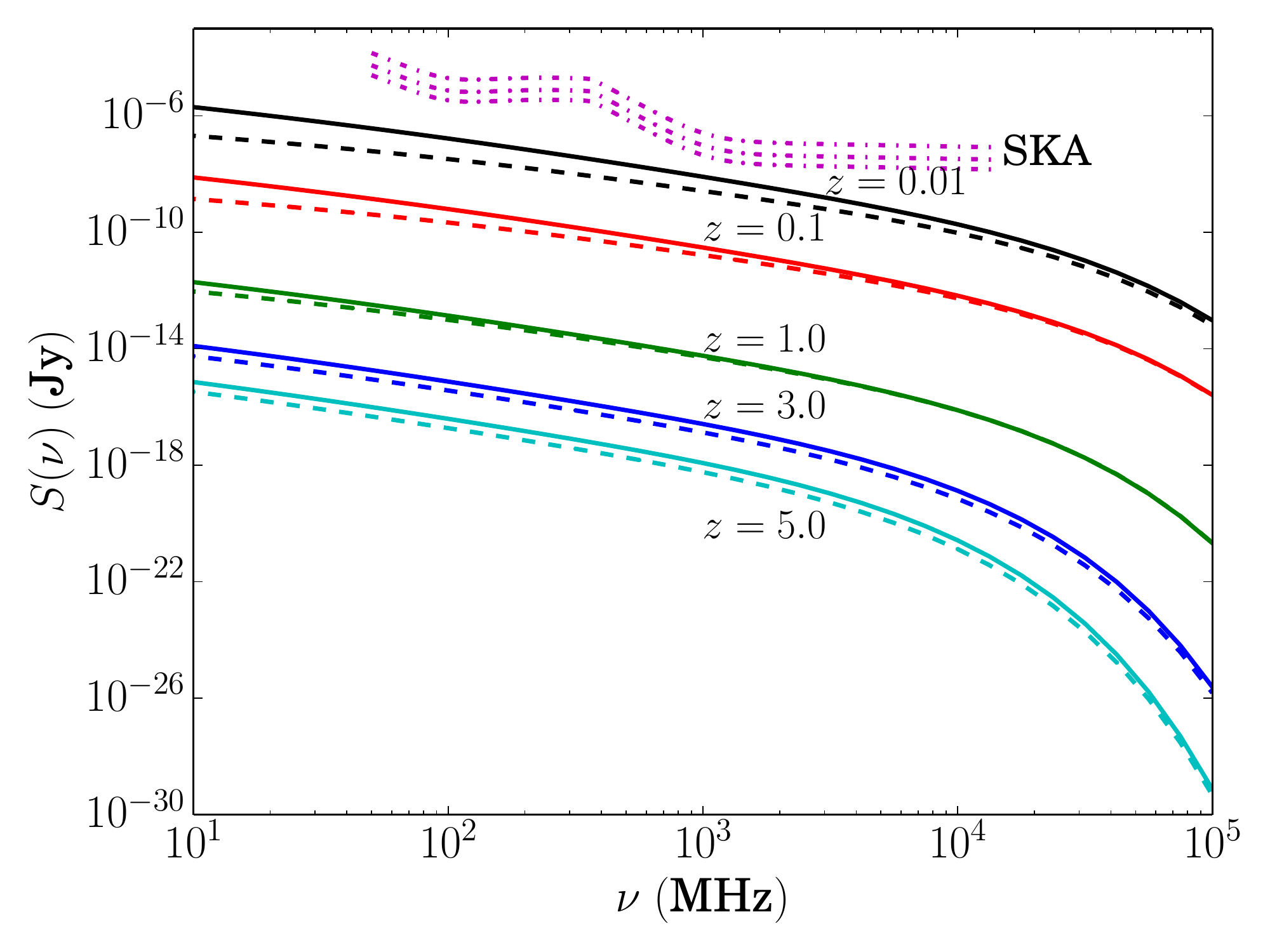}
\includegraphics[scale=0.37]{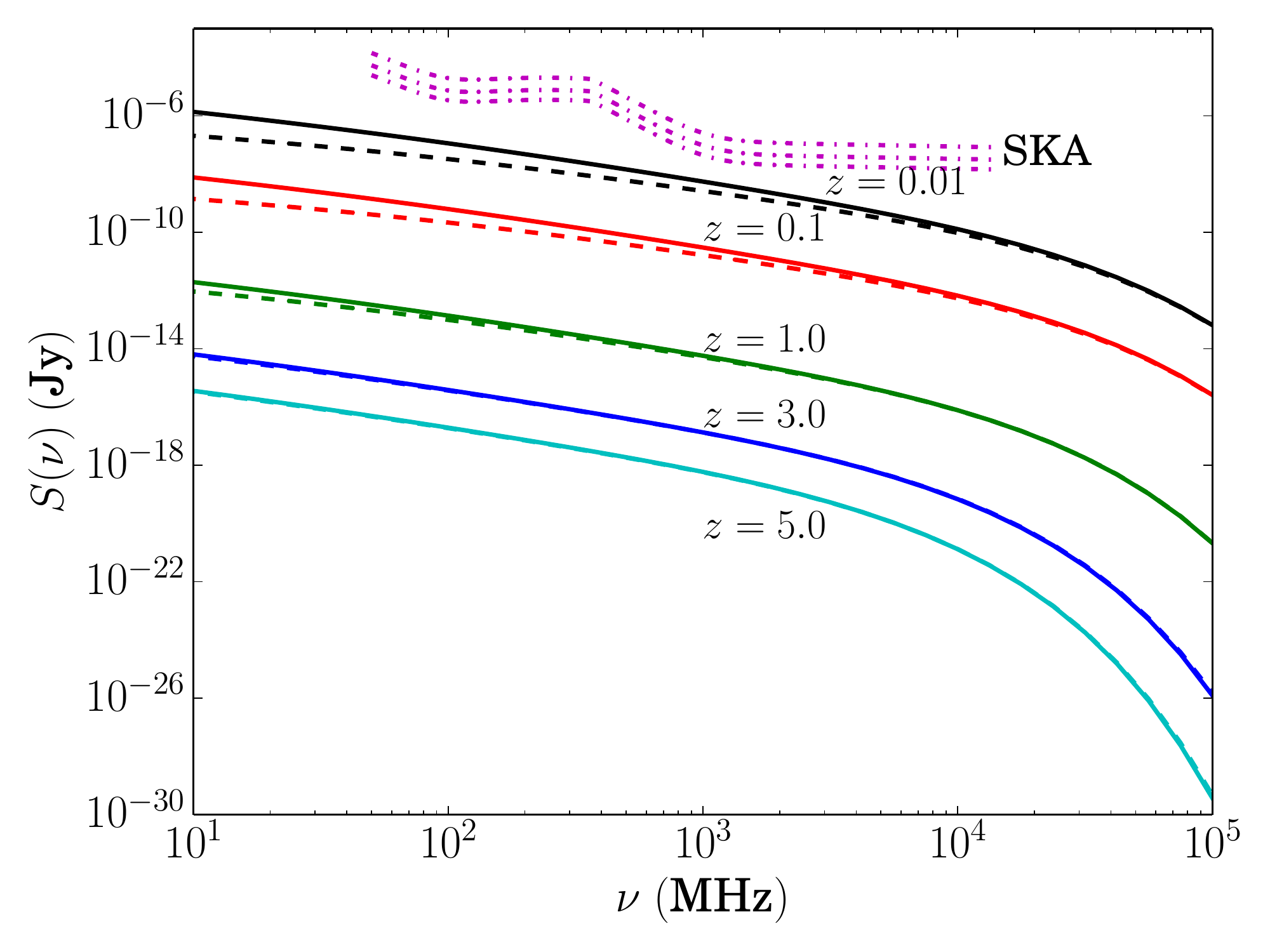}
\caption{Flux densities for dwarf spheroidal galaxies ($M = 10^{7}$M$_{\odot}$), the halo profile is NFW and $\langle B \rangle = 1$ $\mu$G. WIMP mass is $60$ GeV and the composition is $b\overline{b}$. See the caption of Figure~\ref{fig:bb60_m7_b5_nfw} for legend.}
\label{fig:bb60_m7_b1_nfw}
\end{figure}

\begin{figure}[htbp]
\centering
\includegraphics[scale=0.37]{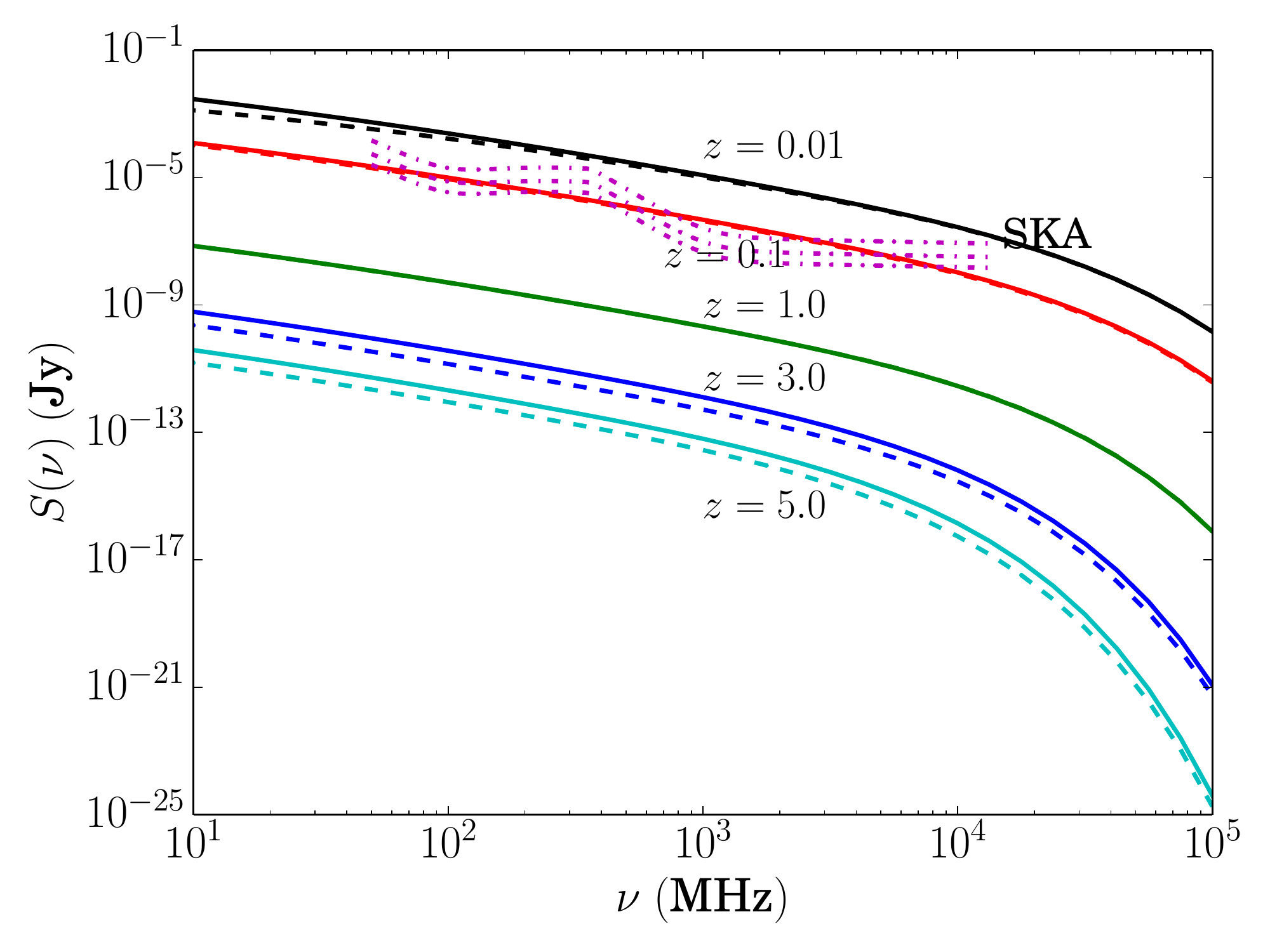}
\includegraphics[scale=0.37]{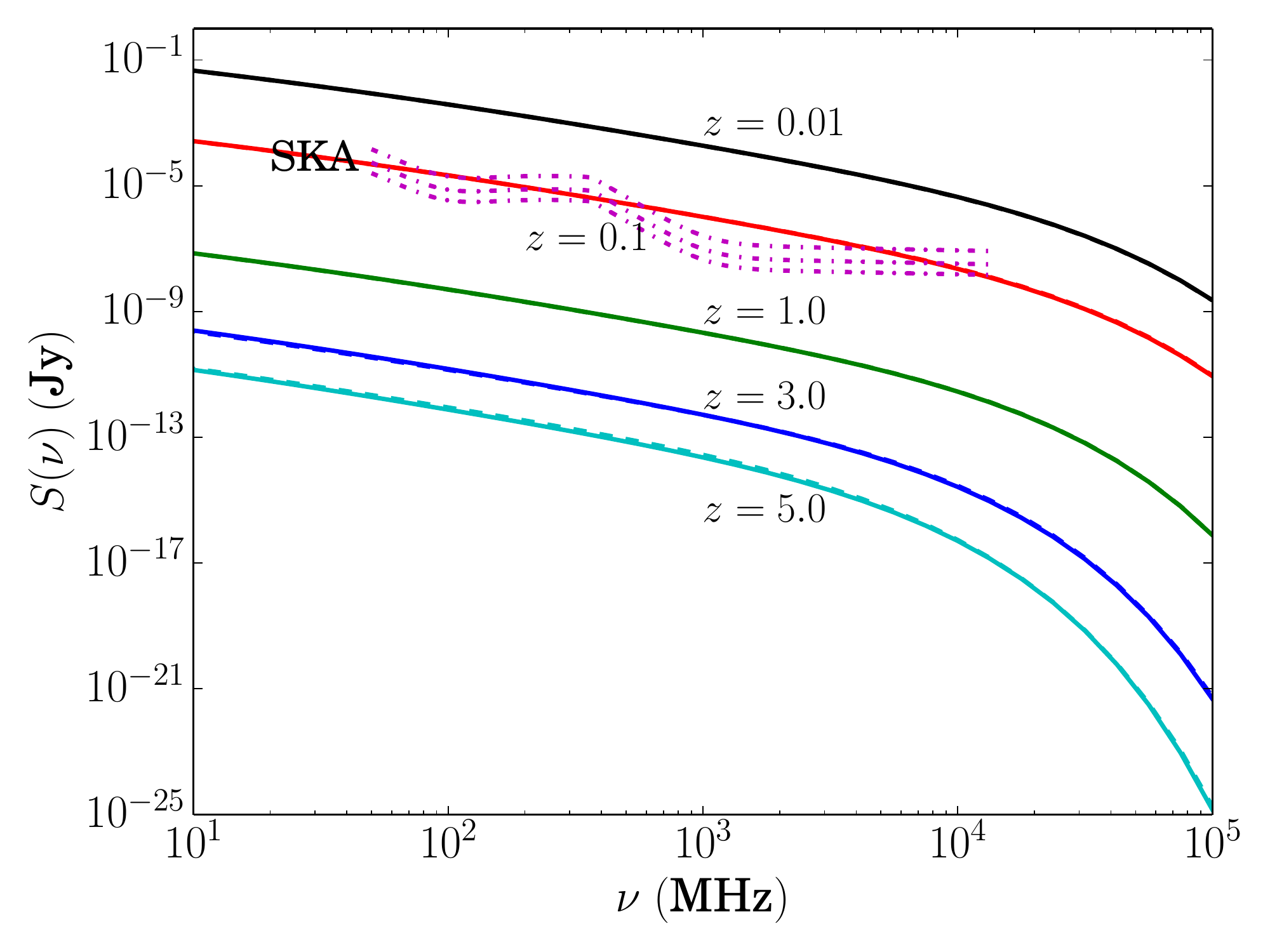}
\caption{Flux densities for galaxies ($M = 10^{12}$M$_{\odot}$), the halo profile is NFW and $\langle B \rangle = 1$ $\mu$G. WIMP mass is $60$ GeV and the composition is $b\overline{b}$. See the caption of Figure~\ref{fig:bb60_m7_b5_nfw} for legend.}
\label{fig:bb60_m12_b1_nfw}
\end{figure}

\begin{figure}[htbp]
\centering
\includegraphics[scale=0.37]{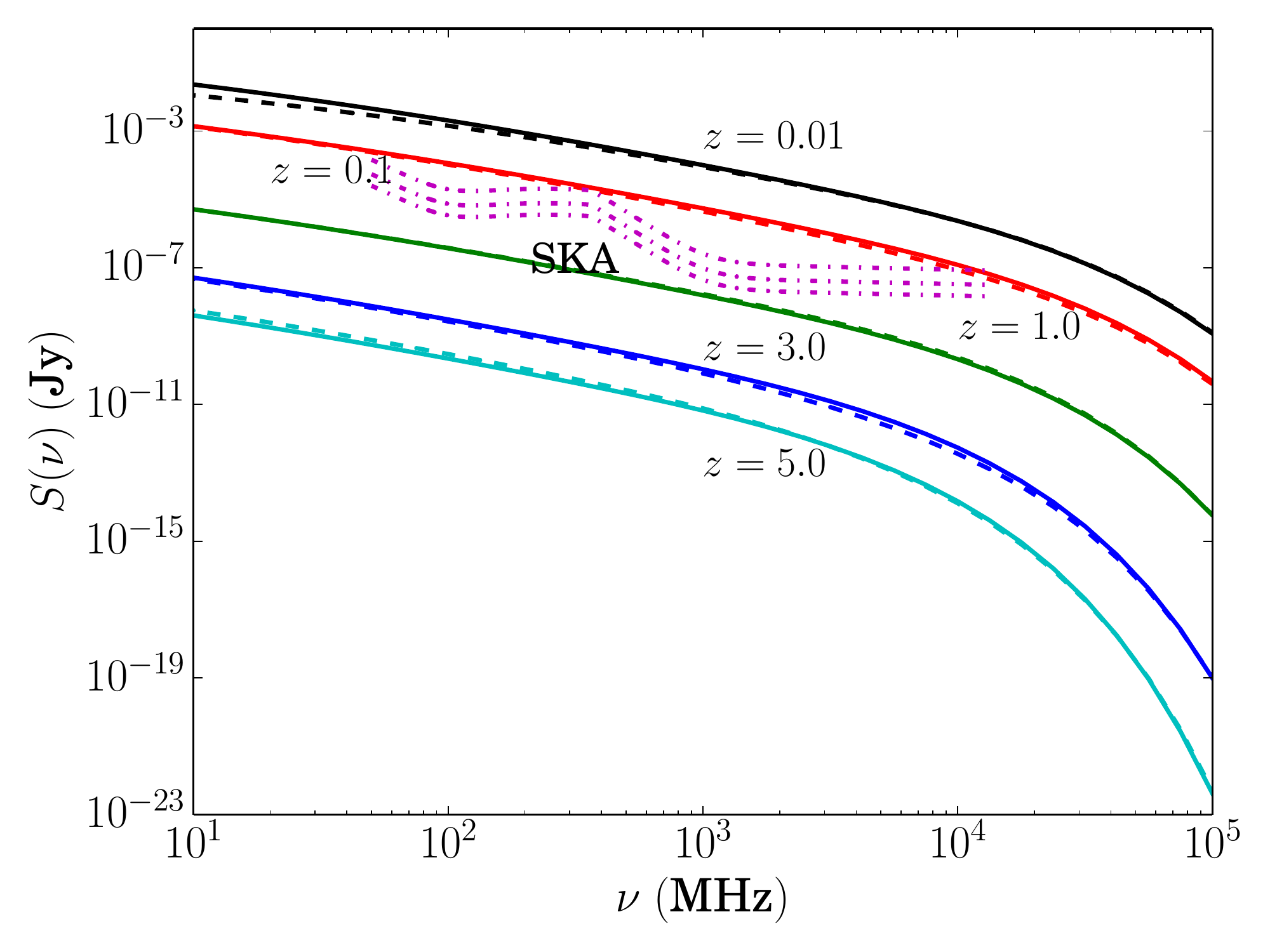}
\includegraphics[scale=0.37]{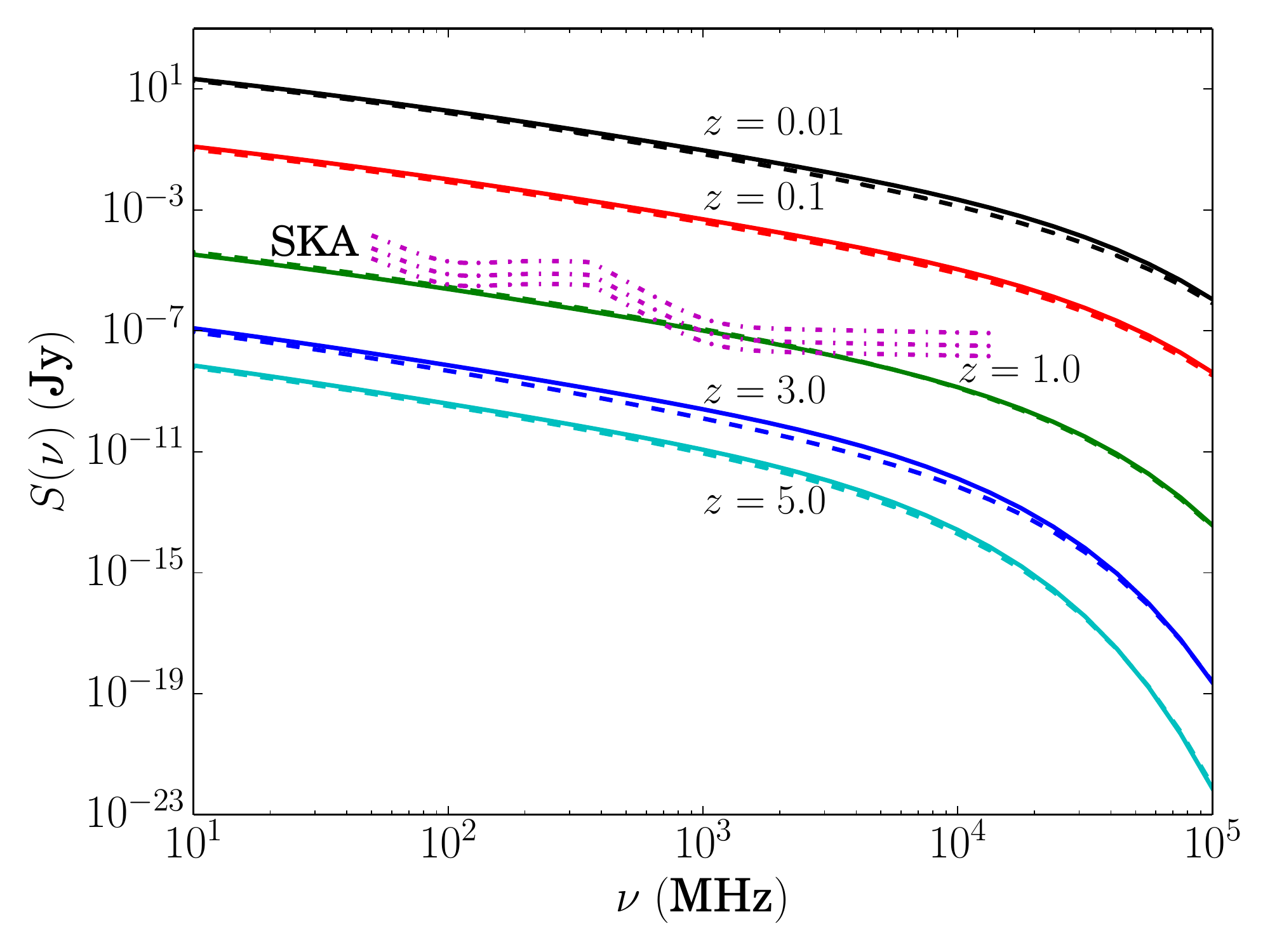}
\caption{Flux densities for galaxy clusters ($M = 10^{15}$M$_{\odot}$), the halo profile is NFW and $\langle B \rangle = 1$ $\mu$G. WIMP mass is $60$ GeV and the composition is $b\overline{b}$. See the caption of Figure~\ref{fig:bb60_m7_b5_nfw} for legend.}
\label{fig:bb60_m15_b1_nfw}
\end{figure}

\begin{figure}[htbp]
\centering
\includegraphics[scale=0.37]{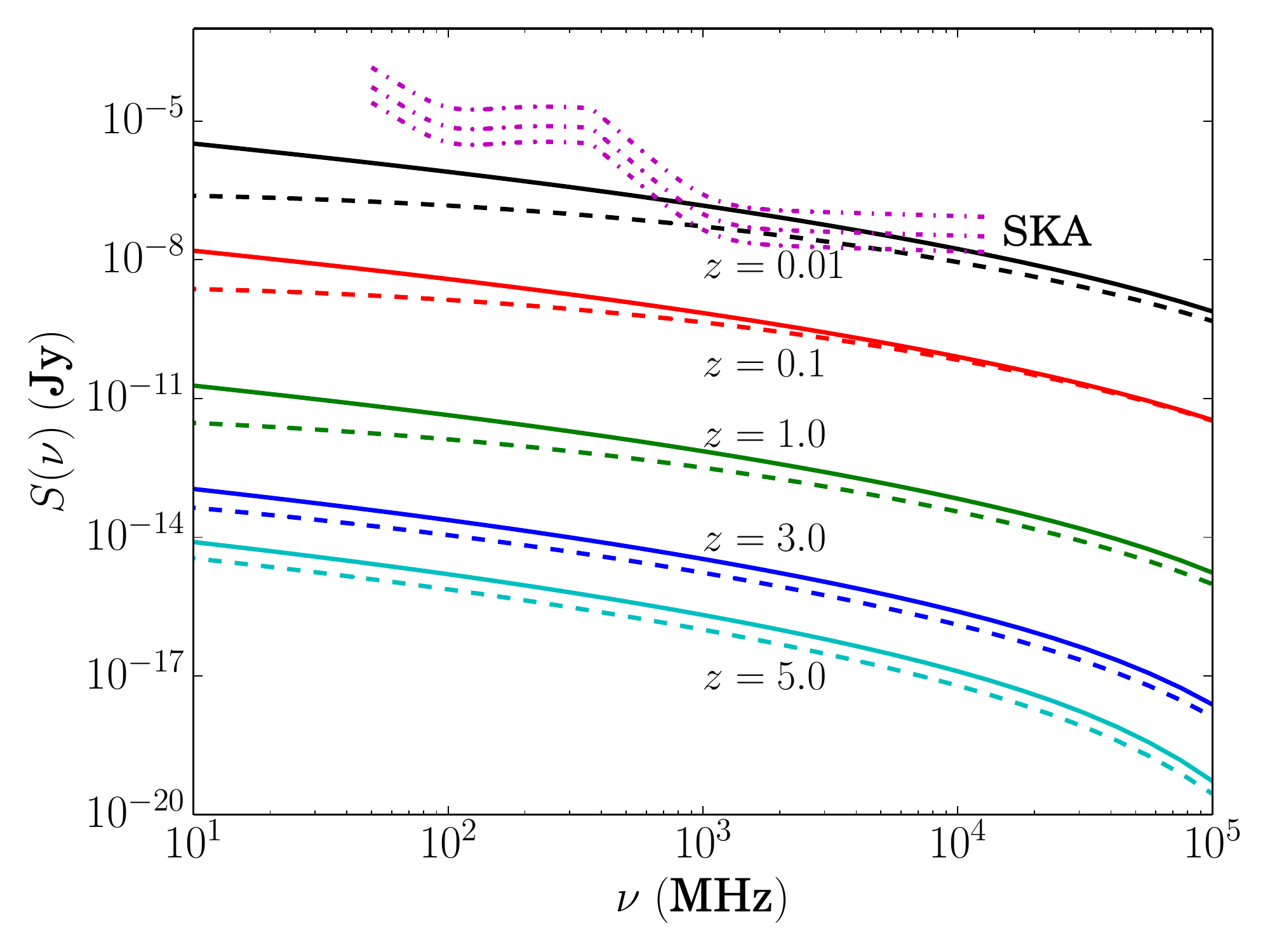}
\includegraphics[scale=0.37]{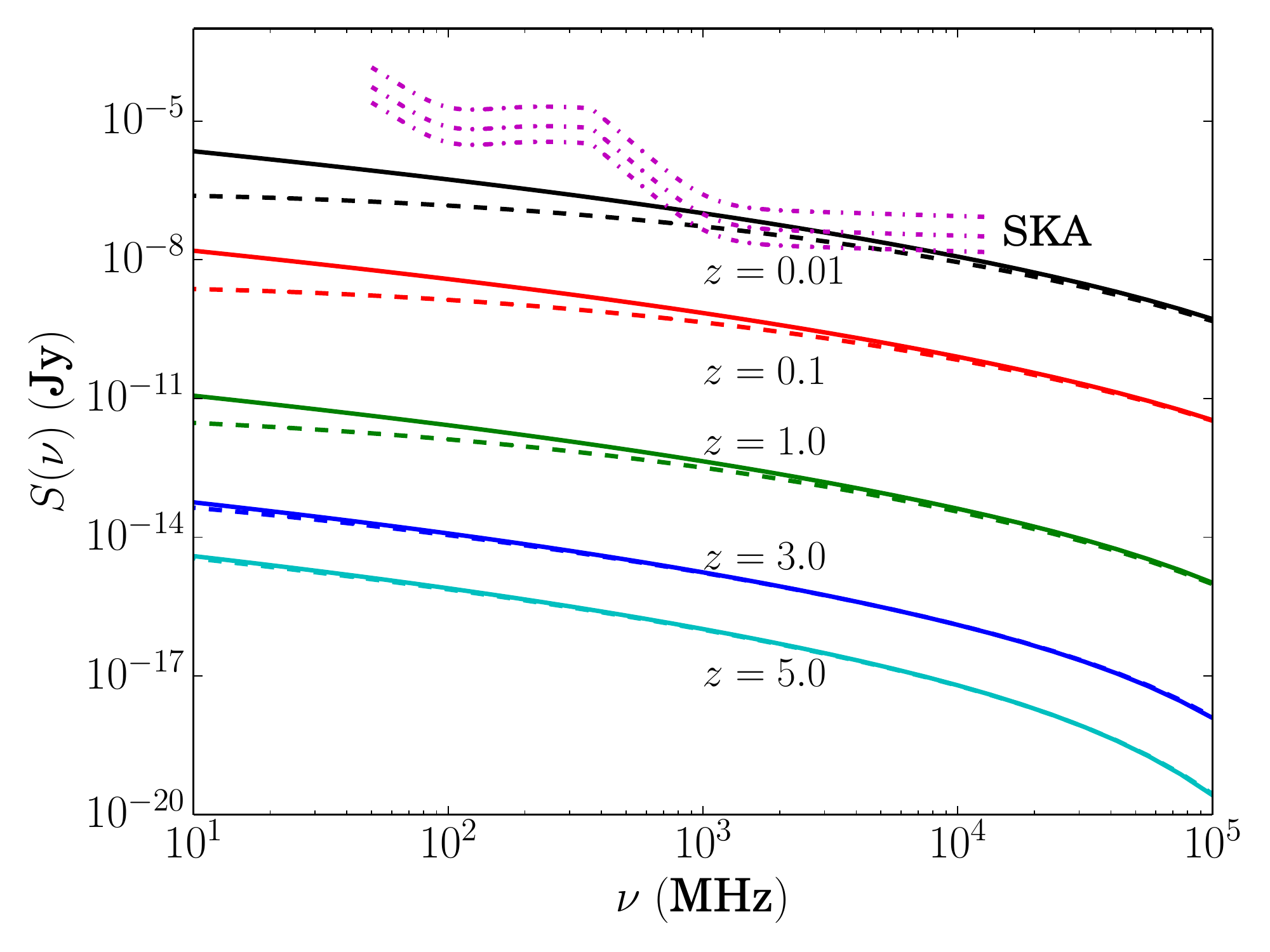}
\caption{Flux densities for dwarf spheroidal galaxies ($M = 10^{7}$M$_{\odot}$), $\langle B \rangle = 5$ $\mu$G. WIMP mass is $60$ GeV and the composition is $\tau^+\tau^-$. See the caption of Figure~\ref{fig:bb60_m7_b5_nfw} for legend.}
\label{fig:tt60_m7_b5_nfw}
\end{figure}

\begin{figure}[htbp]
\centering
\includegraphics[scale=0.37]{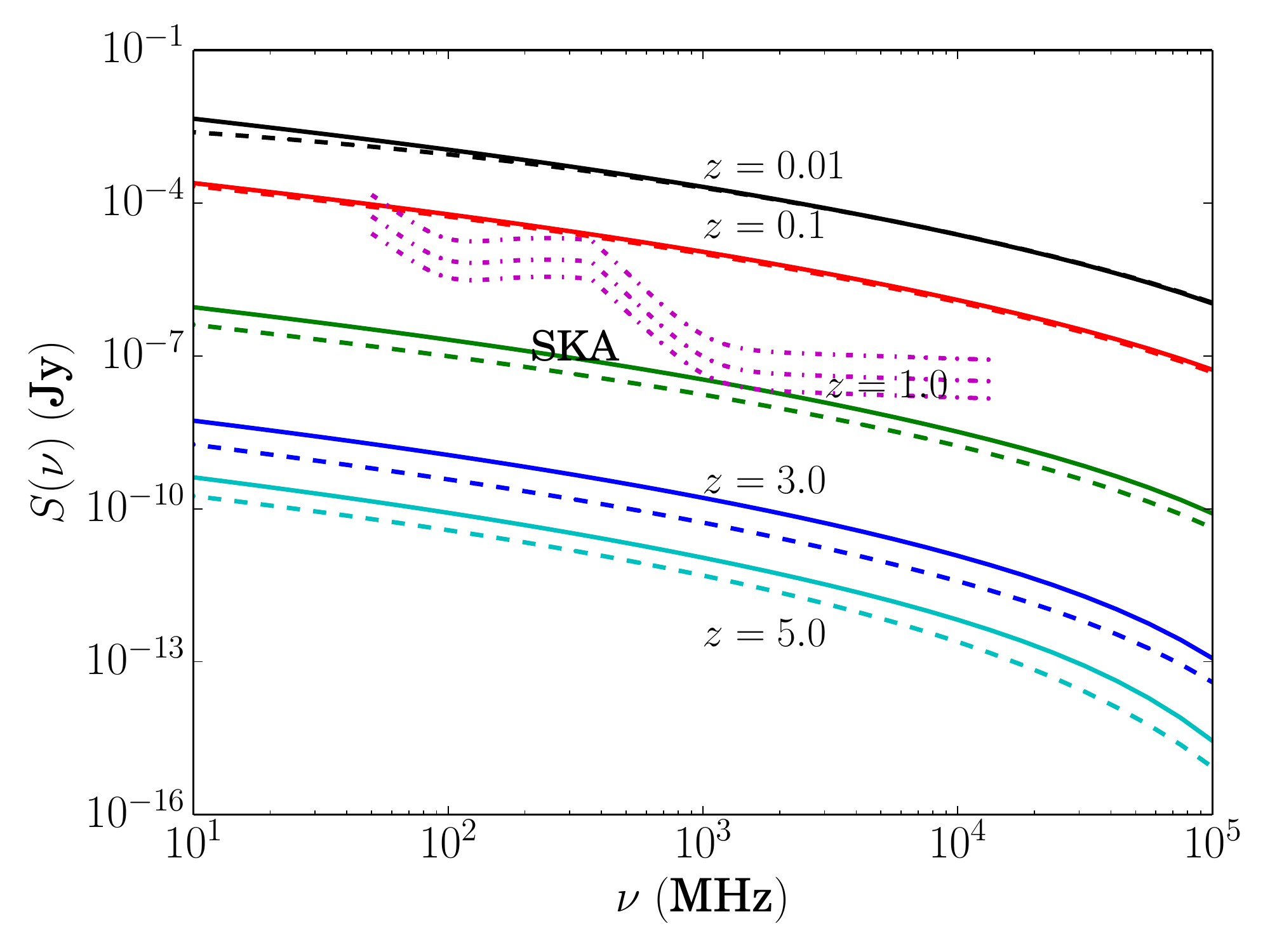}
\includegraphics[scale=0.37]{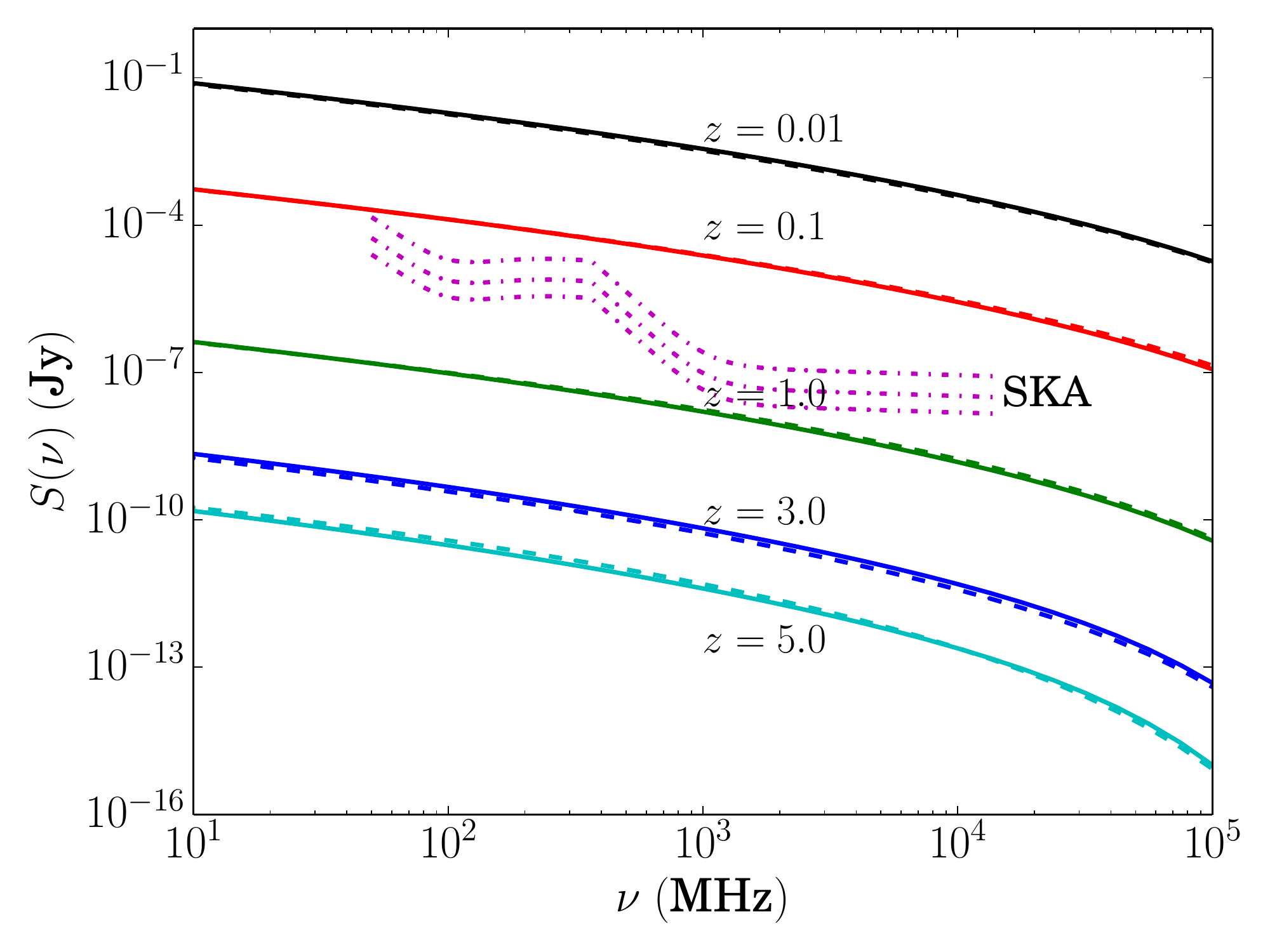}
\caption{Flux densities for galaxies ($M = 10^{12}$M$_{\odot}$), $\langle B \rangle = 5$ $\mu$G. WIMP mass is $60$ GeV and the composition is $\tau^+\tau^-$. See the caption of Figure~\ref{fig:bb60_m7_b5_nfw} for legend.}
\label{fig:tt60_m12_b5_nfw}
\end{figure}

\begin{figure}[htbp]
\centering
\includegraphics[scale=0.37]{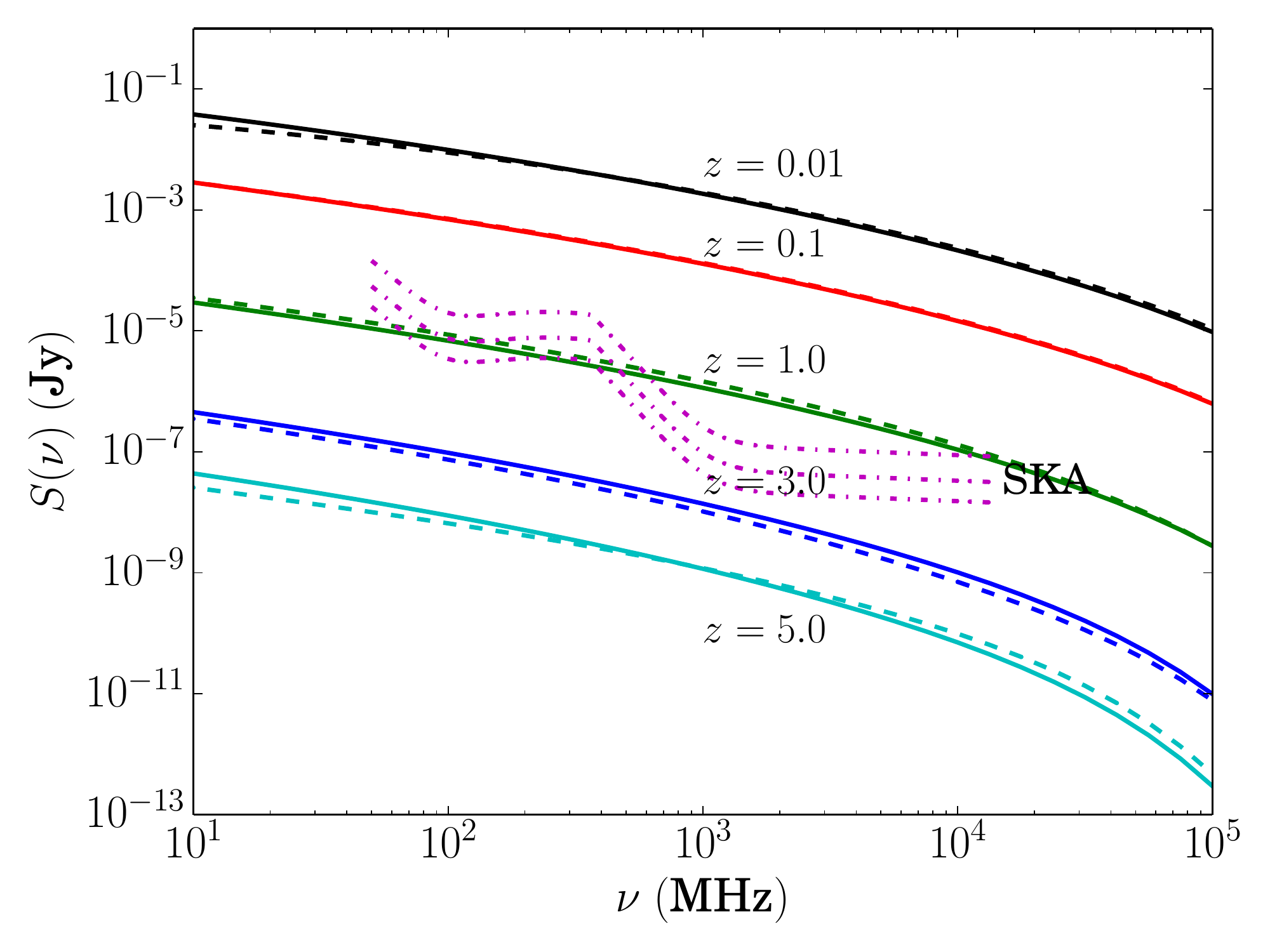}
\includegraphics[scale=0.37]{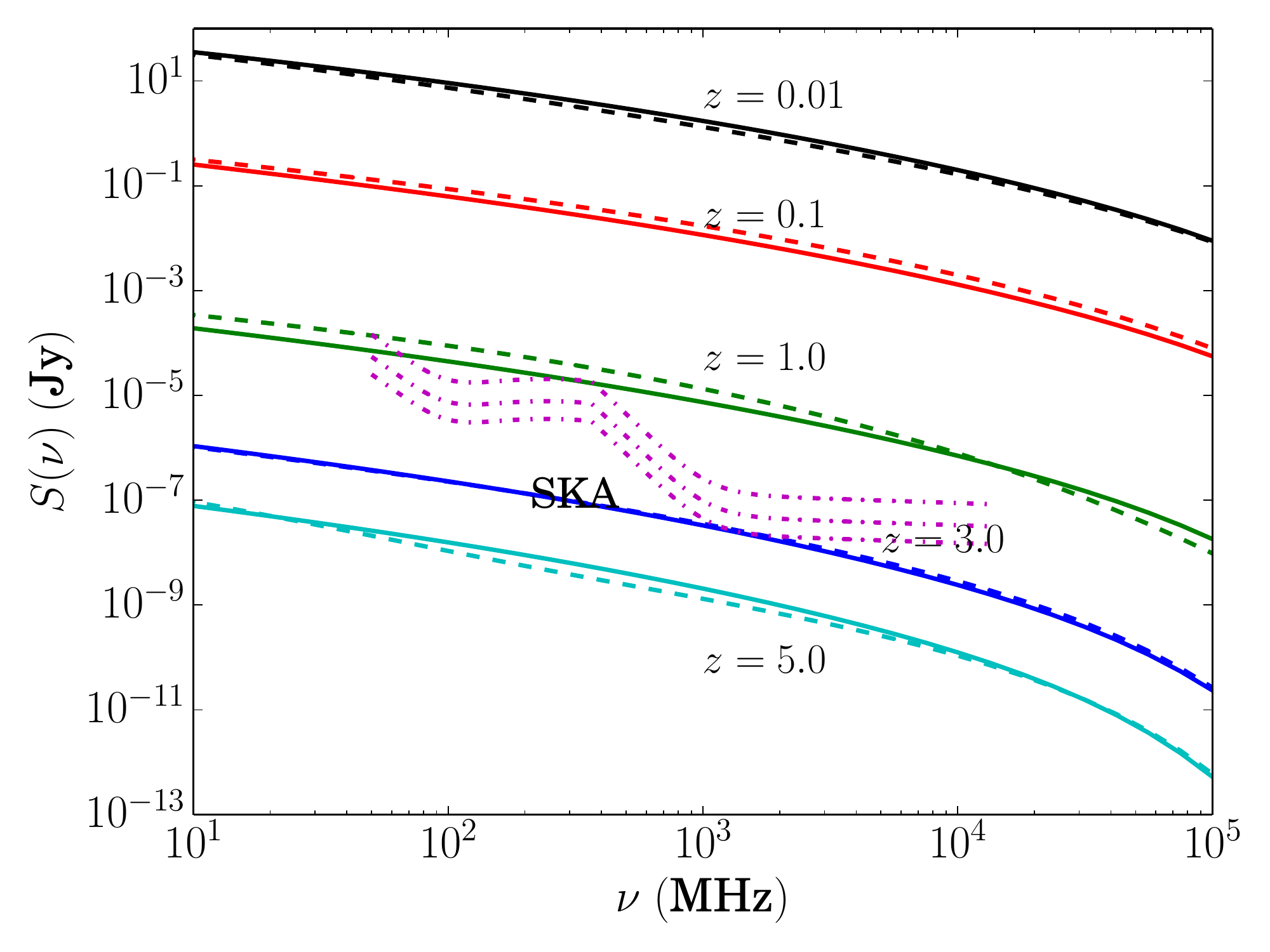}
\caption{Flux densities for galaxy clusters ($M = 10^{15}$M$_{\odot}$), the halo profile is NFW and $\langle B \rangle = 5$ $\mu$G. WIMP mass is $60$ GeV and the composition is $\tau^+\tau^-$. See the caption of Figure~\ref{fig:bb60_m7_b5_nfw} for legend.}
\label{fig:tt60_m15_b5_nfw}
\end{figure}

\begin{figure}[htbp]
\centering
\includegraphics[scale=0.37]{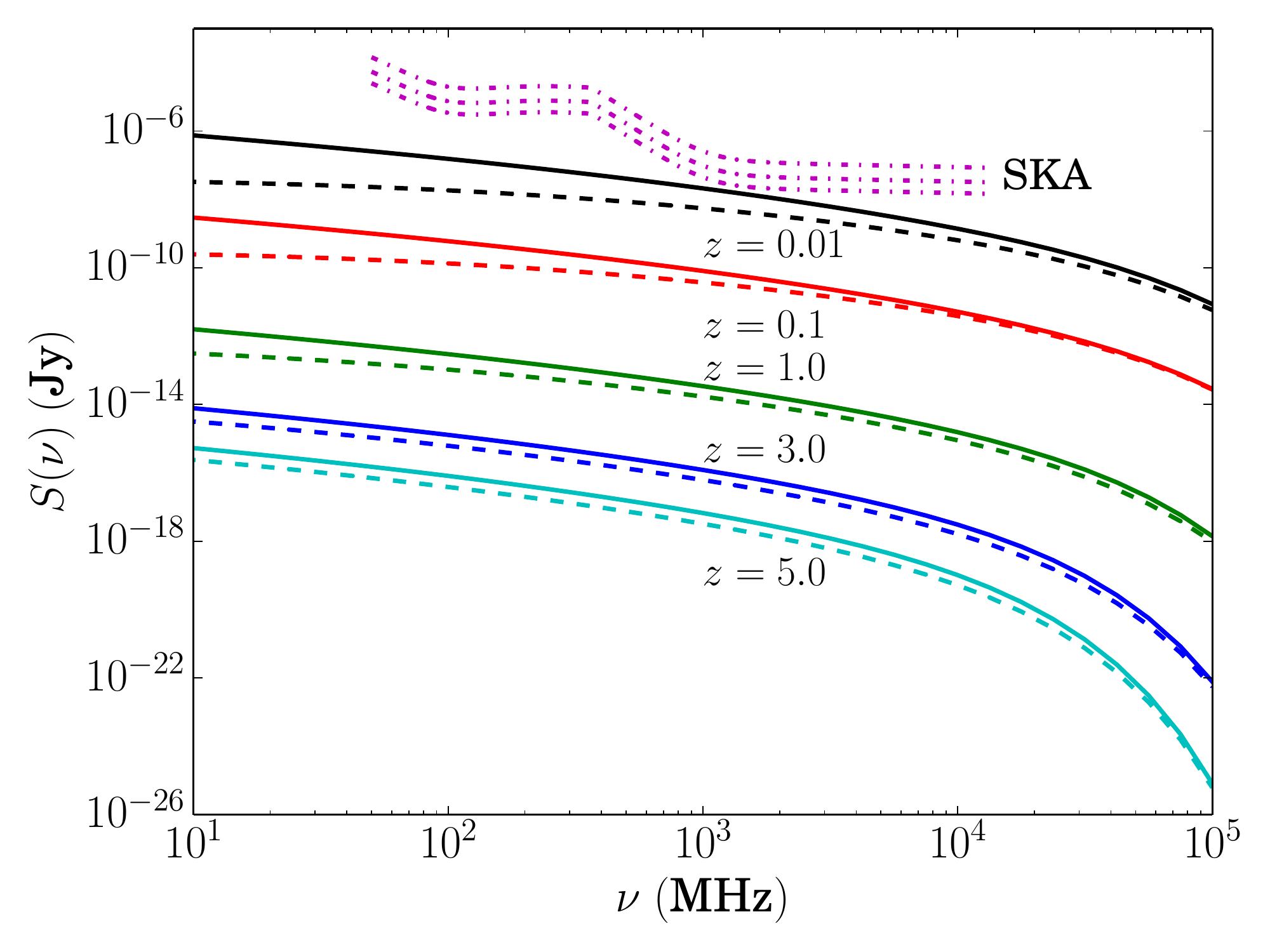}
\includegraphics[scale=0.37]{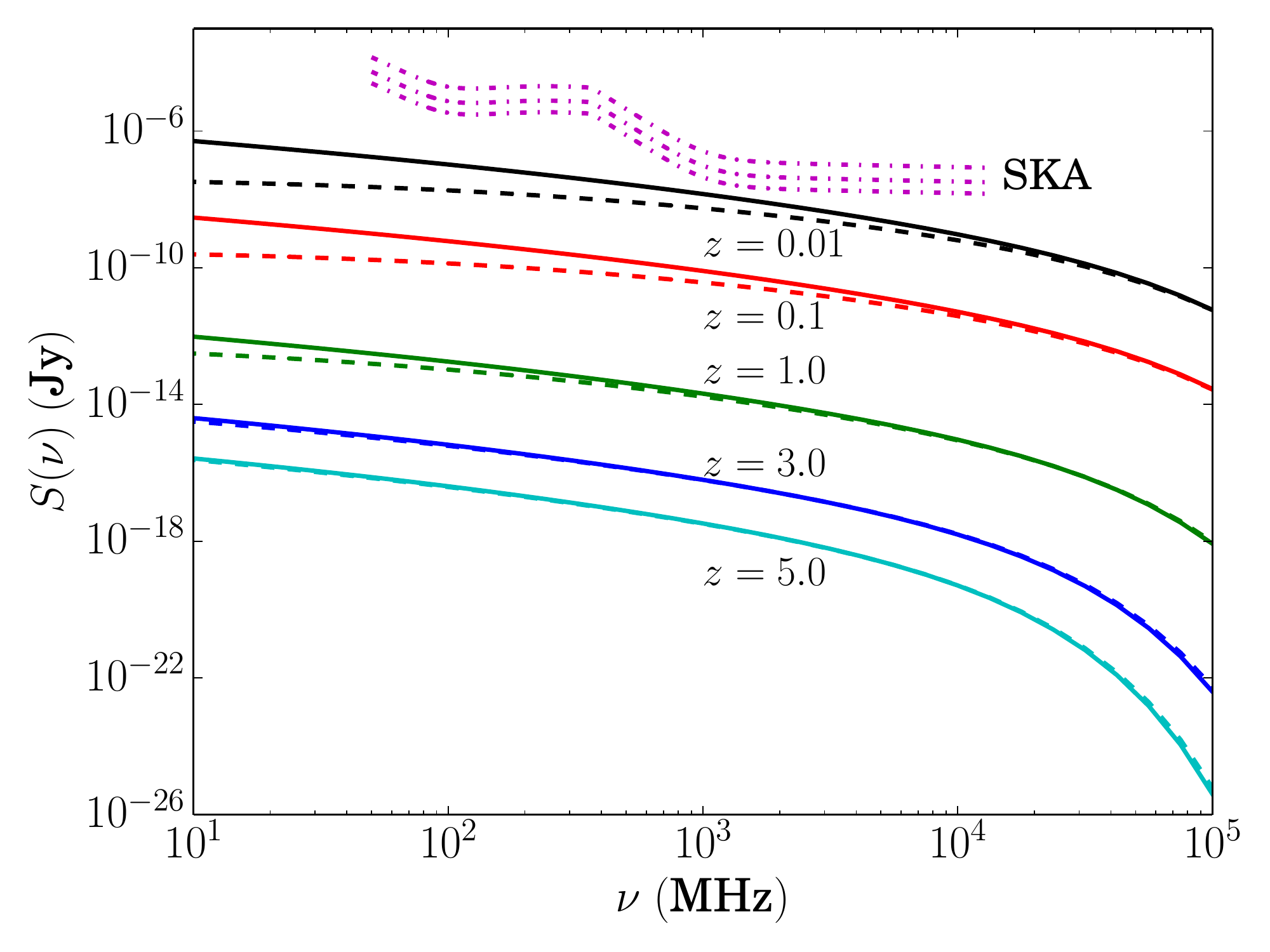}
\caption{Flux densities for dwarf spheroidal galaxies ($M = 10^{7}$M$_{\odot}$), the halo profile is NFW and $\langle B \rangle = 1$ $\mu$G. WIMP mass is $60$ GeV and the composition is $\tau^+\tau^-$. See the caption of Figure~\ref{fig:bb60_m7_b5_nfw} for legend.}
\label{fig:tt60_m7_b1_nfw}
\end{figure}

\begin{figure}[htbp]
\centering
\includegraphics[scale=0.37]{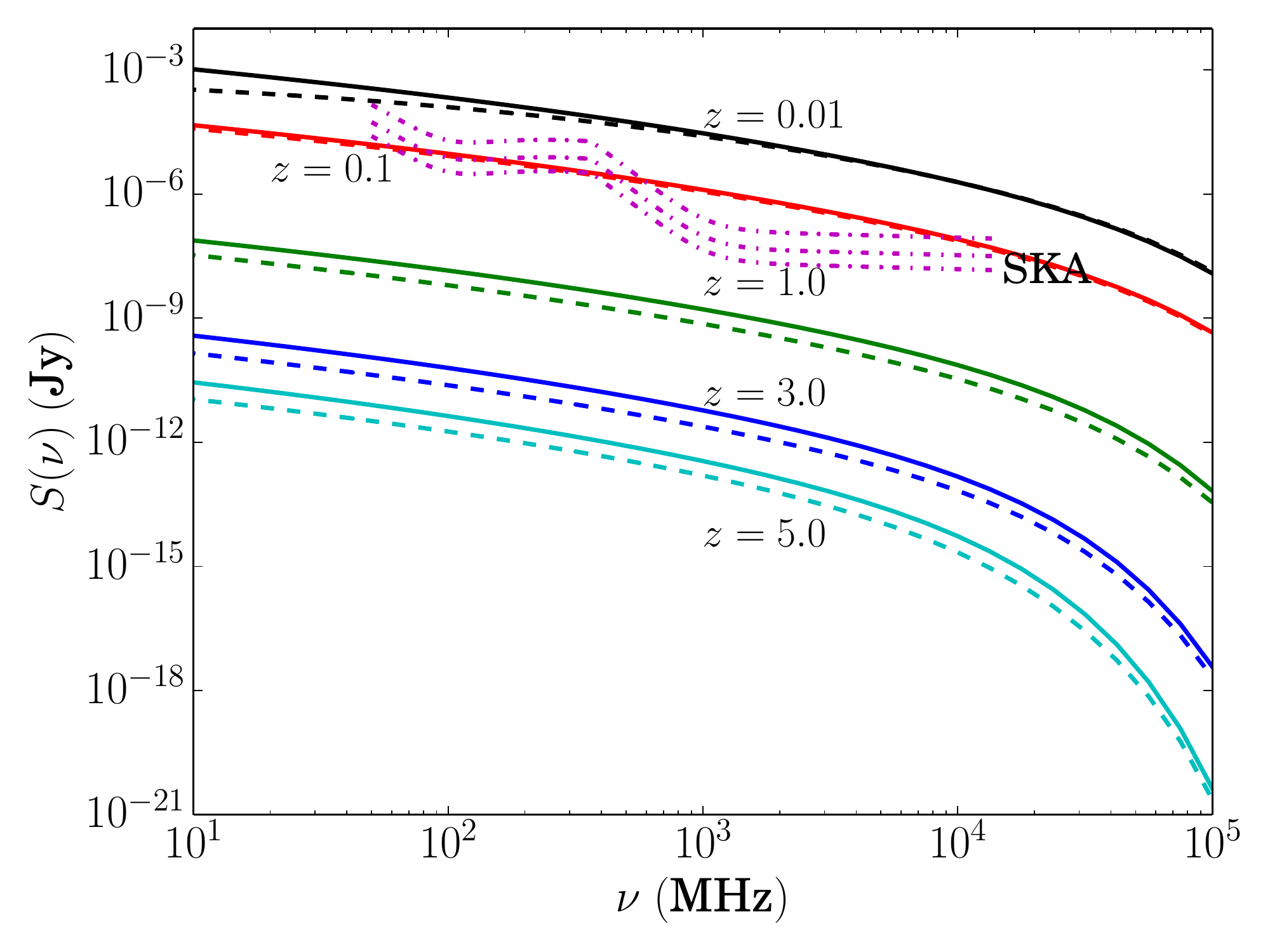}
\includegraphics[scale=0.37]{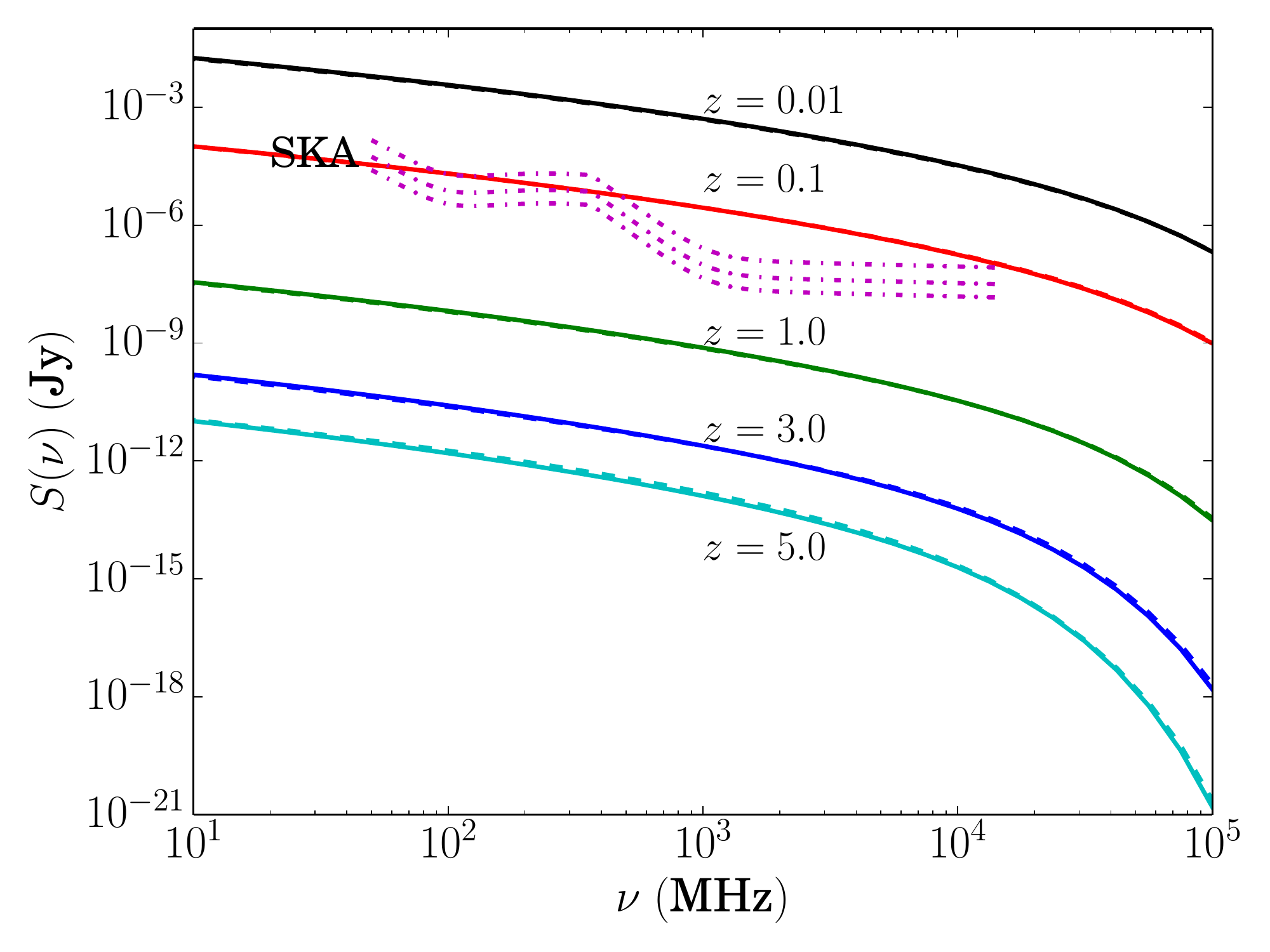}
\caption{Flux densities for galaxies ($M = 10^{12}$M$_{\odot}$), the halo profile is NFW and $\langle B \rangle = 1$ $\mu$G. WIMP mass is $60$ GeV and the composition is $\tau^+\tau^-$. See the caption of Figure~\ref{fig:bb60_m7_b5_nfw} for legend.}
\label{fig:tt60_m12_b1_nfw}
\end{figure}

\begin{figure}[htbp]
\centering
\includegraphics[scale=0.37]{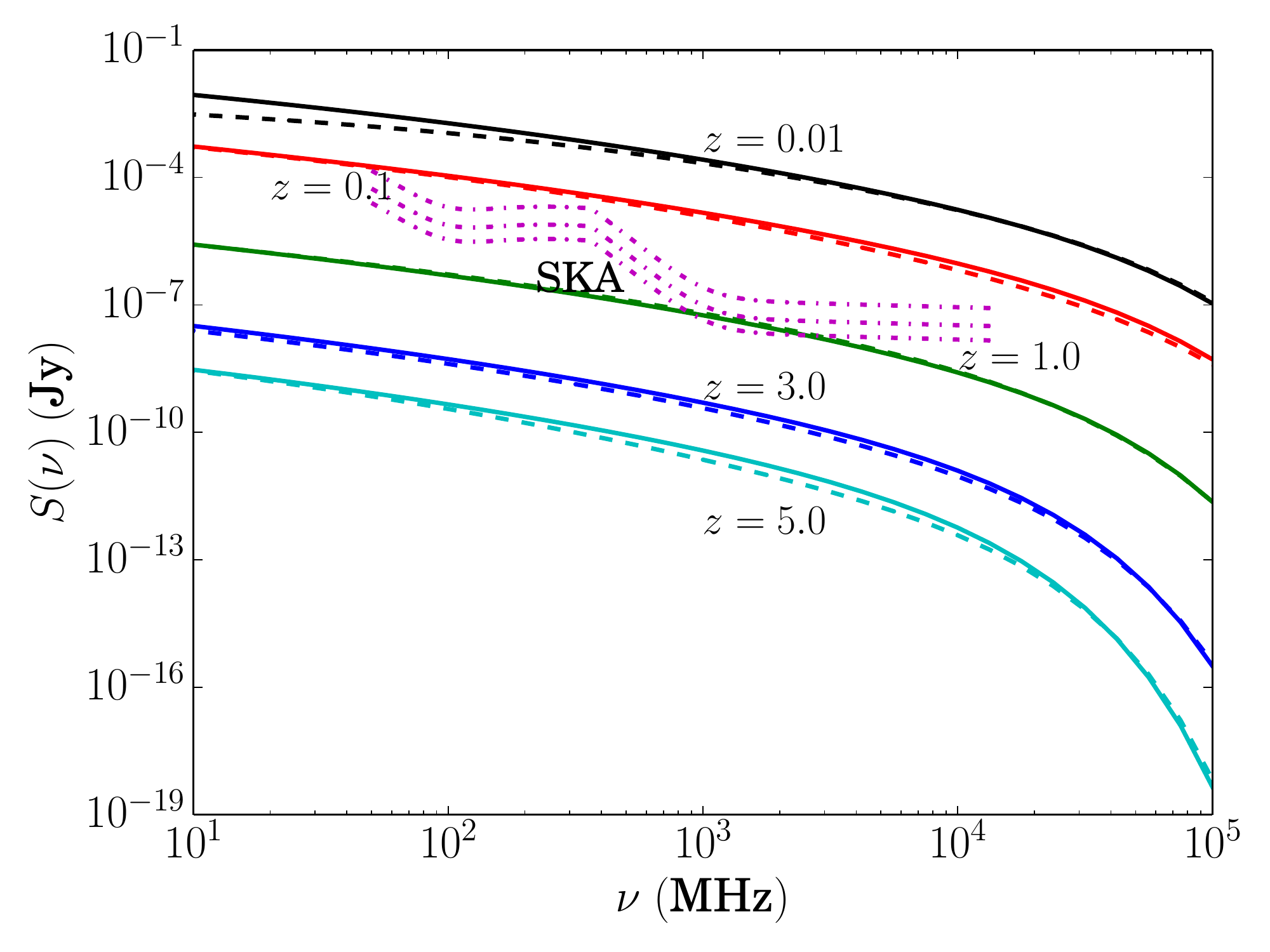}
\includegraphics[scale=0.37]{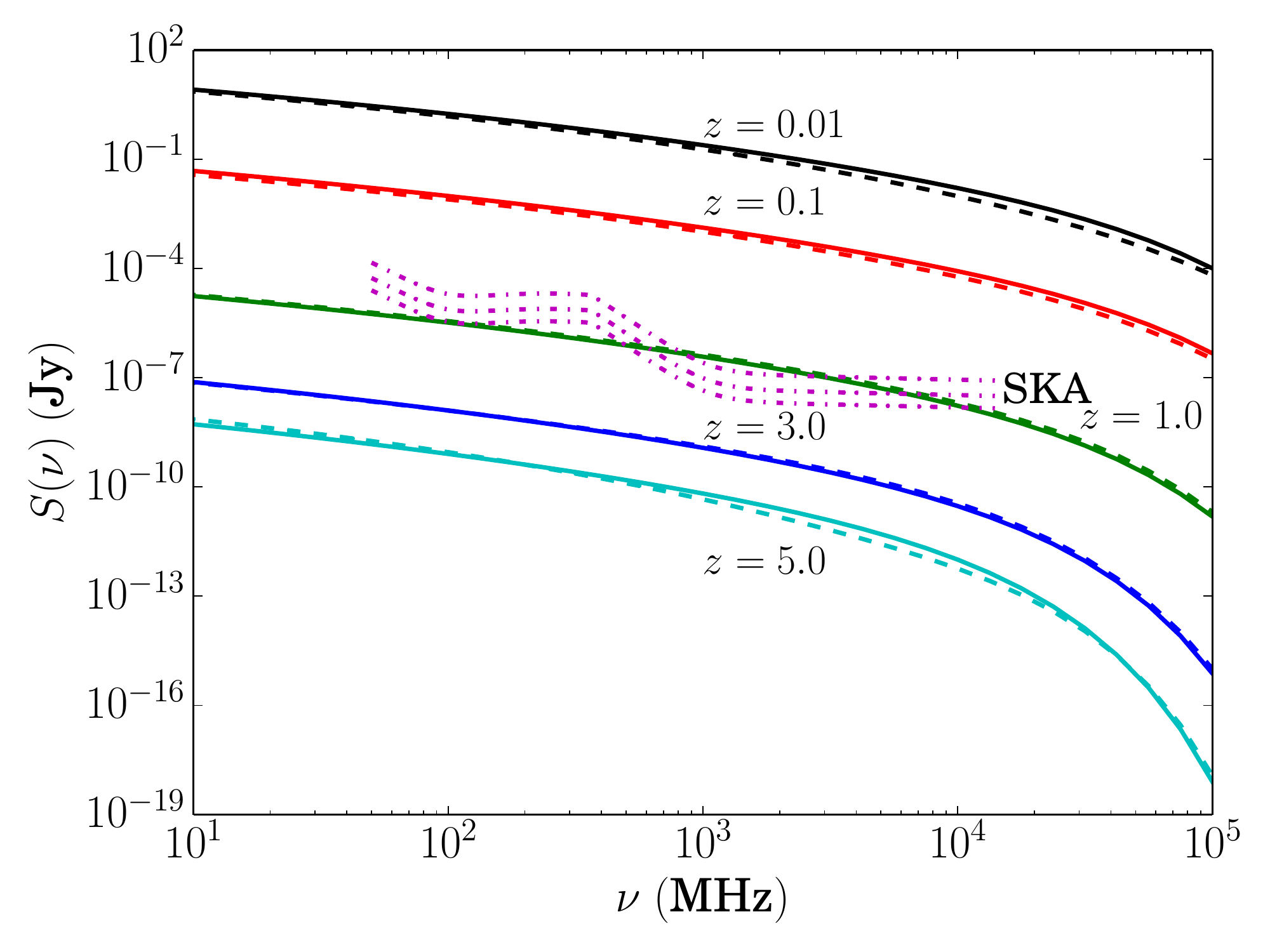}
\caption{Flux densities for galaxy clusters ($M = 10^{15}$M$_{\odot}$), the halo profile is NFW and $\langle B \rangle = 1$ $\mu$G. WIMP mass is $60$ GeV and the composition is $\tau^+\tau^-$. See the caption of Figure~\ref{fig:bb60_m7_b5_nfw} for legend.}
\label{fig:tt60_m15_b1_nfw}
\end{figure}

\begin{figure}[htbp]
\centering
\includegraphics[scale=0.37]{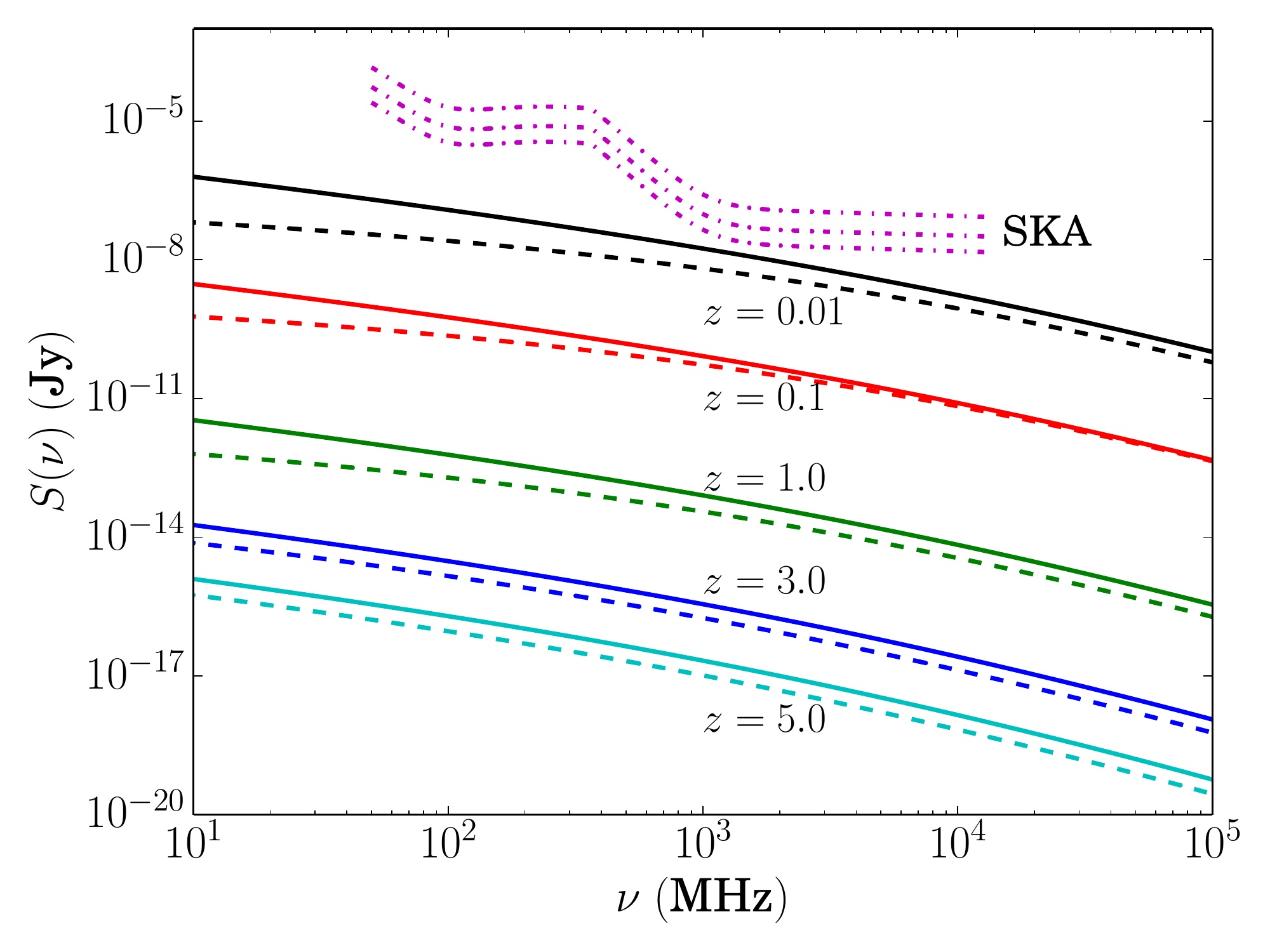}
\includegraphics[scale=0.37]{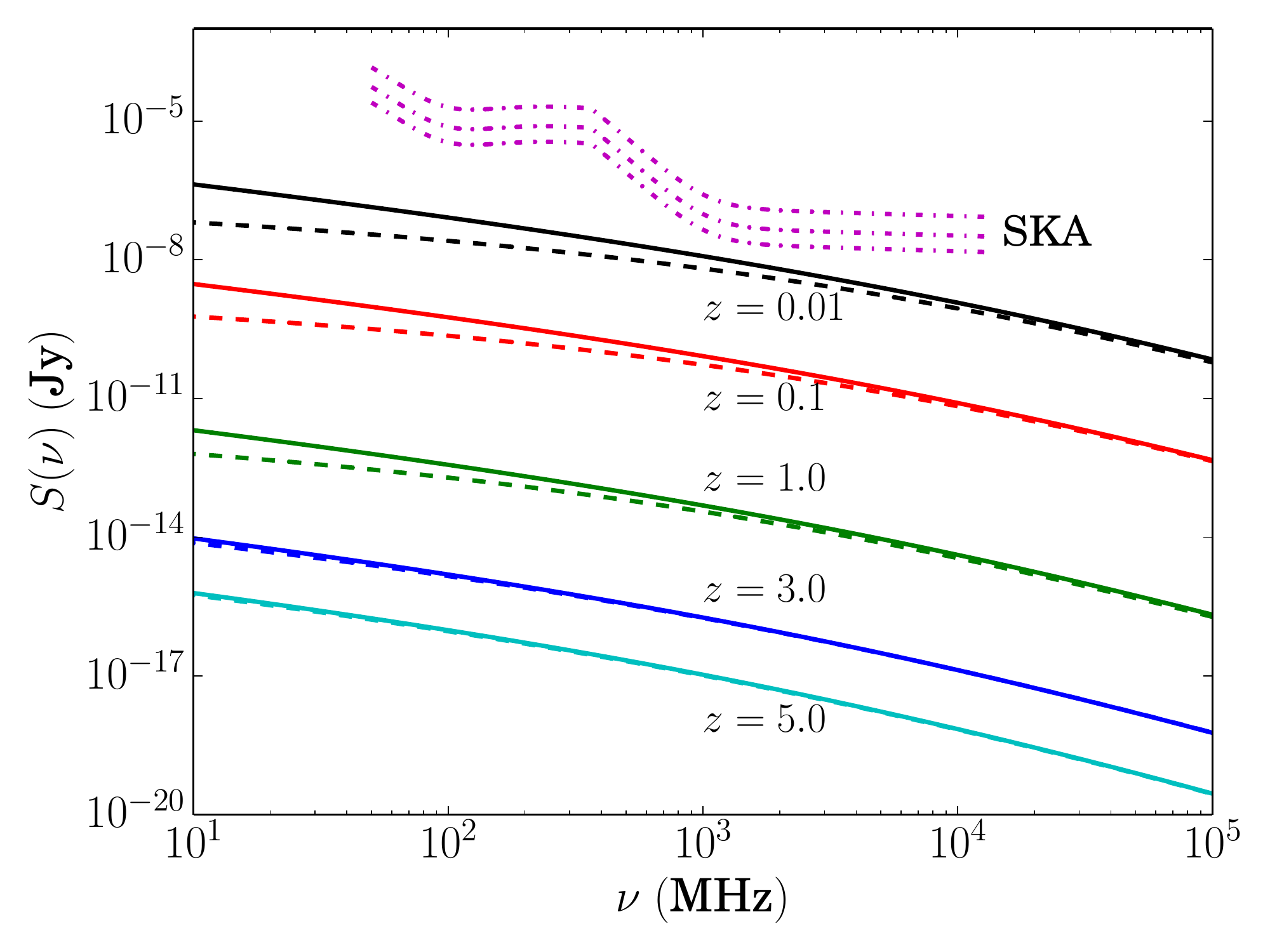}
\caption{Flux densities for dwarf spheroidal galaxies ($M = 10^{7}$M$_{\odot}$), $\langle B \rangle = 5$ $\mu$G. WIMP mass is $500$ GeV and the composition is $b\overline{b}$. See the caption of Figure~\ref{fig:bb60_m7_b5_nfw} for legend.}
\label{fig:bb500_m7_b5_nfw}
\end{figure}

\begin{figure}[htbp]
\centering
\includegraphics[scale=0.37]{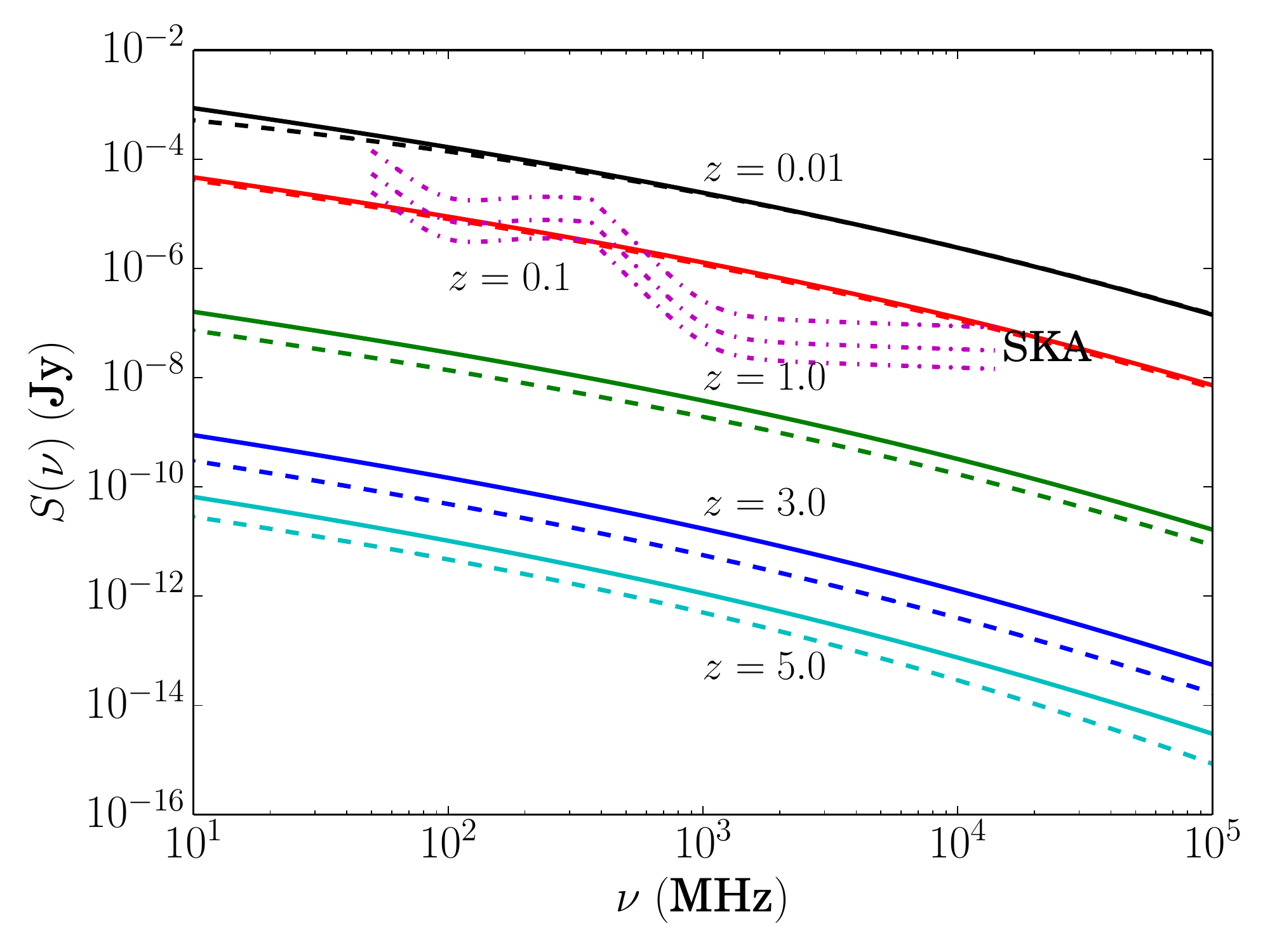}
\includegraphics[scale=0.37]{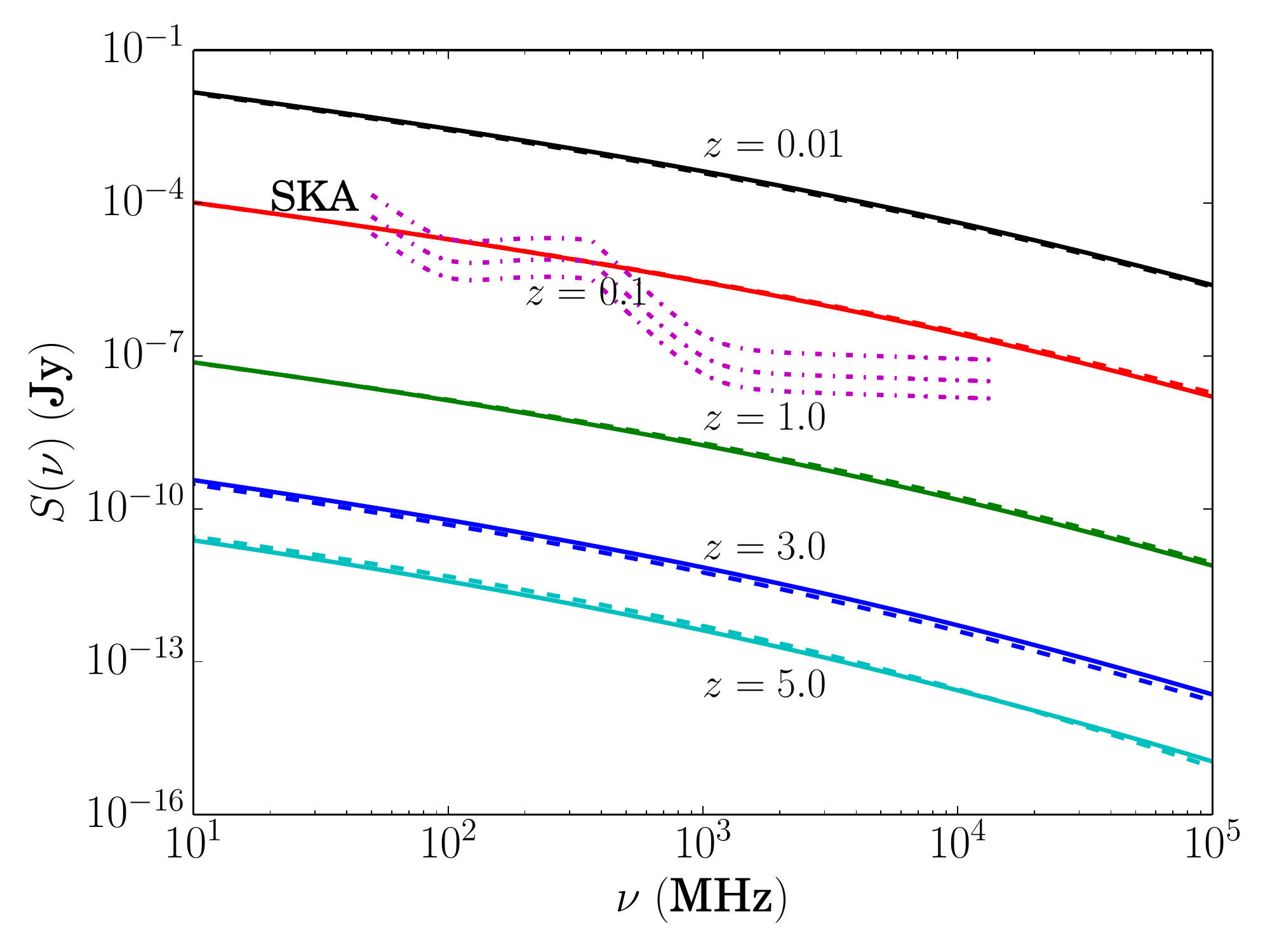}
\caption{Flux densities for galaxies ($M = 10^{12}$M$_{\odot}$), $\langle B \rangle = 5$ $\mu$G. WIMP mass is $500$ GeV and the composition is $b\overline{b}$. See the caption of Figure~\ref{fig:bb60_m7_b5_nfw} for legend.}
\label{fig:bb500_m12_b5_nfw}
\end{figure}

\begin{figure}[htbp]
\centering
\includegraphics[scale=0.37]{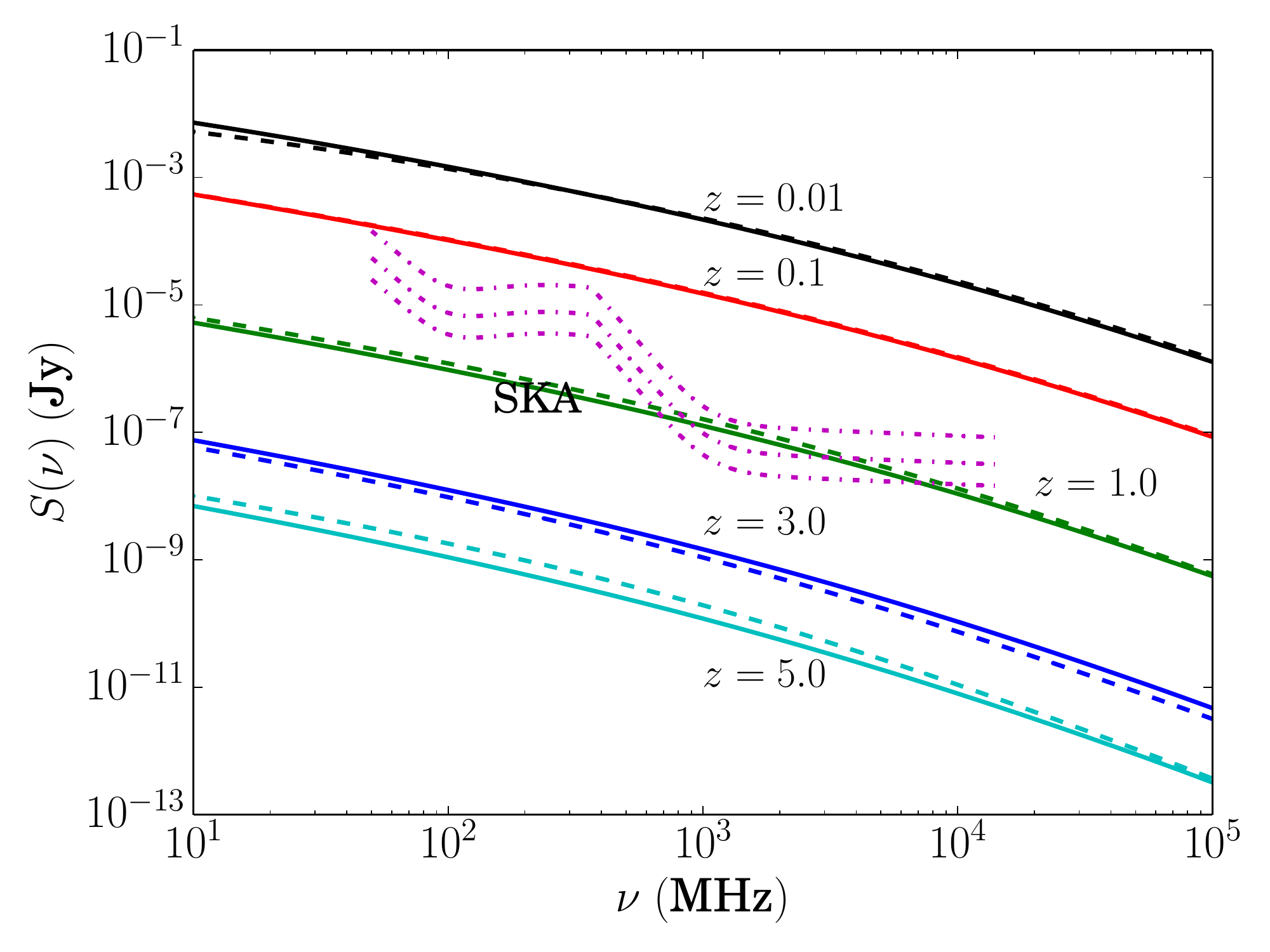}
\includegraphics[scale=0.37]{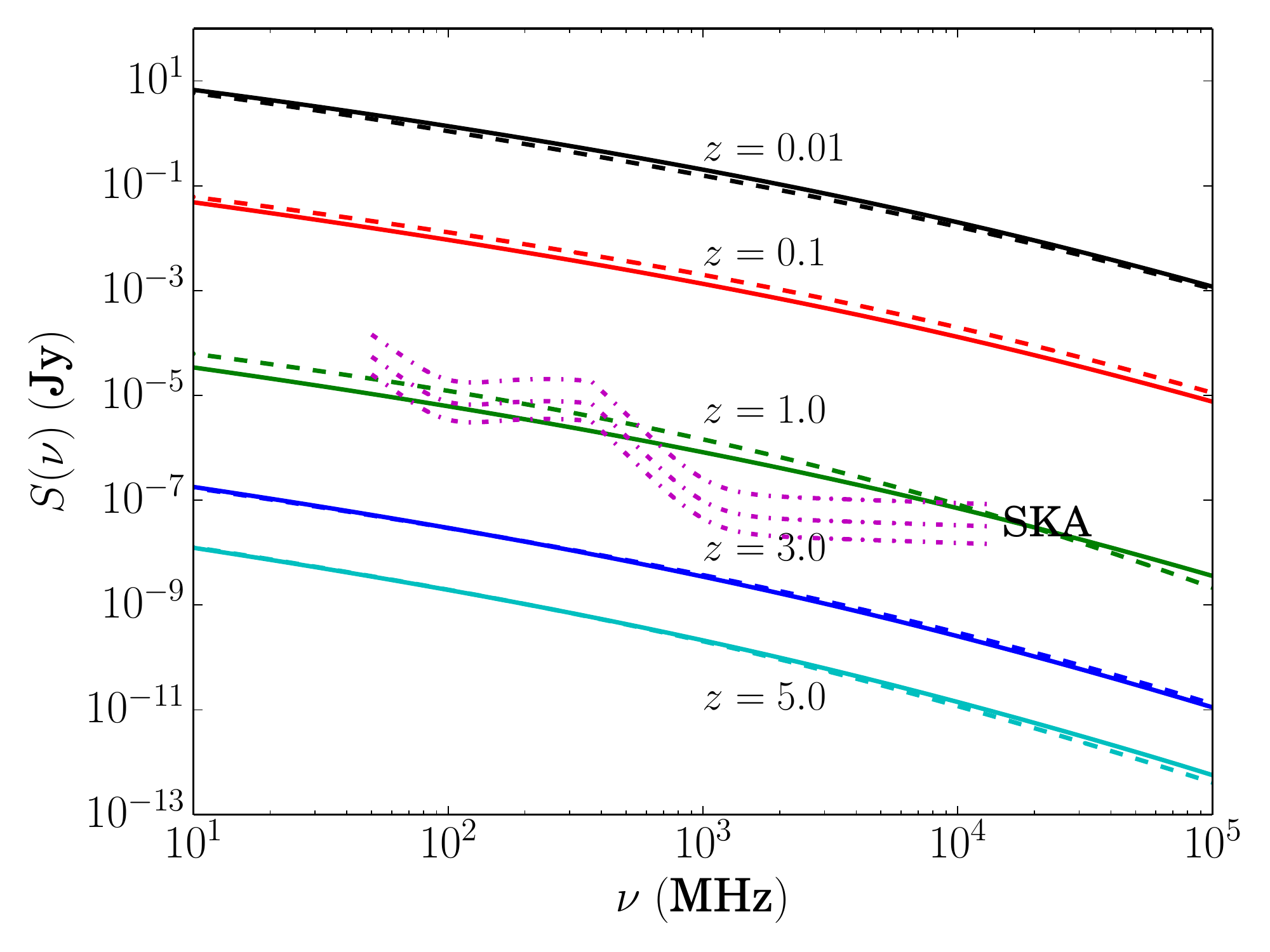}
\caption{Flux densities for galaxy clusters ($M = 10^{15}$M$_{\odot}$), the halo profile is NFW and $\langle B \rangle = 5$ $\mu$G. WIMP mass is $500$ GeV and the composition is $b\overline{b}$. See the caption of Figure~\ref{fig:bb60_m7_b5_nfw} for legend.}
\label{fig:bb500_m15_b5_nfw}
\end{figure}

\begin{figure}[htbp]
\centering
\includegraphics[scale=0.37]{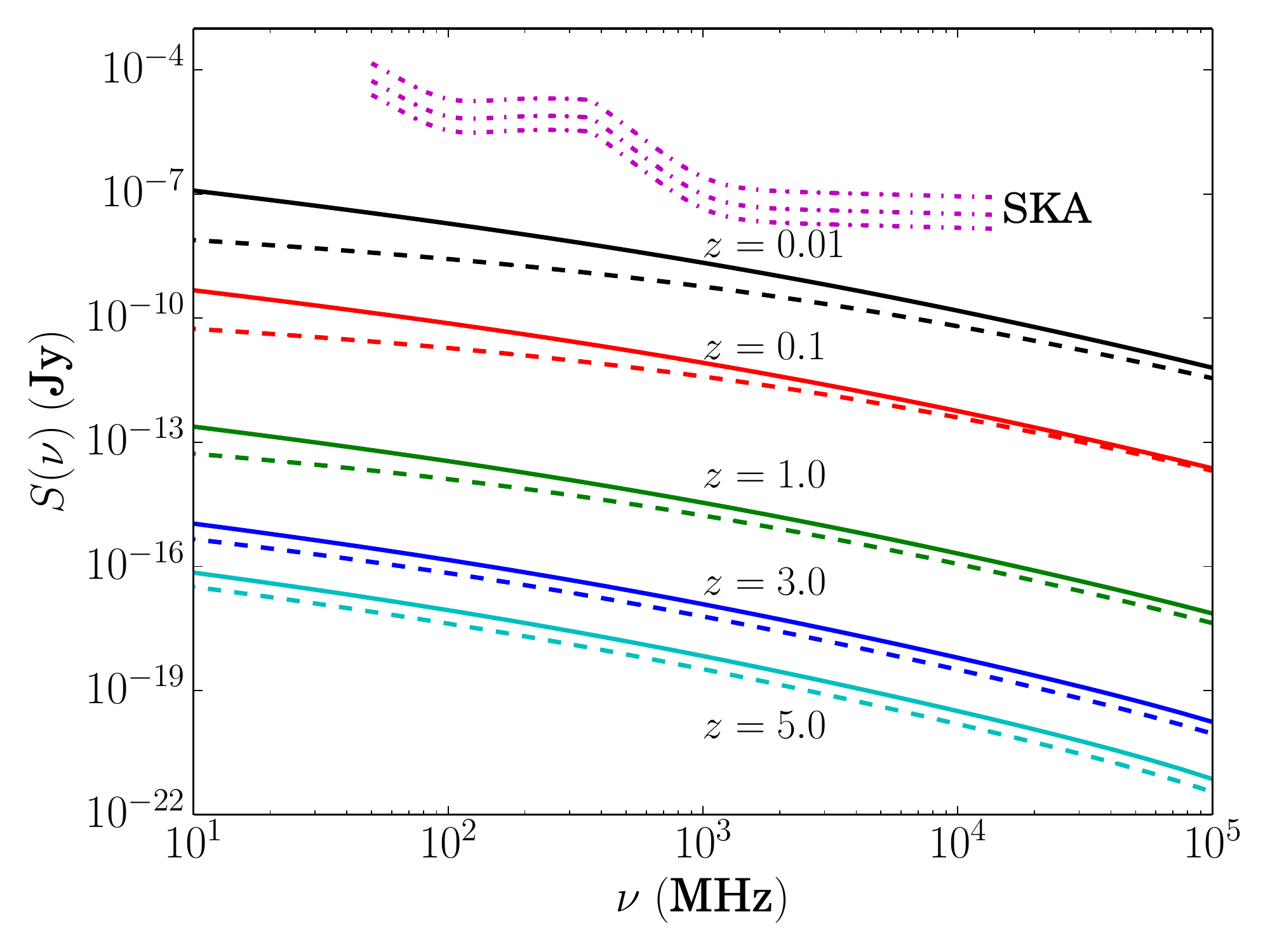}
\includegraphics[scale=0.37]{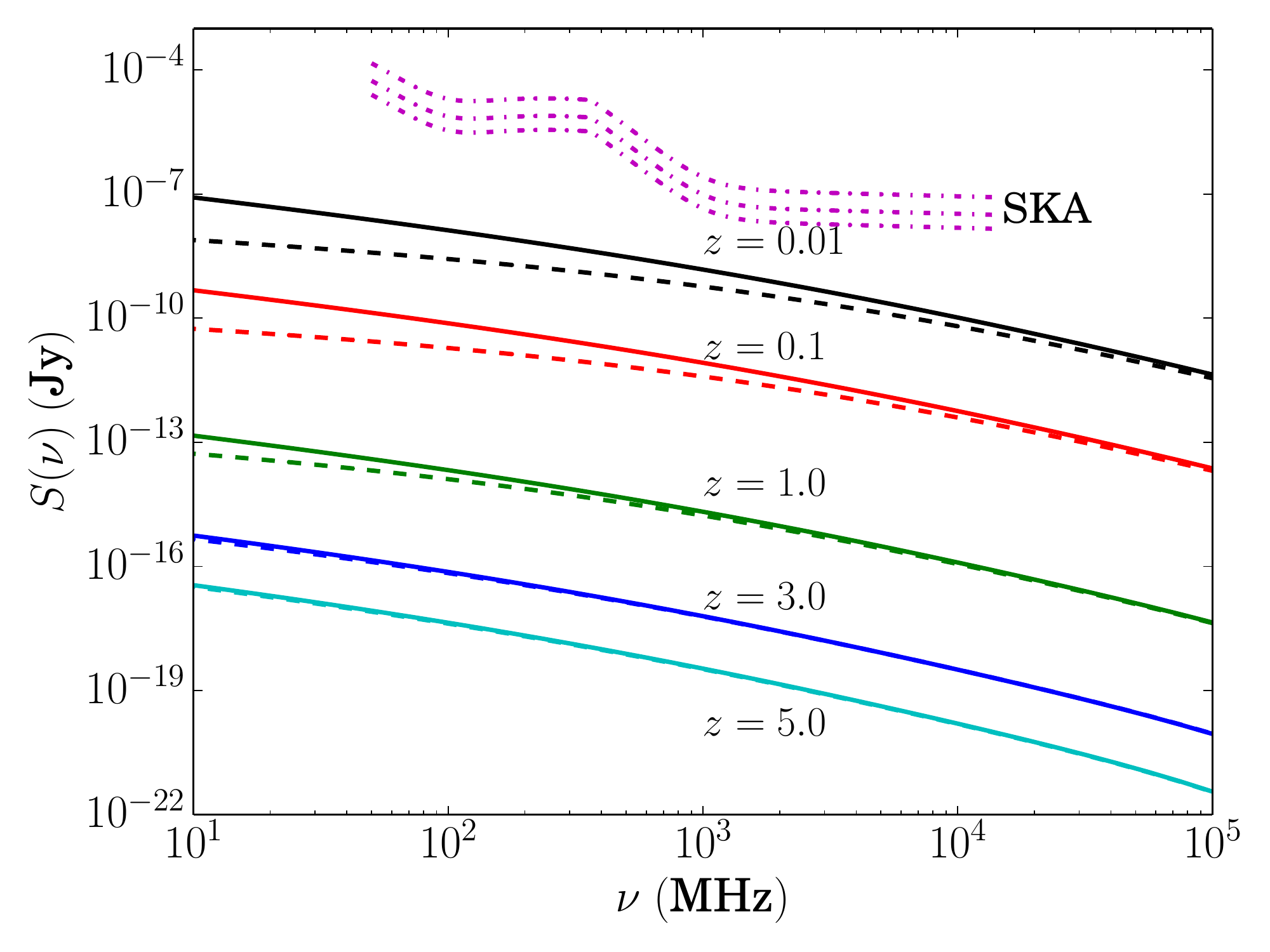}
\caption{Flux densities for dwarf spheroidal galaxies ($M = 10^{7}$M$_{\odot}$), the halo profile is NFW and $\langle B \rangle = 1$ $\mu$G. WIMP mass is $500$ GeV and the composition is $b\overline{b}$. See the caption of Figure~\ref{fig:bb60_m7_b5_nfw} for legend.}
\label{fig:bb500_m7_b1_nfw}
\end{figure}

\begin{figure}[htbp]
\centering
\includegraphics[scale=0.37]{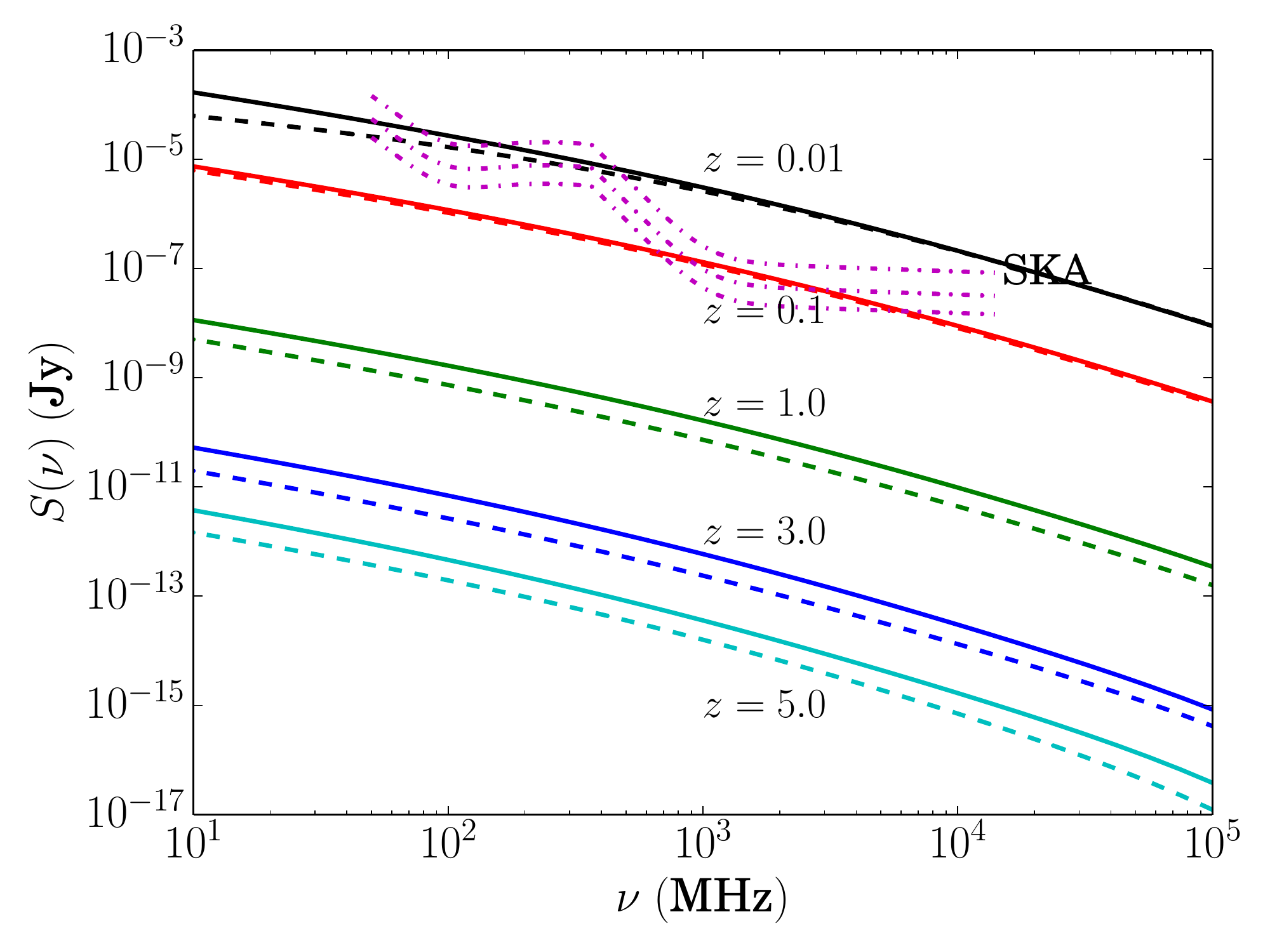}
\includegraphics[scale=0.37]{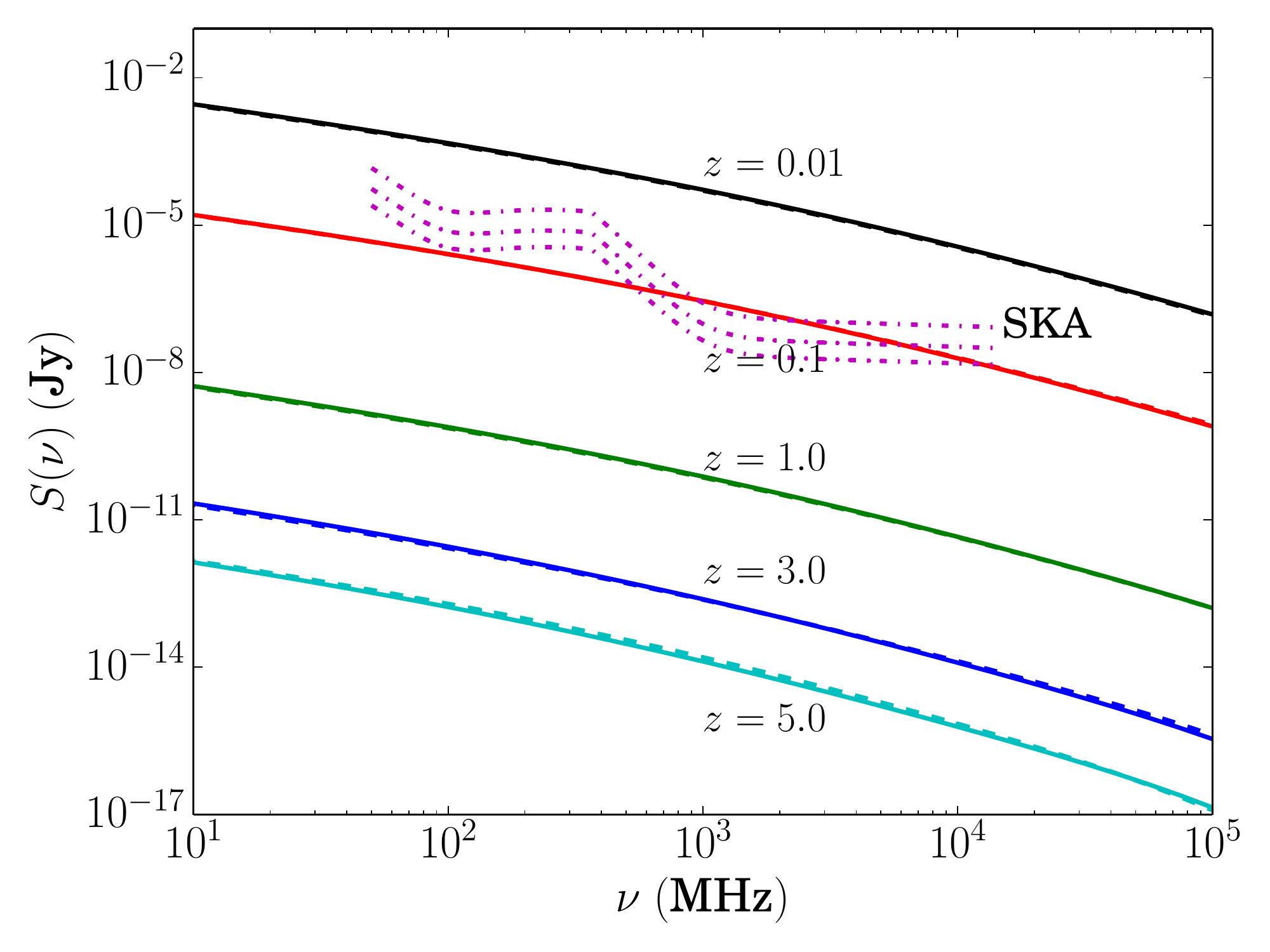}
\caption{Flux densities for galaxies ($M = 10^{12}$M$_{\odot}$), the halo profile is NFW and $\langle B \rangle = 1$ $\mu$G. WIMP mass is $500$ GeV and the composition is $b\overline{b}$. See the caption of Figure~\ref{fig:bb60_m7_b5_nfw} for legend.}
\label{fig:bb500_m12_b1_nfw}
\end{figure}

\begin{figure}[htbp]
\centering
\includegraphics[scale=0.37]{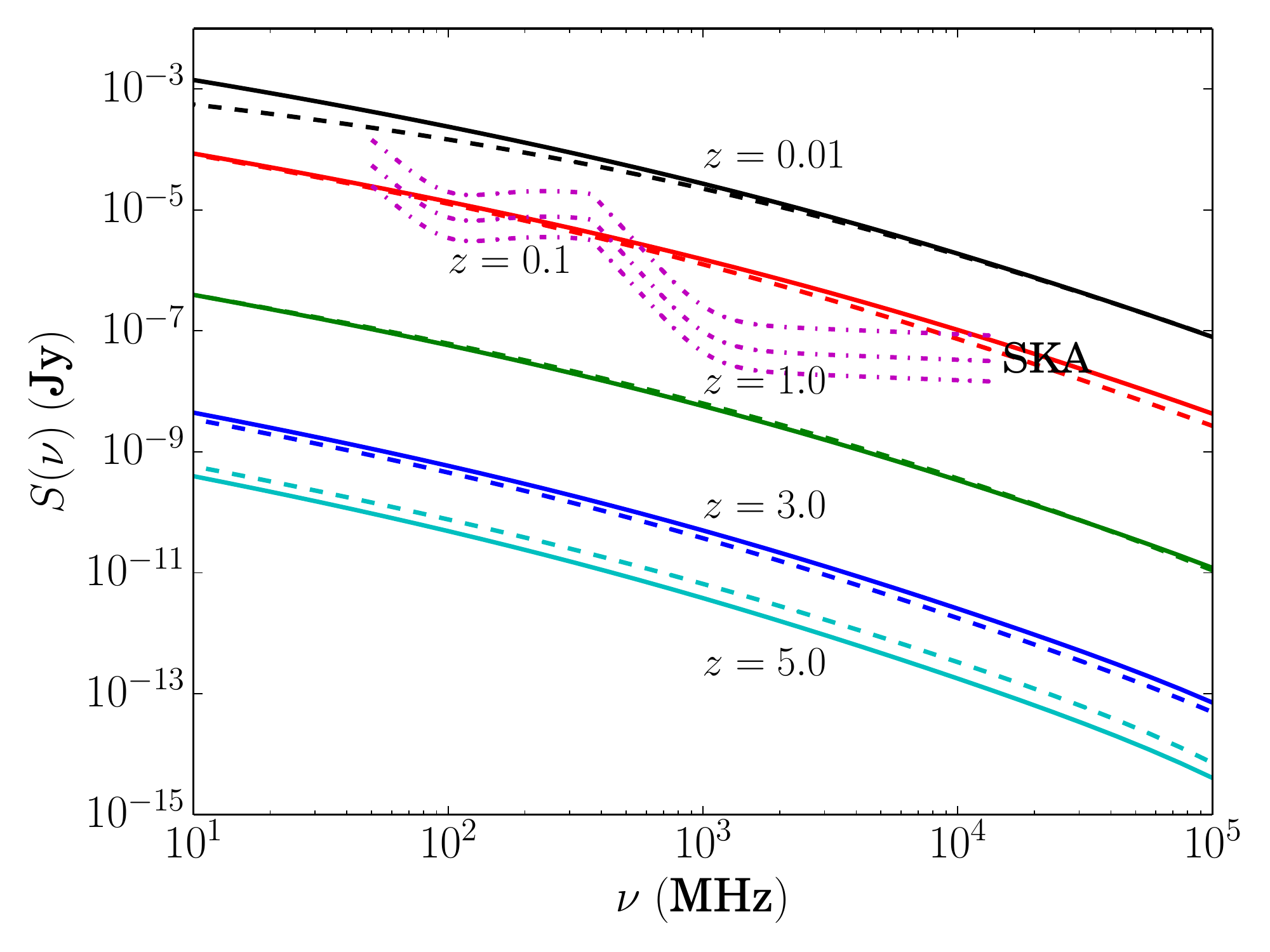}
\includegraphics[scale=0.37]{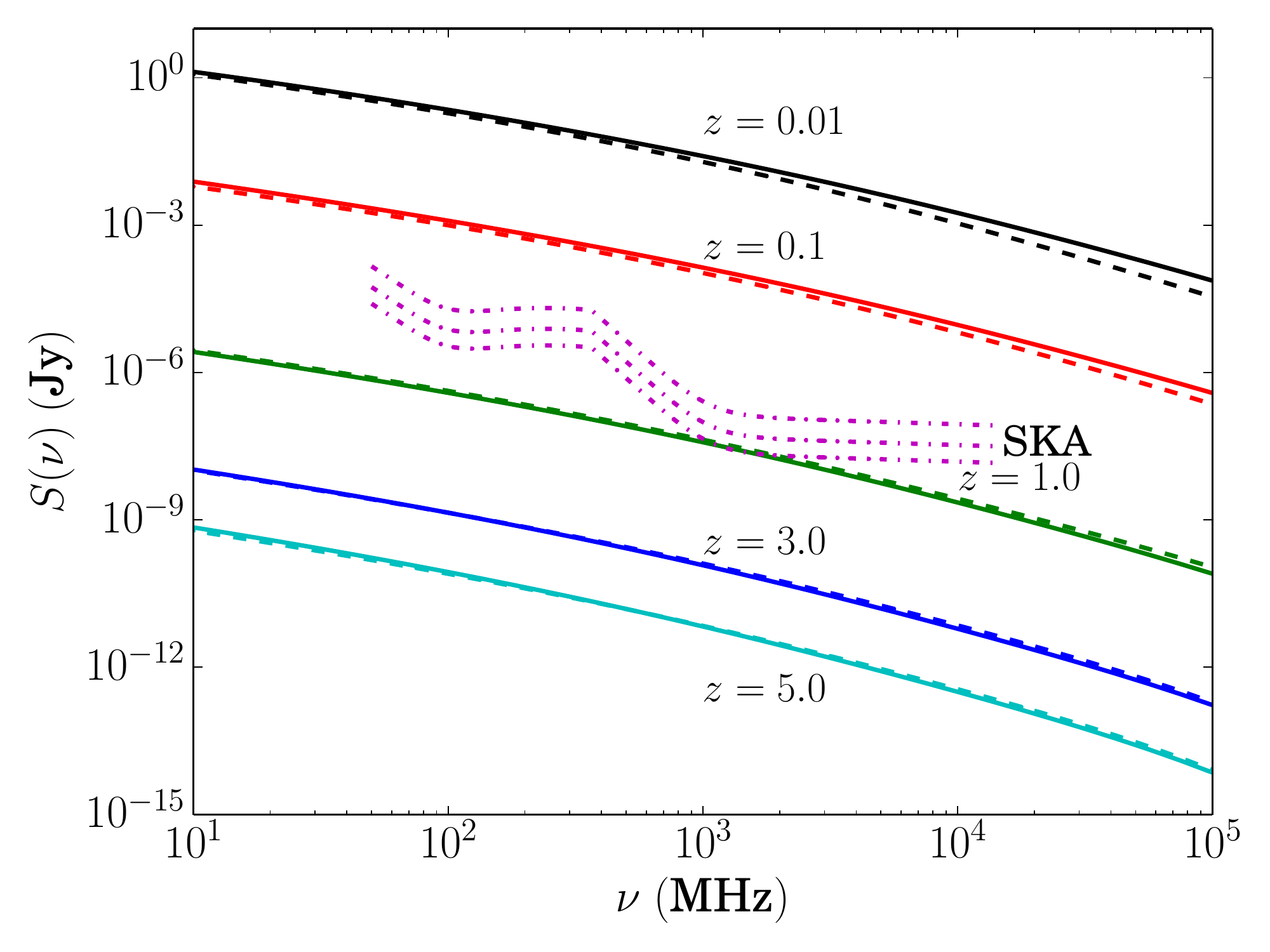}
\caption{Flux densities for galaxy clusters ($M = 10^{15}$M$_{\odot}$), the halo profile is NFW and $\langle B \rangle = 1$ $\mu$G. WIMP mass is $500$ GeV and the composition is $b\overline{b}$. See the caption of Figure~\ref{fig:bb60_m7_b5_nfw} for legend.}
\label{fig:bb500_m15_b1_nfw}
\end{figure}

\begin{figure}[htbp]
\centering
\includegraphics[scale=0.37]{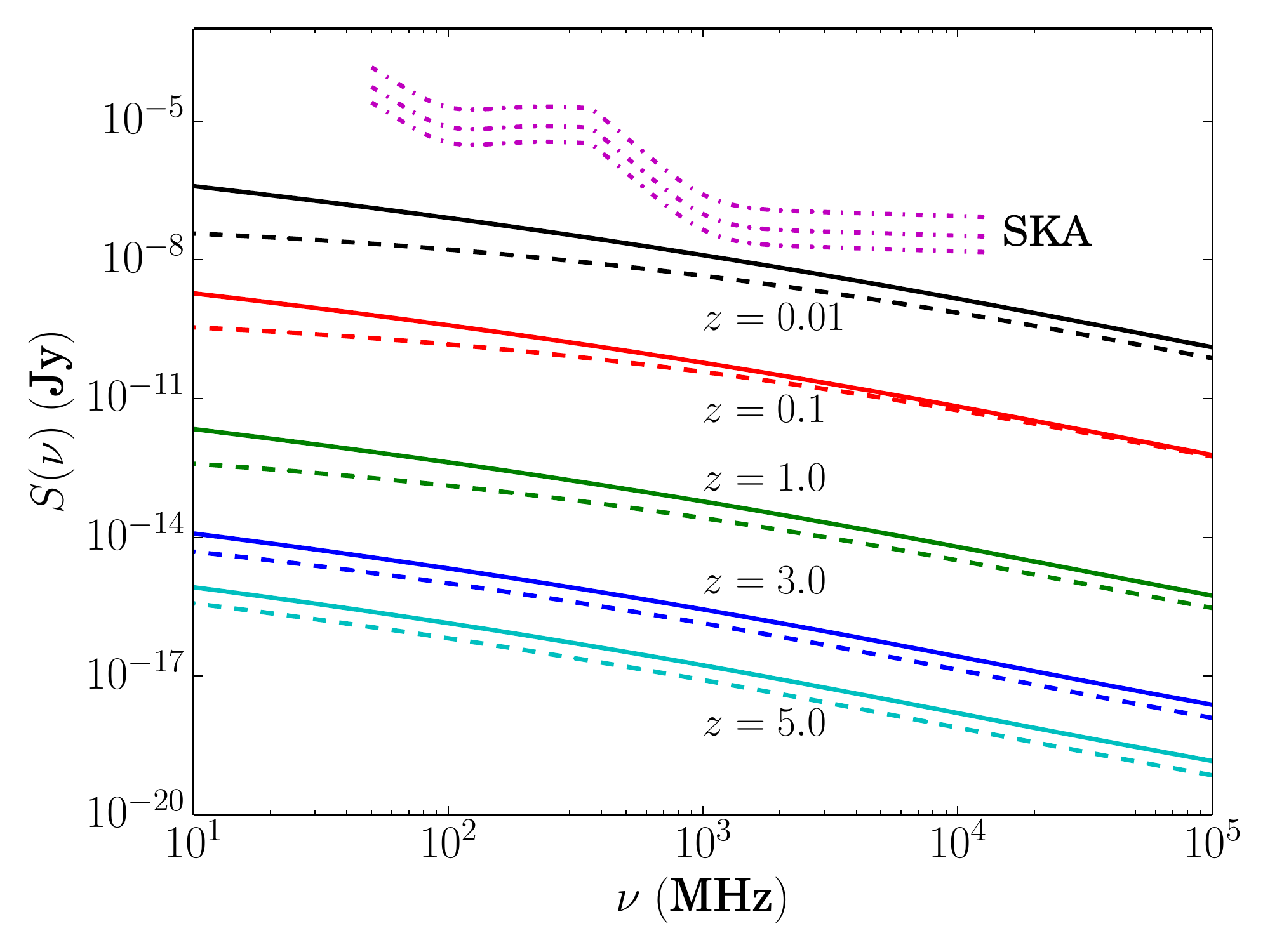}
\includegraphics[scale=0.37]{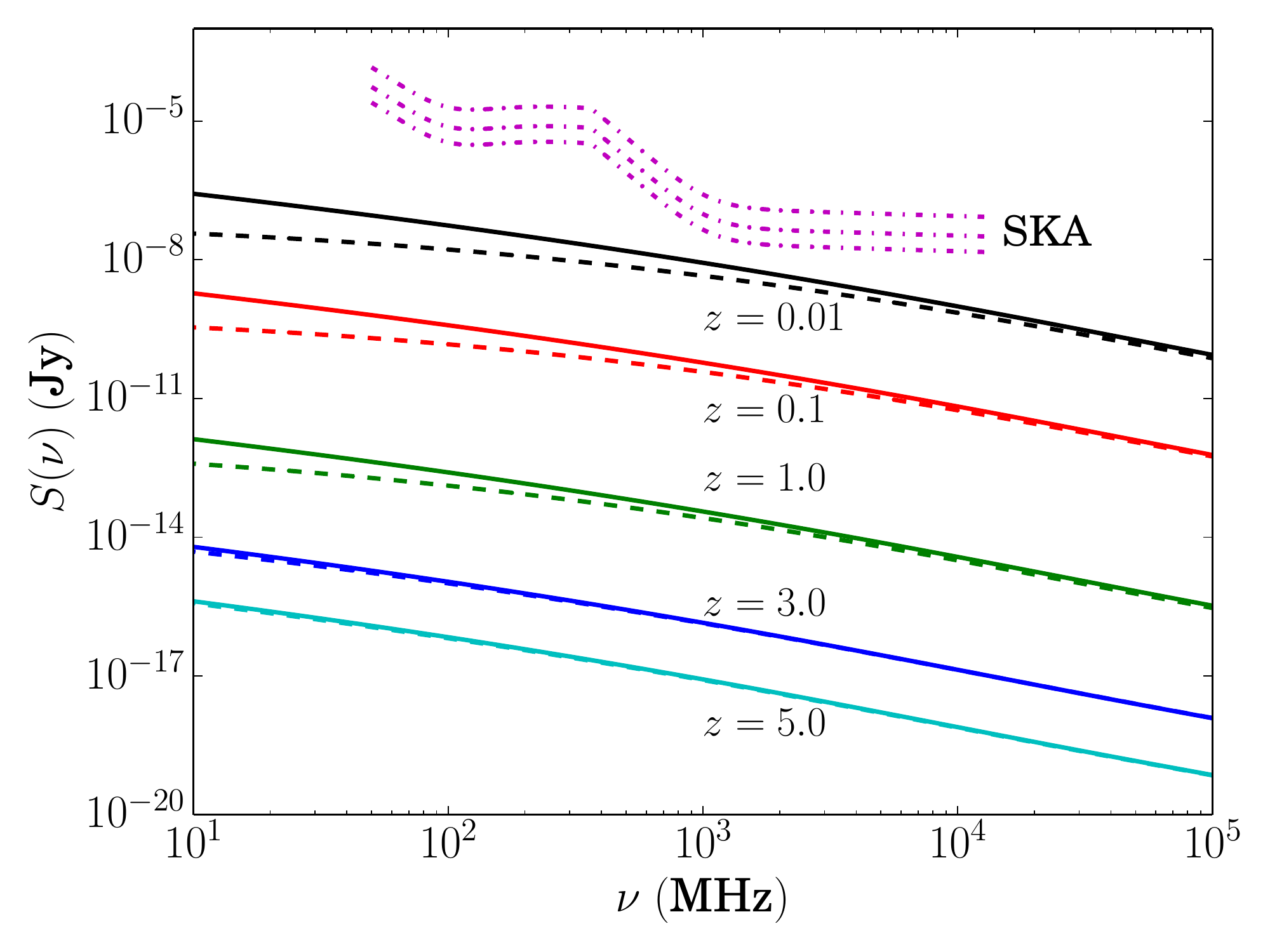}
\caption{Flux densities for dwarf spheroidal galaxies ($M = 10^{7}$M$_{\odot}$), $\langle B \rangle = 5$ $\mu$G. WIMP mass is $500$ GeV and the composition is $W^+ W^-$. See the caption of Figure~\ref{fig:bb60_m7_b5_nfw} for legend.}
\label{fig:ww500_m7_b5_nfw}
\end{figure}

\begin{figure}[htbp]
\centering
\includegraphics[scale=0.37]{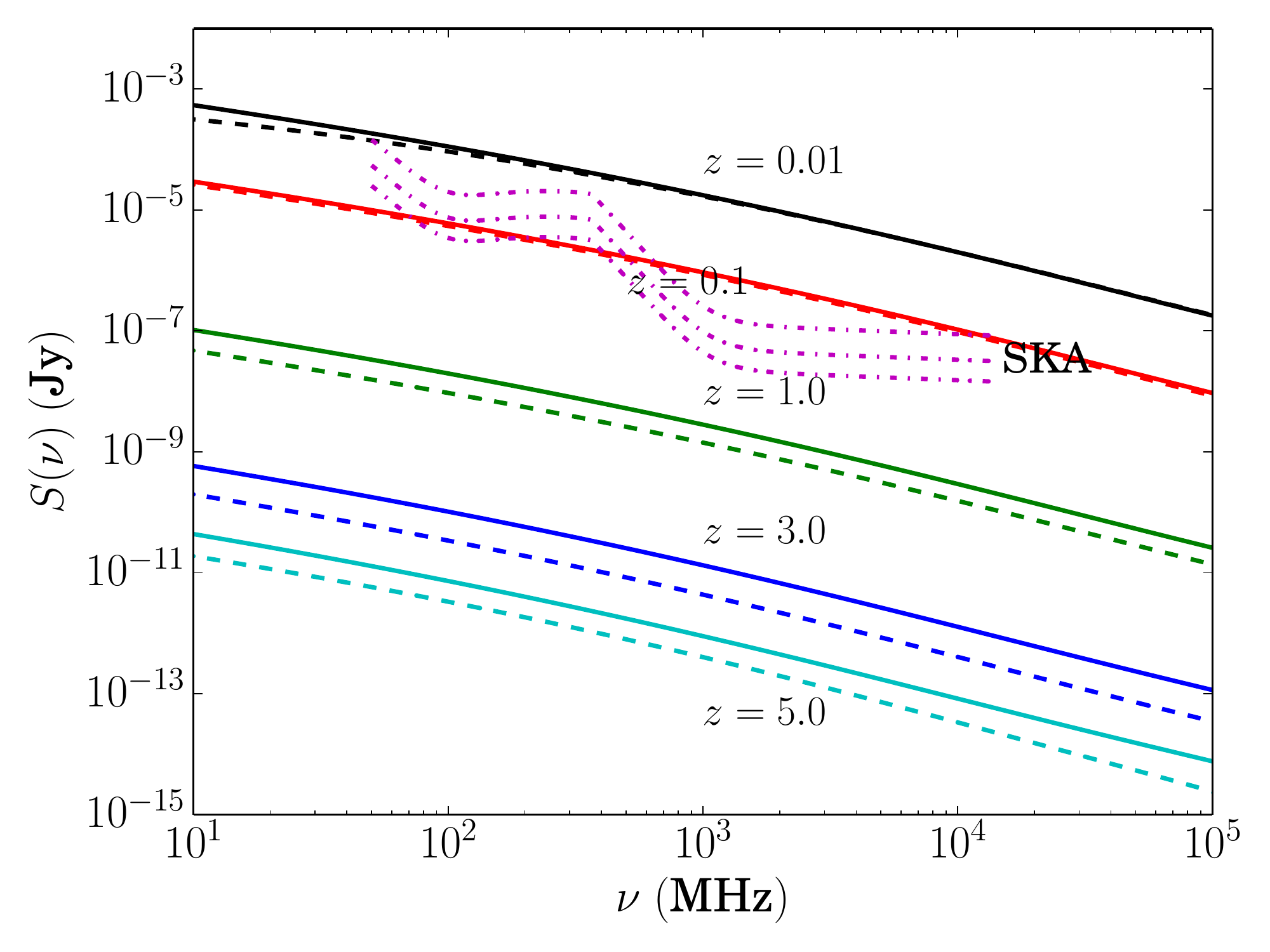}
\includegraphics[scale=0.37]{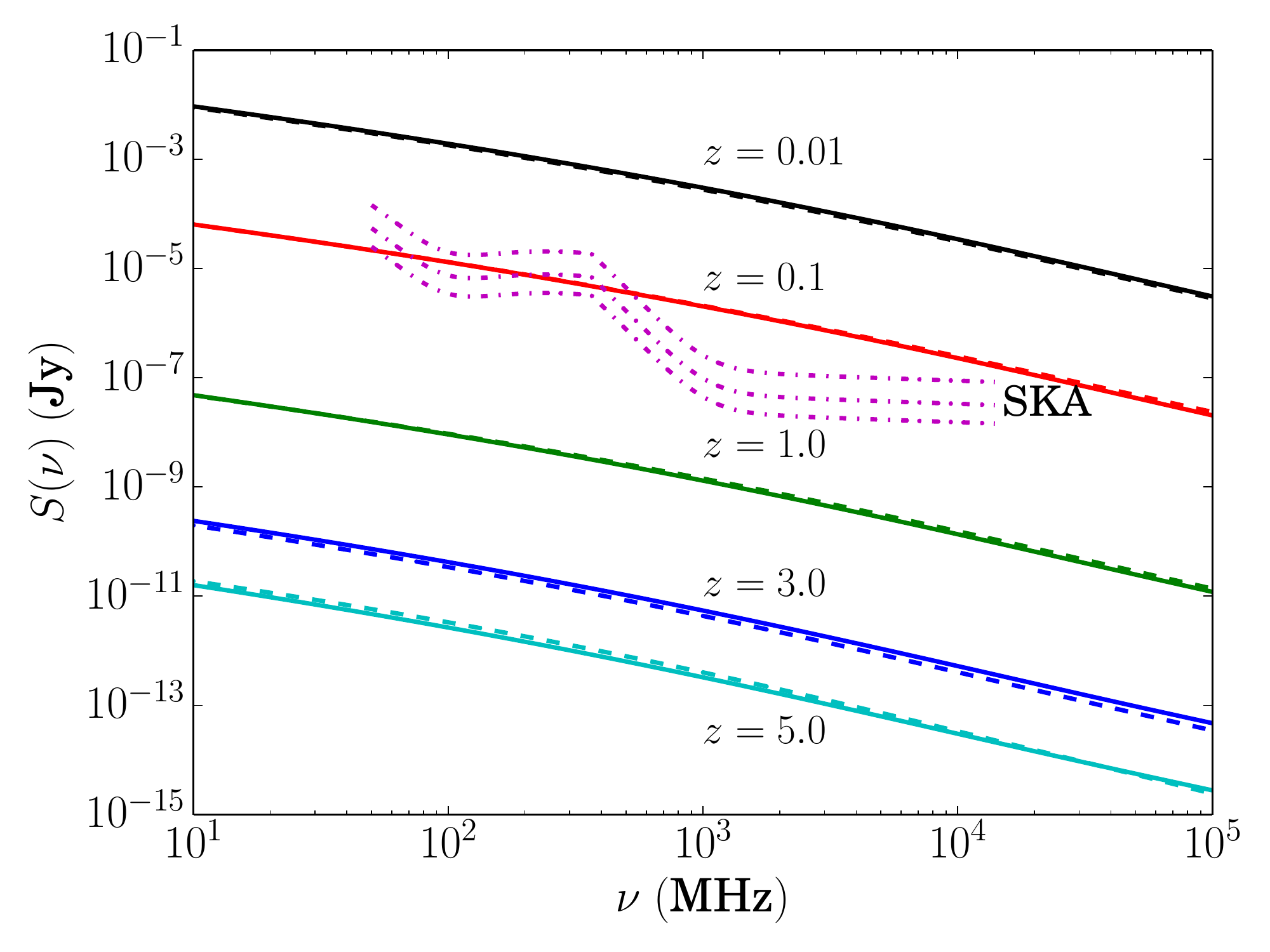}
\caption{Flux densities for galaxies ($M = 10^{12}$M$_{\odot}$), $\langle B \rangle = 5$ $\mu$G. WIMP mass is $500$ GeV and the composition is $W^+ W^-$. See the caption of Figure~\ref{fig:bb60_m7_b5_nfw} for legend.}
\label{fig:ww500_m12_b5_nfw}
\end{figure}

\begin{figure}[htbp]
\centering
\includegraphics[scale=0.37]{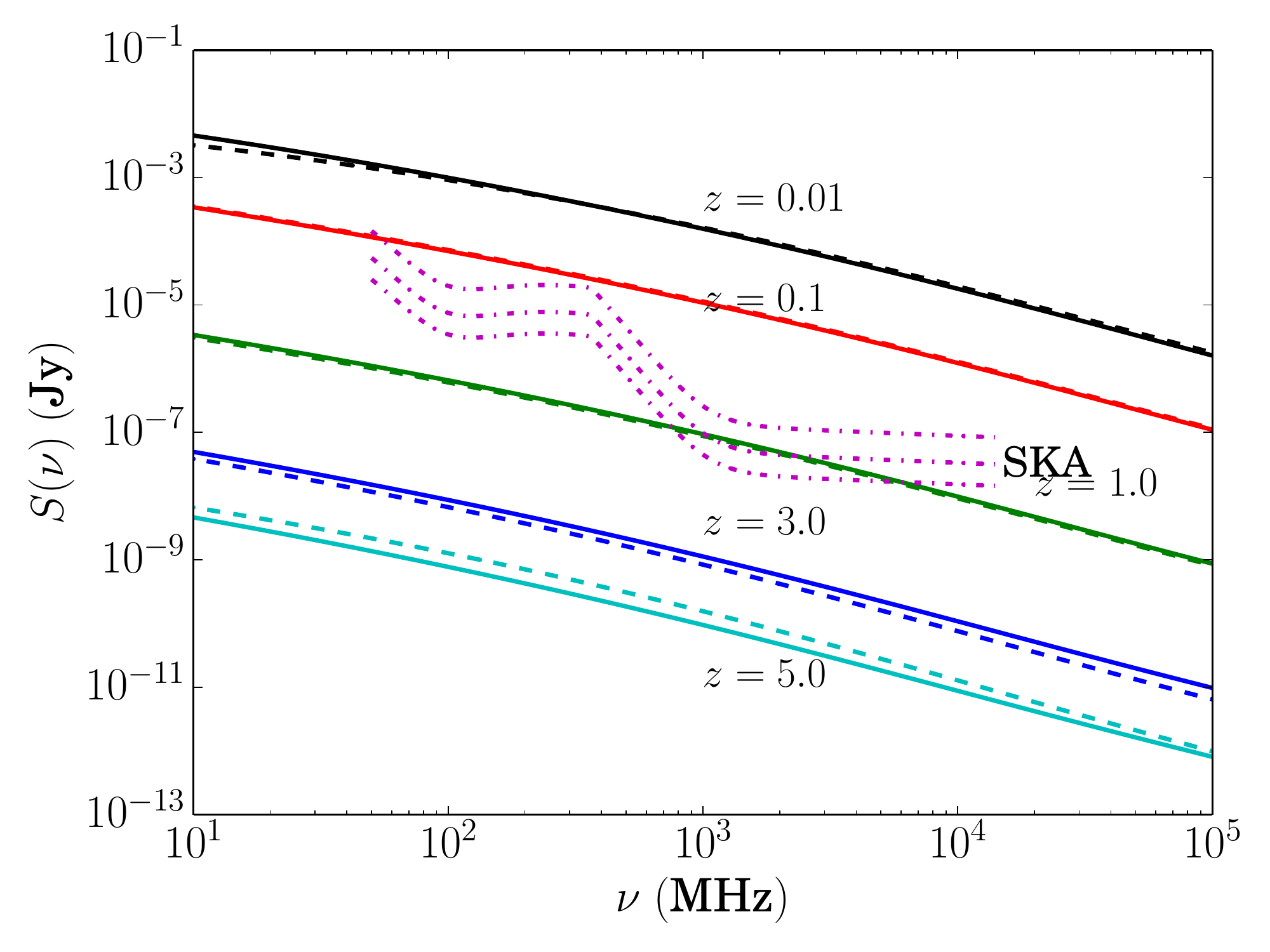}
\includegraphics[scale=0.37]{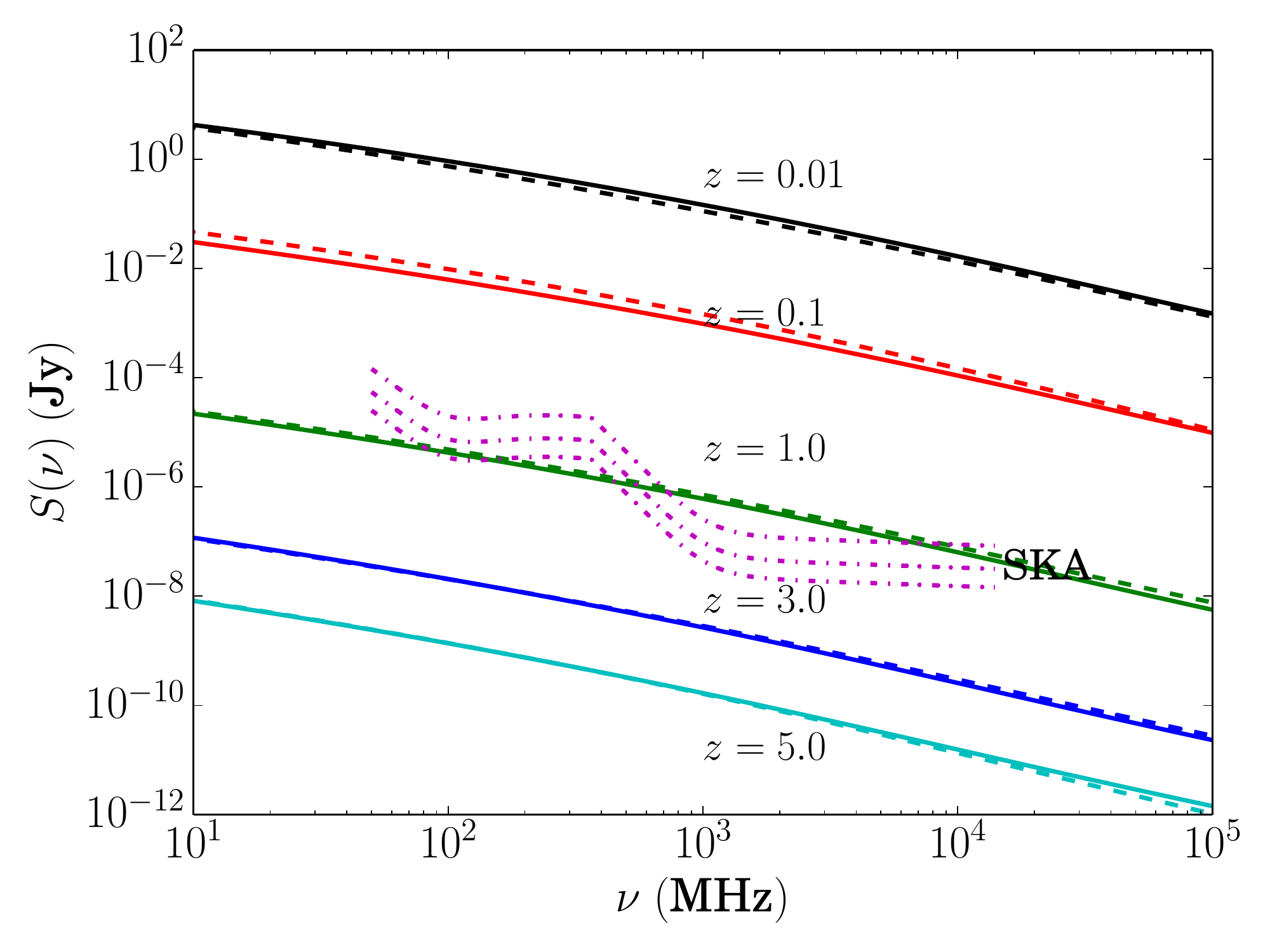}
\caption{Flux densities for galaxy clusters ($M = 10^{15}$M$_{\odot}$), the halo profile is NFW and $\langle B \rangle = 5$ $\mu$G. WIMP mass is $500$ GeV and the composition is $W^+ W^-$. See the caption of Figure~\ref{fig:bb60_m7_b5_nfw} for legend.}
\label{fig:ww500_m15_b5_nfw}
\end{figure}

\begin{figure}[htbp]
\centering
\includegraphics[scale=0.37]{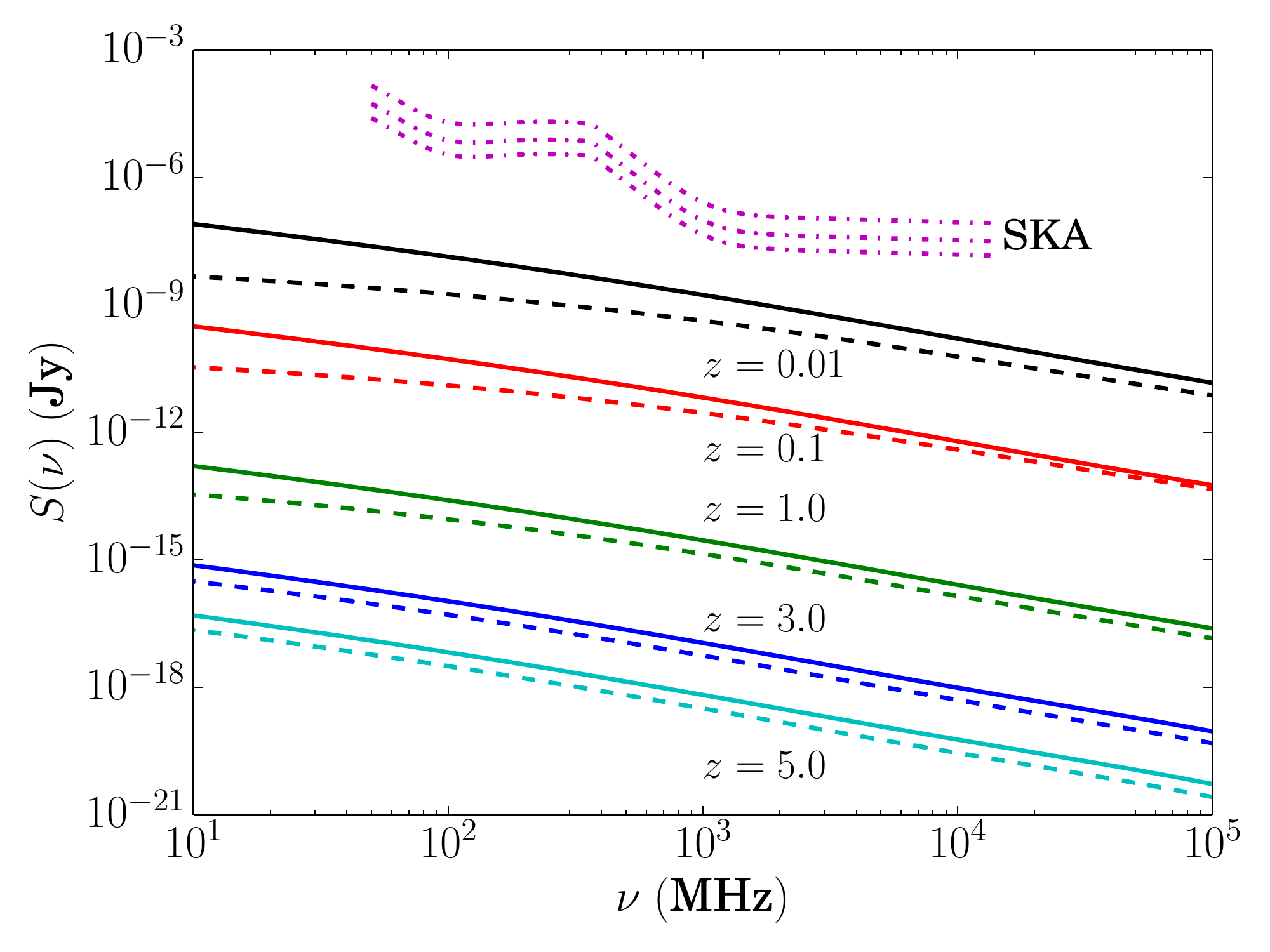}
\includegraphics[scale=0.37]{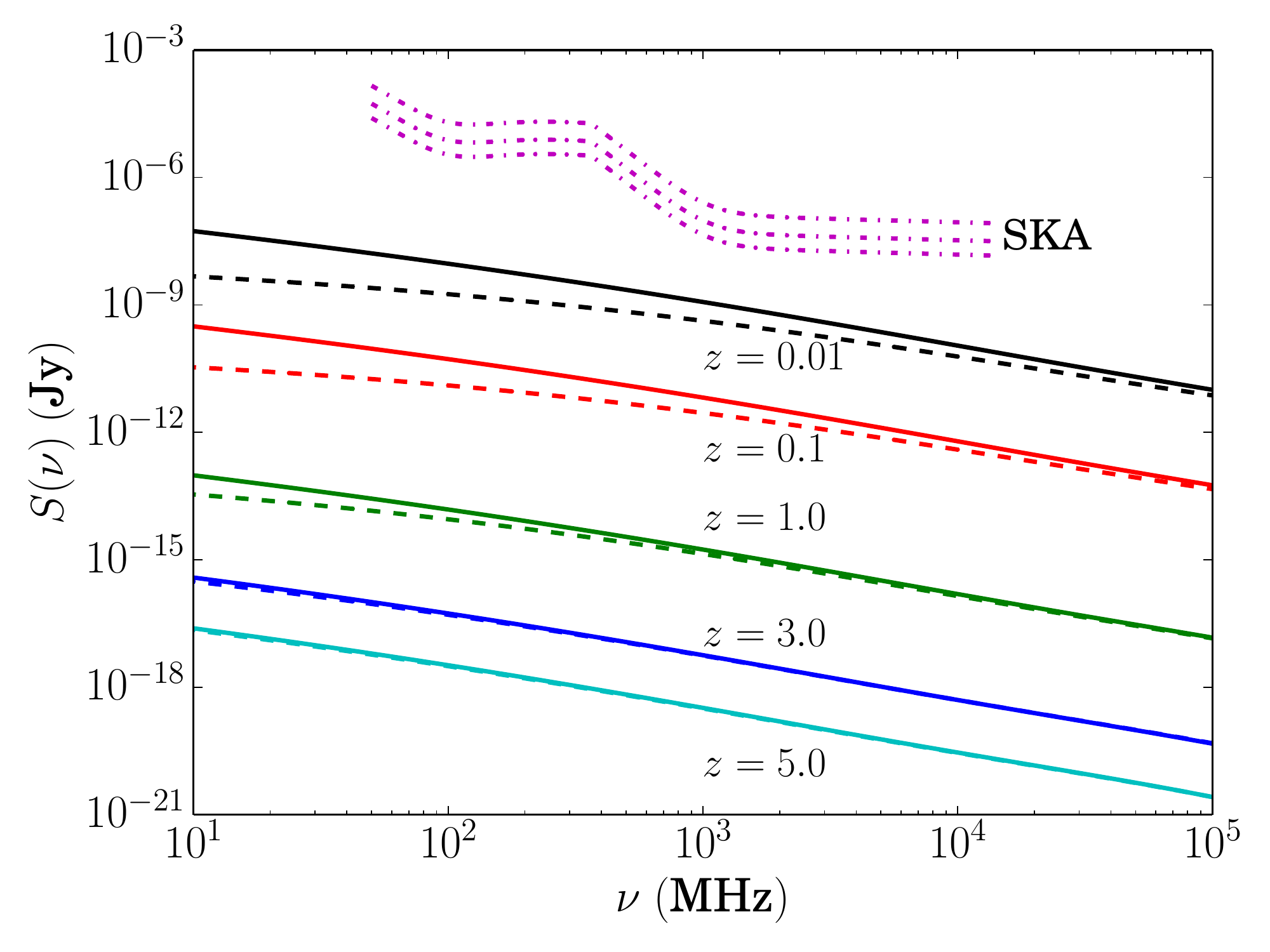}
\caption{Flux densities for dwarf spheroidal galaxies ($M = 10^{7}$M$_{\odot}$), the halo profile is NFW and $\langle B \rangle = 1$ $\mu$G. WIMP mass is $500$ GeV and the composition is $W^+ W^-$. See the caption of Figure~\ref{fig:bb60_m7_b5_nfw} for legend.}
\label{fig:ww500_m7_b1_nfw}
\end{figure}

\begin{figure}[htbp]
\centering
\includegraphics[scale=0.37]{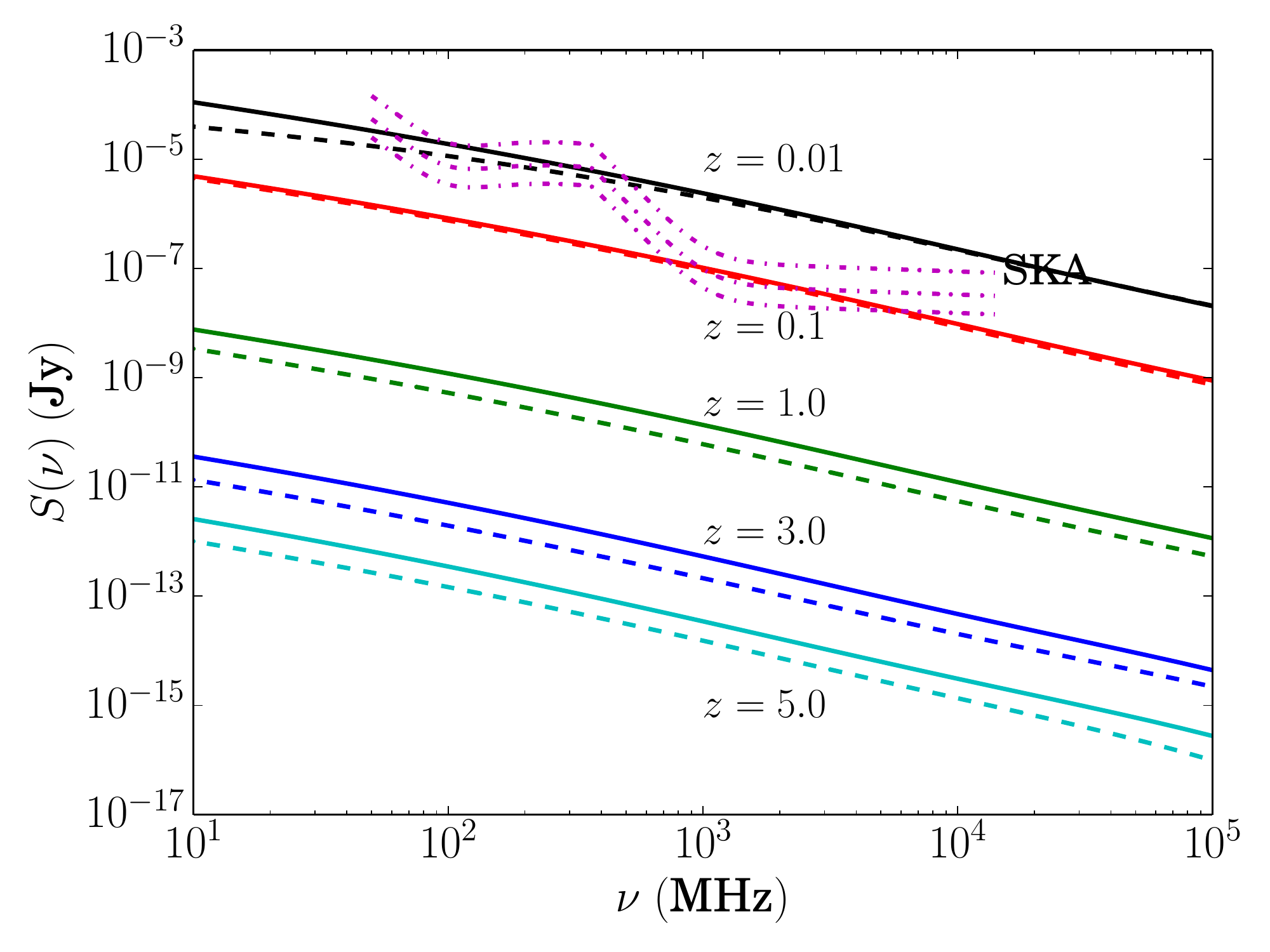}
\includegraphics[scale=0.37]{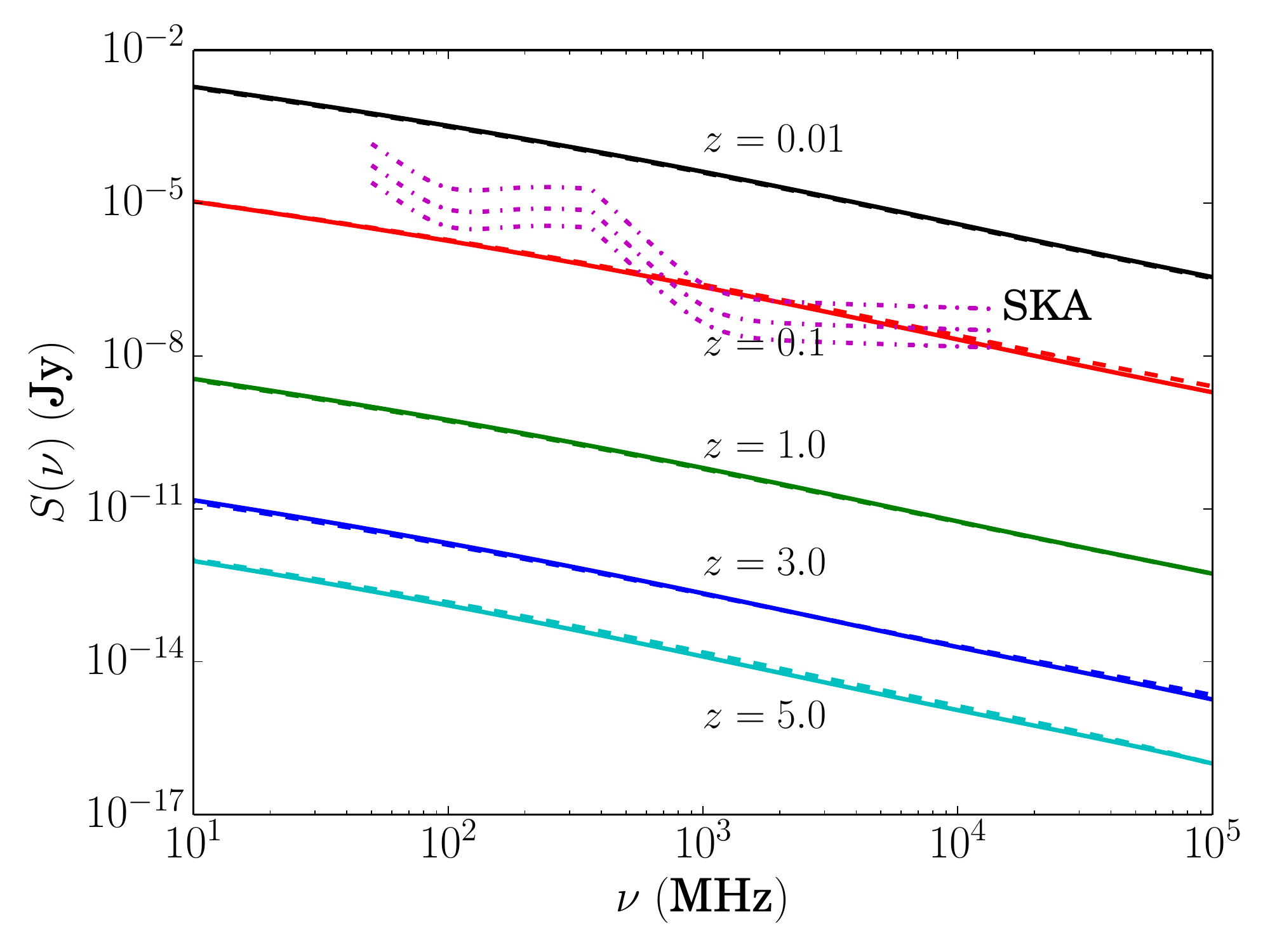}
\caption{Flux densities for galaxies ($M = 10^{12}$M$_{\odot}$), the halo profile is NFW and $\langle B \rangle = 1$ $\mu$G. WIMP mass is $500$ GeV and the composition is $W^+ W^-$. See the caption of Figure~\ref{fig:bb60_m7_b5_nfw} for legend.}
\label{fig:ww500_m12_b1_nfw}
\end{figure}

\begin{figure}[htbp]
\centering
\includegraphics[scale=0.37]{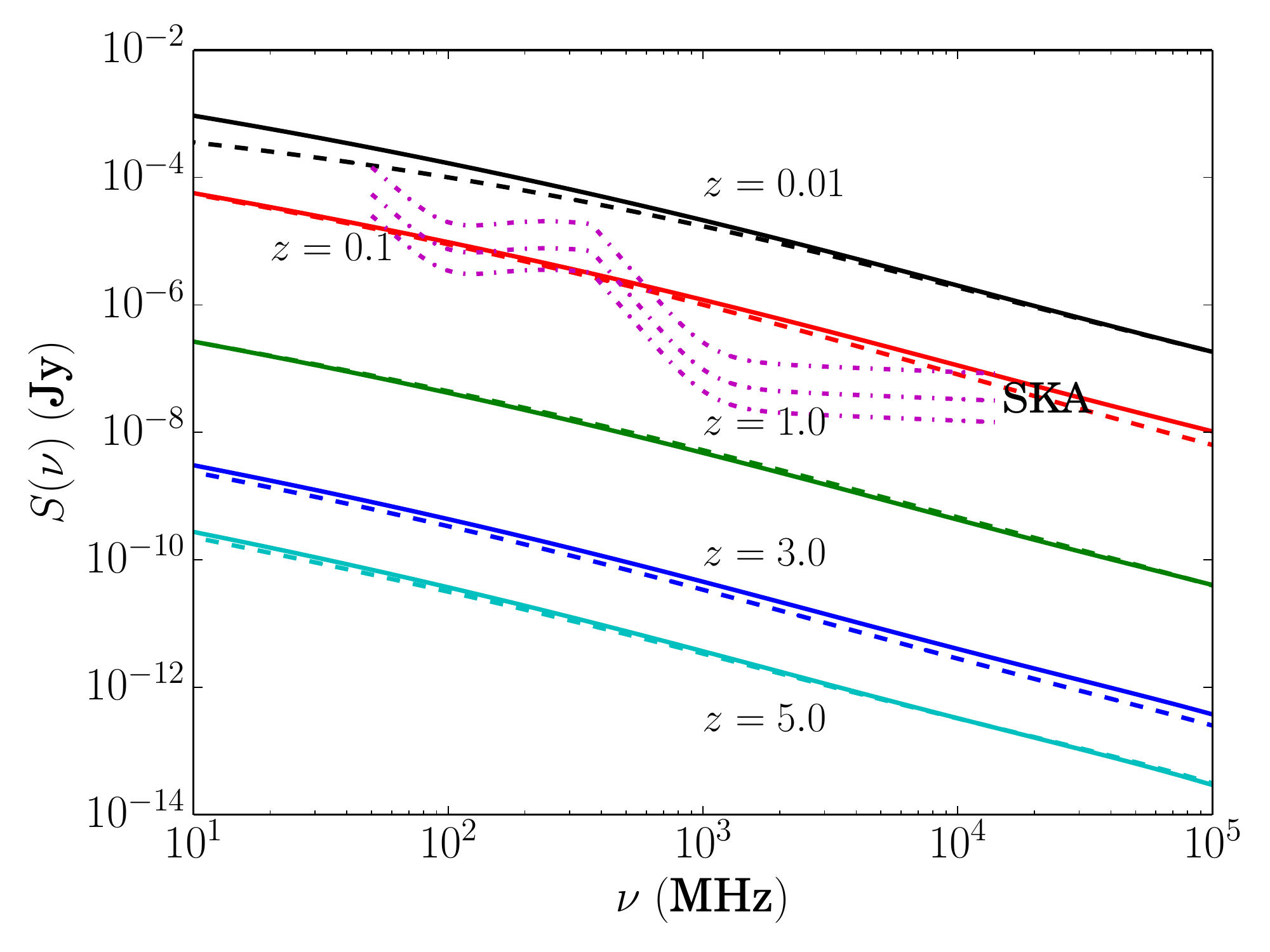}
\includegraphics[scale=0.37]{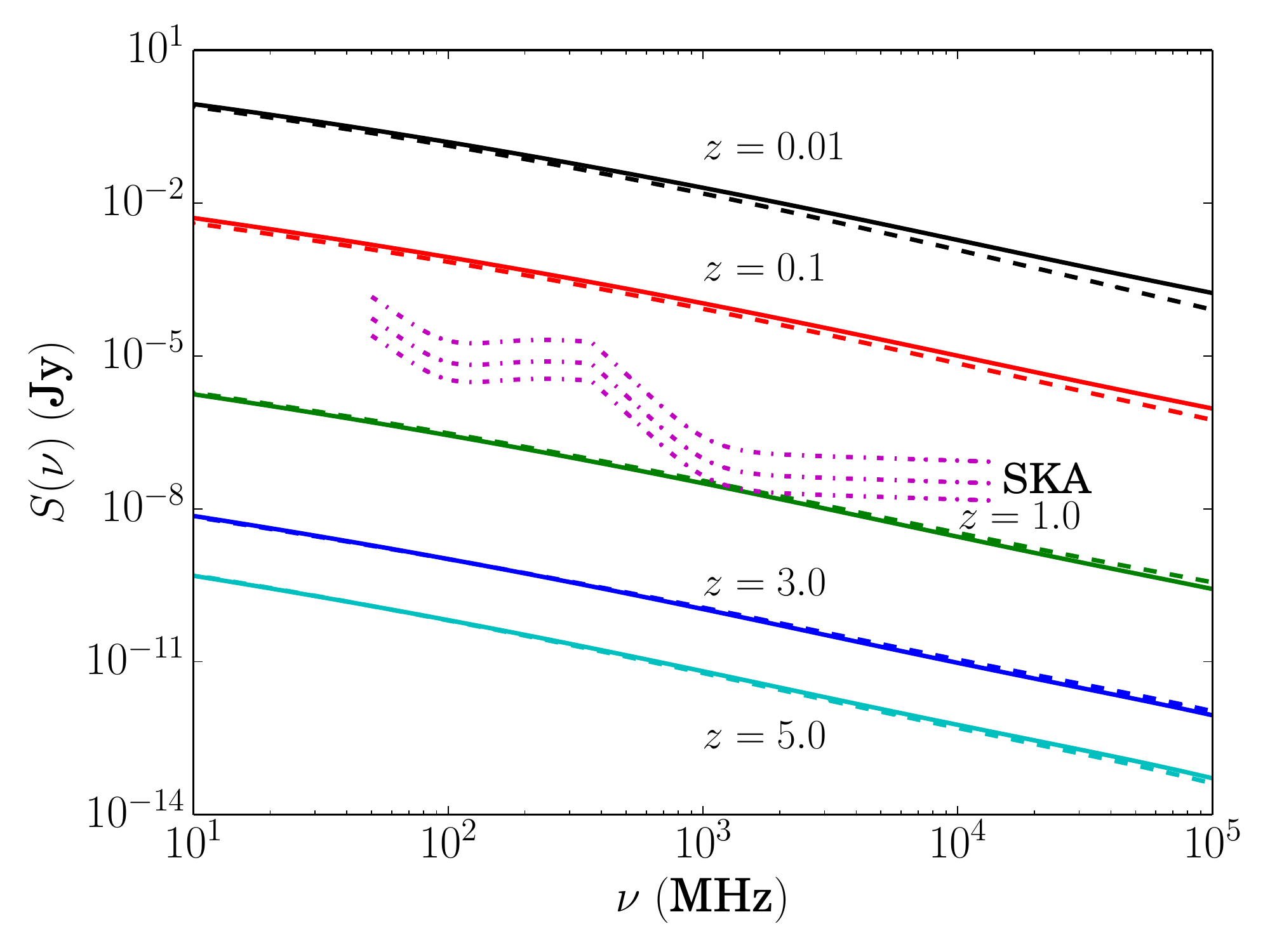}
\caption{Flux densities for galaxy clusters ($M = 10^{15}$M$_{\odot}$), the halo profile is NFW and $\langle B \rangle = 1$ $\mu$G. WIMP mass is $500$ GeV and the composition is $W^+ W^-$. See the caption of Figure~\ref{fig:bb60_m7_b5_nfw} for legend.}
\label{fig:ww500_m15_b1_nfw}
\end{figure}

In Figures~\ref{fig:bb60_m7_b5_nfw} to \ref{fig:ww500_m15_b1_nfw} we show the results of an exhaustive study of the neutralino induced synchrotron emissions from dark matter halos with three different masses ($10^7$, $10^{12}$ and $10^{15}$ M$_{\odot}$), two different magnetic field values ($1 \mu$G and $5\mu$G), and five different redshifts ($0.01$, $0.1$, $1.0$, $3.0$ and $5.0$). We also investigate the $b\overline{b}$, $\tau^+ \tau^-$, and $W^+W^-$ annihilation channels for neutralinos of mass $60$ and $500$ GeV. As a reference, a thermally-averaged, zero-temperature cross section value of
$\langle \sigma V \rangle = 3 \times 10^{-26}$ cm$^3$ s$^{-1}$ is assumed. This value is chosen to coincide with the thermal relic abundance upper bound given by~\cite{jungman1996}
\begin{equation}
\langle \sigma V\rangle \le \frac{3 \times 10^{-27}}{\Omega_{\chi}h^2} \; \mbox{cm$^3$ s$^{-1}$} \; ,
\end{equation}
where $\Omega_{\chi}$ is the present day density parameter for neutralino dark matter, $\Omega_{\chi} h^2 \approx 0.1$, and $h$ is the reduced Hubble constant in units of $100$ km s$^{-1}$ Mpc$^{-1}$.

A common theme in these results is that diffusion only plays a role at very small redshift and for very small halo masses, as can be seen in Figs.~\ref{fig:bb60_m7_b5_nfw}, \ref{fig:tt60_m7_b5_nfw}, \ref{fig:bb500_m7_b5_nfw}, and \ref{fig:ww500_m7_b5_nfw} for the calculation of flux within the virial radius. This conclusion is very nicely reinforced by Figure~\ref{fig:loss} which demonstrates that the diffusion time-scale is only comparable to that of energy-loss processes for the small mass halos at small redshift. 

In general, from Figs.~\ref{fig:bb60_m7_b5_nfw}, \ref{fig:tt60_m7_b5_nfw}, \ref{fig:bb500_m7_b5_nfw}, and \ref{fig:ww500_m7_b5_nfw} we note that dwarf galaxy halos with masses around $10^7$ M$_{\odot}$ will at most be marginally detectable at the $1\sigma$ c.l. with the sensitivity of the SKA for 1000 hours of integration time at redshift $z = 0.01$. This place restrictions on mass-redshift combination of halos that can be used for robust detection/constraint of dark matter annihilation processes via the SKA. Larger halos, however, show promising detection of fluxes from within redshifts below one, as can be seen in Figs.~\ref{fig:bb60_m12_b5_nfw}, \ref{fig:tt60_m12_b5_nfw}, \ref{fig:bb500_m12_b5_nfw}, \ref{fig:ww500_m12_b5_nfw}, \ref{fig:bb60_m15_b5_nfw}, \ref{fig:tt60_m15_b5_nfw}, \ref{fig:bb500_m15_b5_nfw}, and \ref{fig:ww500_m15_b5_nfw}. For higher redshifts ($z \ge 3$) we note that all the halos fall below the $1\sigma$ c.l. detection with the SKA sensitivity. 

In the case of the weaker field, as in Figs.~\ref{fig:bb60_m7_b1_nfw}, \ref{fig:bb500_m12_b1_nfw} and, \ref{fig:ww500_m15_b1_nfw}, we see an overall reduction in flux from the similar Figs.~\ref{fig:bb60_m7_b5_nfw}, \ref{fig:bb500_m12_b5_nfw} and \ref{fig:ww500_m15_b5_nfw}. This is due to the fact that the synchrotron power depends upon the magnetic field strength. However, the weaker field has a particular effect that is prominent for the lighter neutralino masses: this is the expected suppression of the production of electrons with energies above a threshold dictated by the neutralino mass and magnetic field strength which results in a dramatic spectral steepening at high frequencies. The cut-off frequency, resulting from the suppression of higher energy electron production, then exhibits a trivial redshift dependence, as seen in Figs.~\ref{fig:tt60_m7_b1_nfw}, \ref{fig:tt60_m12_b1_nfw} and \ref{fig:tt60_m15_b1_nfw}.

It is worth noting that there is little difference between the emissions of the Einasto and NFW profiles as seen in the comparison of Figures~\ref{fig:bb60_m7_b5_ein}, \ref{fig:bb60_m12_b5_ein} (for the Einasto case) and \ref{fig:bb60_m15_b5_ein} with their NFW case counterparts in Figs.\ref{fig:bb60_m7_b5_nfw}, \ref{fig:bb60_m12_b5_nfw} and \ref{fig:bb60_m15_b5_nfw}. This similarity is due to the choice of $\alpha$, however, the effect of the sub-halos results in a slightly higher magnitude for the spectra of halos with Einasto profile.We also show in Figure~\ref{fig:bb60_m7_b5_burkert} that in the case of dwarf galaxies there is little change to our results from the adoption of a cored profile, in this case a Burkert profile (see eq.\eqref{eq:nfw}). The exploration of this case is significant since dwarf galaxy kinematics are also compatible with shallower cored profiles~\cite{Walker2009,Adams2014}.

\begin{figure}[htbp]
\centering
\includegraphics[scale=0.37]{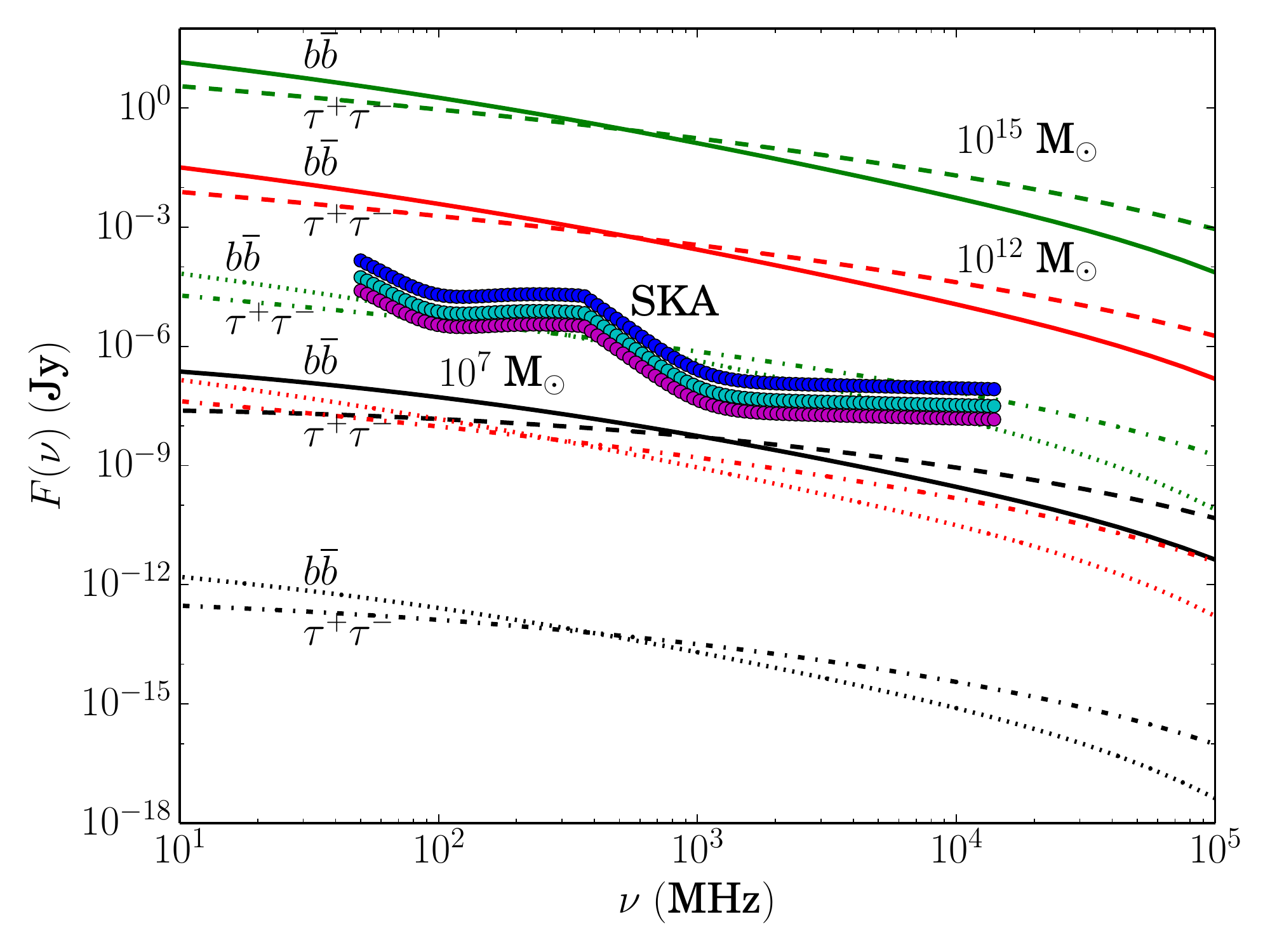}
\includegraphics[scale=0.37]{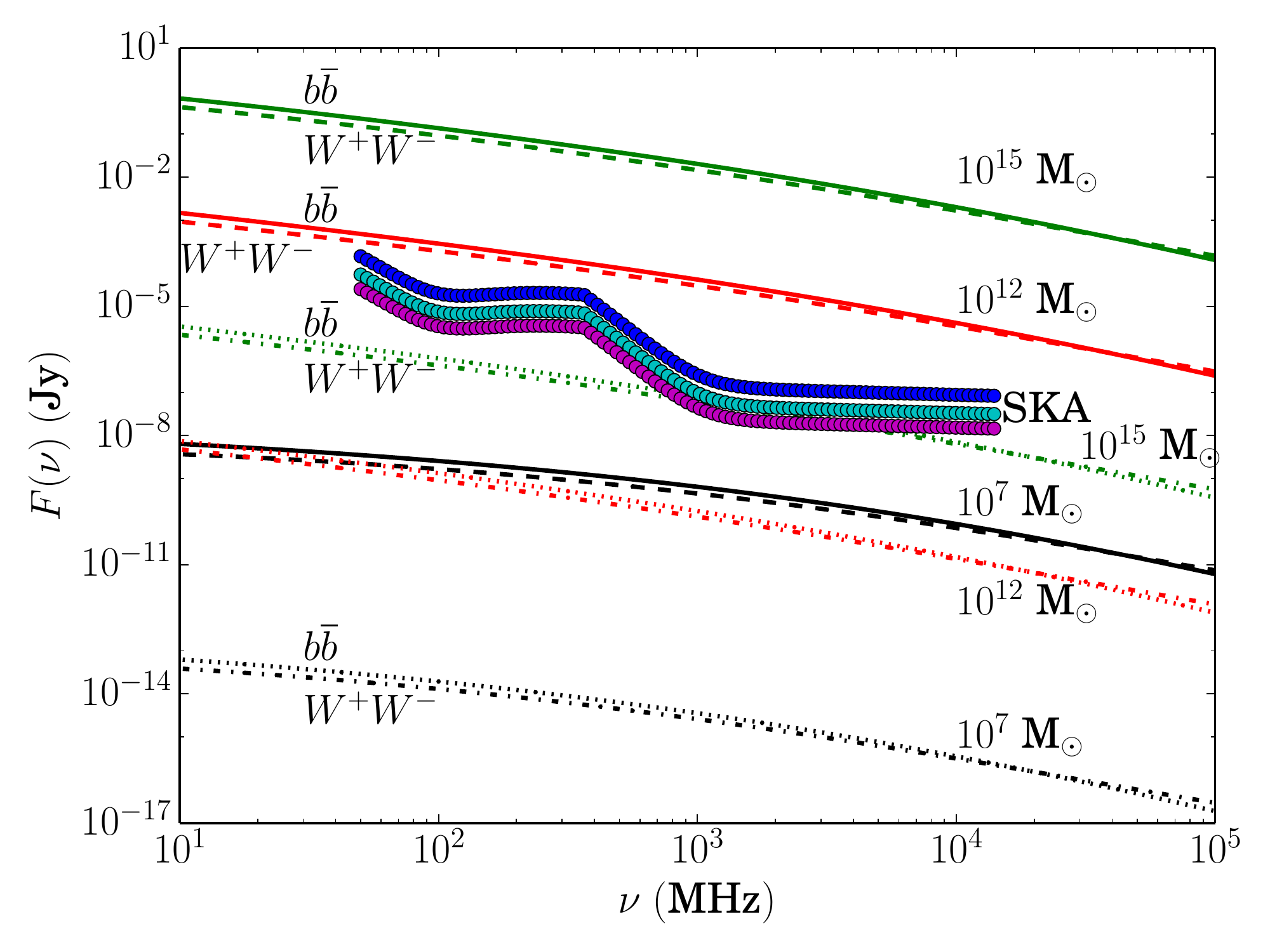}
\caption{Comparison of fluxes from different annihilation channels. Green lines are for halos of mass $10^{15}$ $M_{\odot}$, red for $10^{12}$ $M_{\odot}$ and black for $10^{7}$ $M_{\odot}$. Left: $m_{\chi} = 60$ GeV, solid lines correspond to the $b\bar{b}$ channel with $z=0.01$, dotted lines to the same channel with $z=1.0$, and dashed lines to $\tau^+\tau-$ with $z=0.01$ while dash-dotted lines are for $z=1.0$. Right: $m_{\chi} = 500$ GeV, solid lines correspond to the $b\bar{b}$ channel with $z=0.01$, dotted lines to the same channel with $z=1.0$, and dashed lines to $W^+W-$ with $z=0.01$ while dash-dotted lines are for $z=1.0$. The circle points are SKA 1$\sigma$ sensitivity limits for 30, 240, and 1000 hour integration times respectively.}
\label{fig:spectral-sig}
\end{figure}

Importantly, the different choices of neutralino annihilation channel affect the shape of the spectra. In the 60 GeV case comparing the $\tau^+\tau^-$ and $b\overline{b}$ spectra we find that the former is shallower and there is a crossing of the spectra typically at frequencies around 1 GHz, evident in Figure~\ref{fig:spectral-sig} or through comparison of Figs.~\ref{fig:bb60_m12_b5_nfw} and \ref{fig:tt60_m12_b5_nfw} (or any similar pairing). The differences between the two spectra can amount to several orders of magnitude in the extremes of the frequency range shown. In the 500 GeV case, comparing the $b\overline{b}$ and $W^+ W^-$ channels embodied by Figs.~\ref{fig:bb500_m12_b5_nfw} and \ref{fig:ww500_m12_b5_nfw} respectively, we see very similar spectra, with a slightly larger magnitude in the $b\overline{b}$ case but also a steeper slope at high frequency than that of $W^+ W^-$. It is important to note that, in the $\nu$ range we consider,  the differences in spectral shape resulting from differing annihilation channels are found to be redshift independent and do not depend upon the mass of the halo or the mass of the neutralino. However, the overall spectral shape is trivially dependent on redshift as previously discussed, thus the redshift dependency of the flux spectrum is influenced only by the neutralino mass and the magnetic field strength. This means that the dominant annihilation channels can be deduced from the shape of the observed spectrum, regardless of the redshift of the source or the mass of the neutralino in question. For sources at redshifts of $z = 1$ or lower it is evident that the optimal frequency for comparison of annihilation channel is around 1 GHz or lower, placing it well within the low frequency range of the SKA-MID.

\begin{figure}[htbp]
\centering
\includegraphics[scale=0.37]{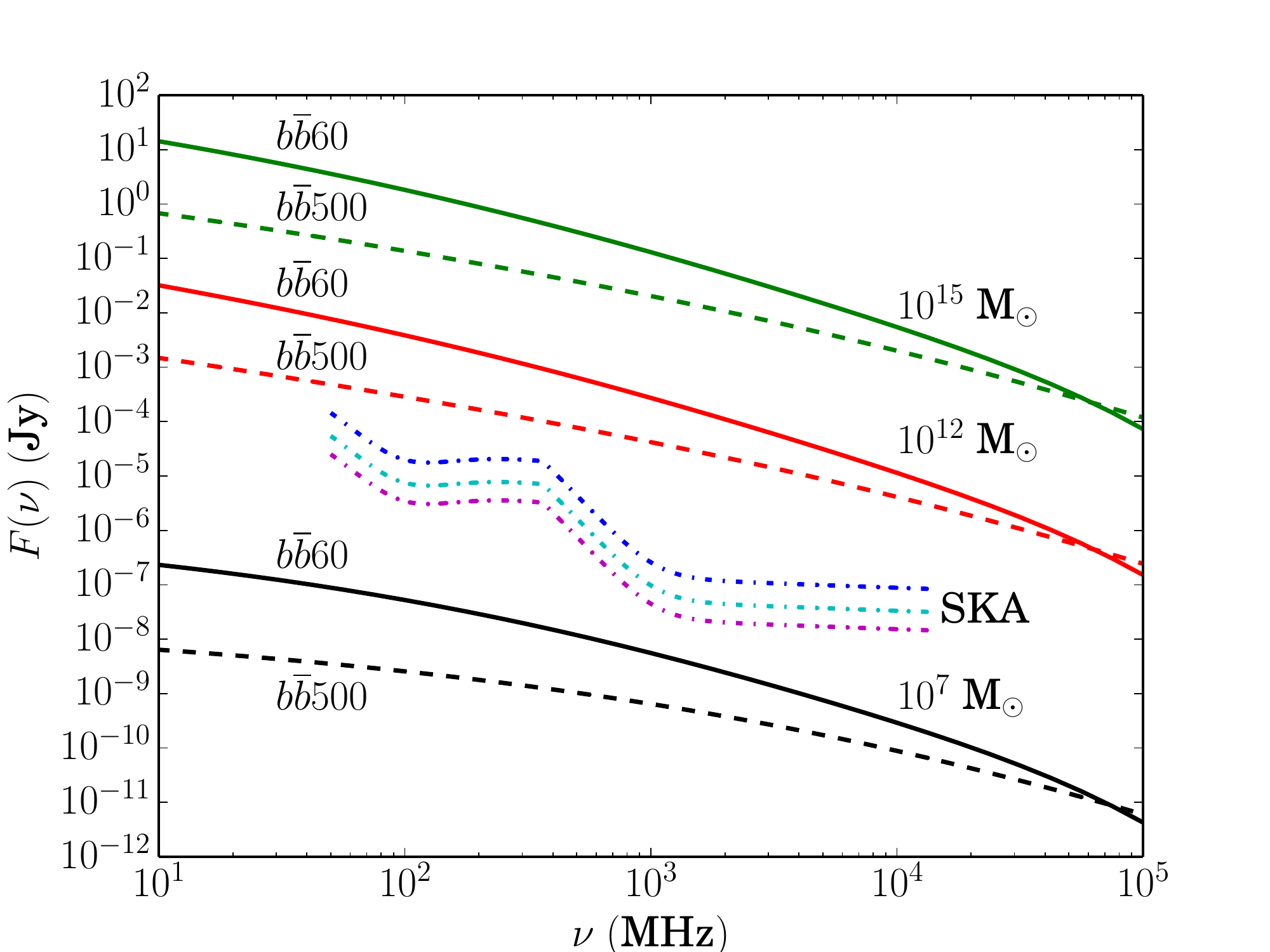}
\includegraphics[scale=0.37]{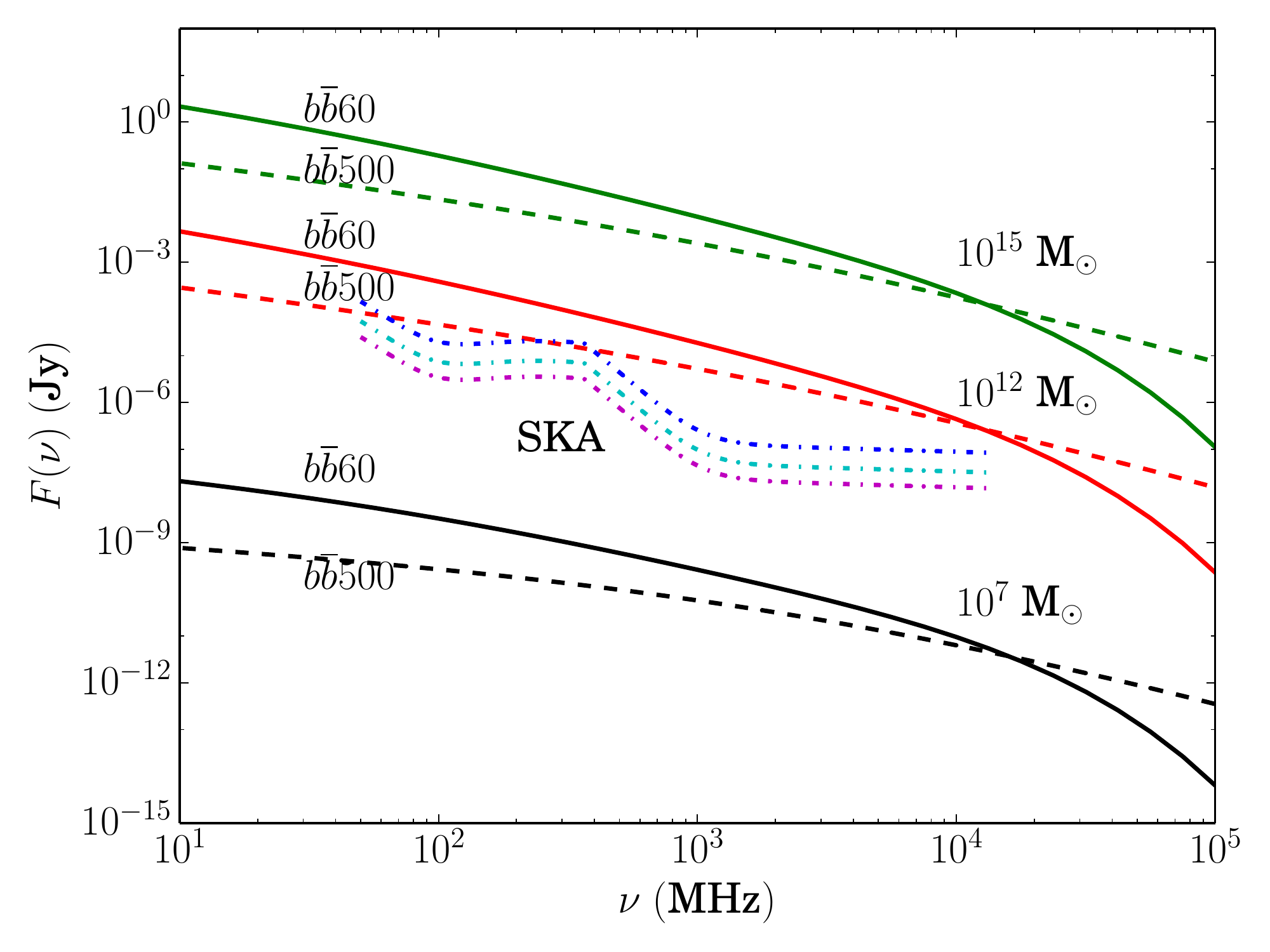}
\caption{Comparison of fluxes from different WIMP masses in the $b\bar{b}$ channel. Solid lines correspond to $m_{\chi} = 60$ GeV and dashed lines to $m_{\chi} = 500$ GeV. Green lines are for halos of mass $10^{15}$ $M_{\odot}$, red for $10^{12}$ $M_{\odot}$ and black for $10^{7}$ $M_{\odot}$. The pink dash-dotted lines are SKA 1$\sigma$ sensitivity limits for 30, 240, and 1000 hour integration times respectively. Left: $B = 5$ $\mu$G. Right: $B = 1$ $\mu$G.}
\label{fig:spectral-sig-2}
\end{figure}

The different choices of neutralino mass have an effect on the magnitude of the spectrum, as is evident in Figure~\ref{fig:spectral-sig-2}. More importantly, the choice of neutralino mass, together with magnetic field strength, dictates the cut-off behaviour of the higher frequency regions of emission spectrum, as can be seen in the aforementioned Figure~\ref{fig:spectral-sig-2} or from comparison of Figs.~\ref{fig:bb500_m7_b1_nfw}, \ref{fig:bb500_m12_b1_nfw} and \ref{fig:bb500_m15_b1_nfw} with their counterparts in Figs.~\ref{fig:bb60_m7_b1_nfw}, \ref{fig:bb60_m12_b1_nfw} and \ref{fig:bb60_m15_b1_nfw}. In the former set of figures, the larger neutralino mass means there is a far shallower slope to the spectrum than in the latter trio despite the fact that both use the same conditions otherwise. However, this high-frequency spectral steepening is also seen to be affected by the magnetic field strength and thus higher-frequency observation must be accompanied by studies of the magnetic field in target sources in order to lift this degeneracy. The mass dependence of the cut-off frequency means that higher frequency observations between 5 and 20 GHz, while somewhat less optimal for cross-section constraints, are still important to exploring the mass axis of the neutralino parameter space.

\clearpage

\begin{figure}[htbp]
\centering
\includegraphics[scale=0.37]{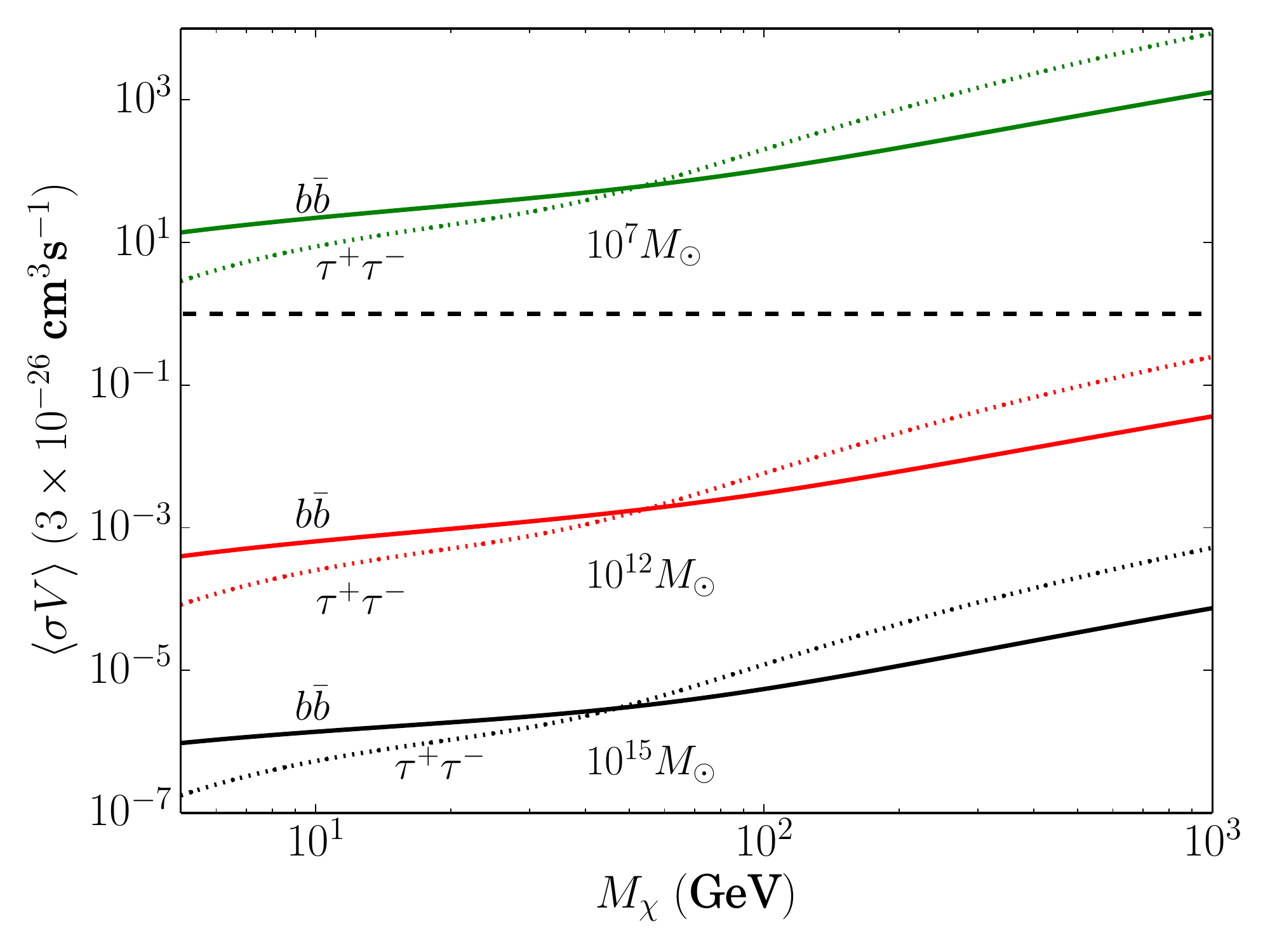}
\includegraphics[scale=0.37]{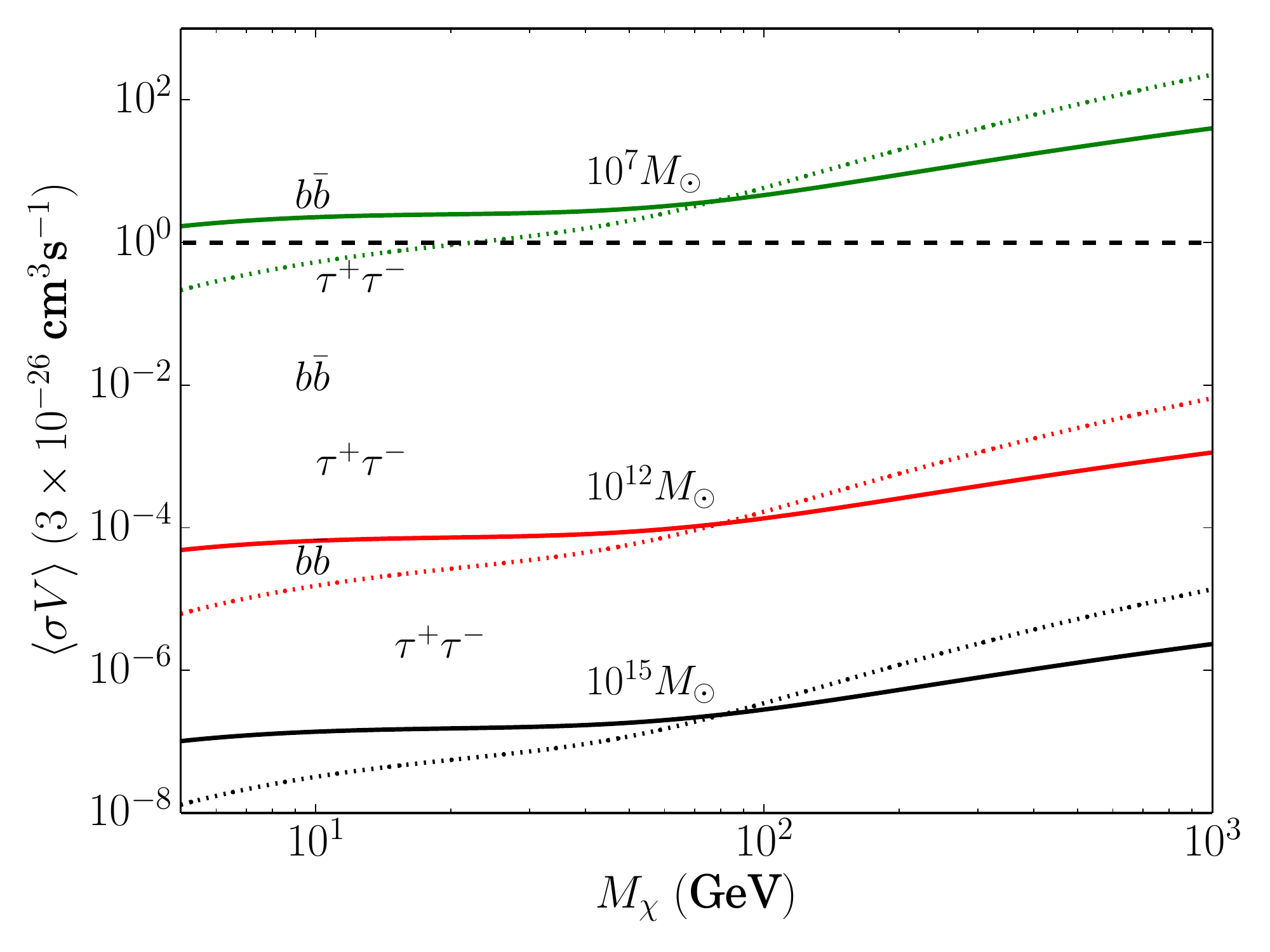}
\caption{Minimum detectable cross-section ($\langle \sigma V\rangle$) from 30 hour SKA integration time for $z = 0.01$ at 300 MHz (left) and 1 GHz (right) as the neutralino mass $M_{\chi}$ is varied with annihilation channel $b\overline{b}$ in solid lines and $\tau^+\tau^-$ in dotted lines. Black lines correspond to halos with mass $10^{15}$ M$_{\odot}$, red lines to $10^{12}$ M$_{\odot}$ and green lines to $10^{7}$ M$_{\odot}$. The black dashed line is $\langle \sigma V\rangle = 3\times 10^{-26}$ cm$^3$ s$^{-1}$.}
\label{fig:sigv_z0}
\end{figure}

\begin{figure}[htbp]
\centering
\includegraphics[scale=0.37]{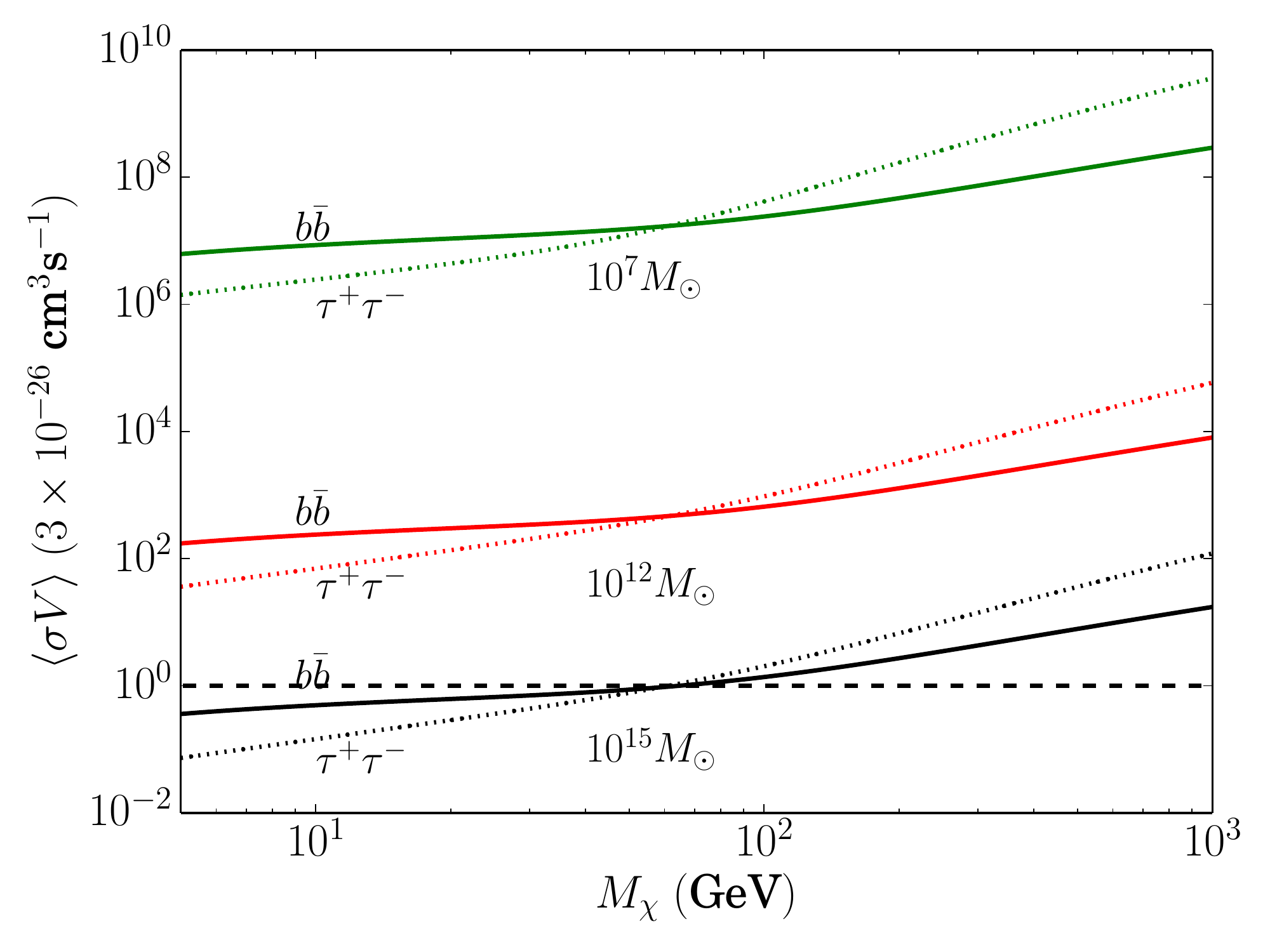}
\includegraphics[scale=0.37]{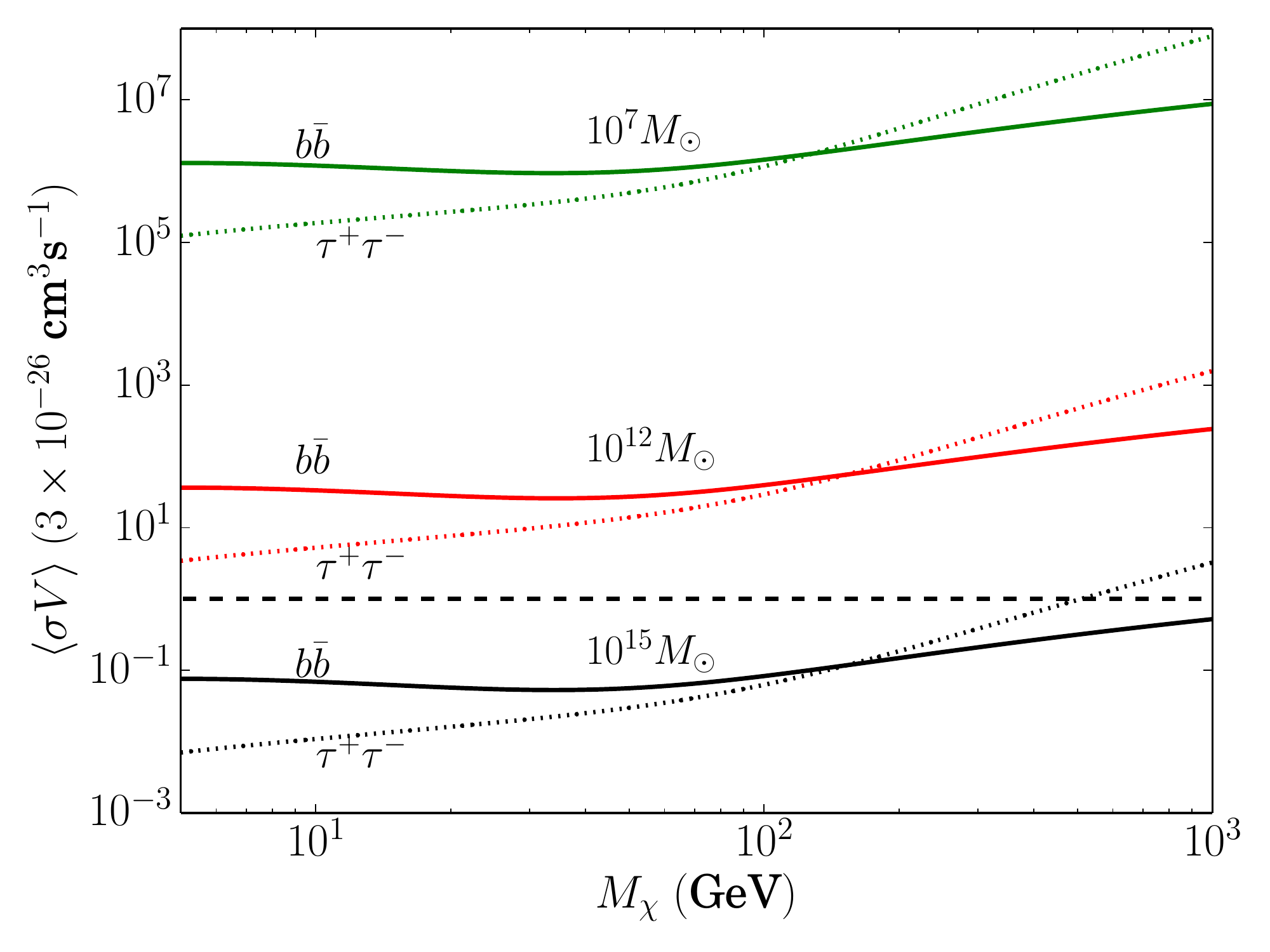}
\caption{Minimum detectable cross-section ($\langle \sigma V\rangle$) from 30 hour SKA integration time for $z = 1$ at 300 MHz (left) and 1 GHz (right) as the neutralino mass $M_{\chi}$ is varied with annihilation channel $b\overline{b}$ in solid lines and $\tau^+\tau^-$ in dotted lines. Black lines correspond to halos with mass $10^{15}$ M$_{\odot}$, red lines to $10^{12}$ M$_{\odot}$ and green lines to $10^{7}$ M$_{\odot}$. The black dashed line is $\langle \sigma V\rangle = 3\times 10^{-26}$ cm$^3$ s$^{-1}$}
\label{fig:sigv_z1}
\end{figure}

\begin{figure}[htbp]
\centering
\includegraphics[scale=0.6]{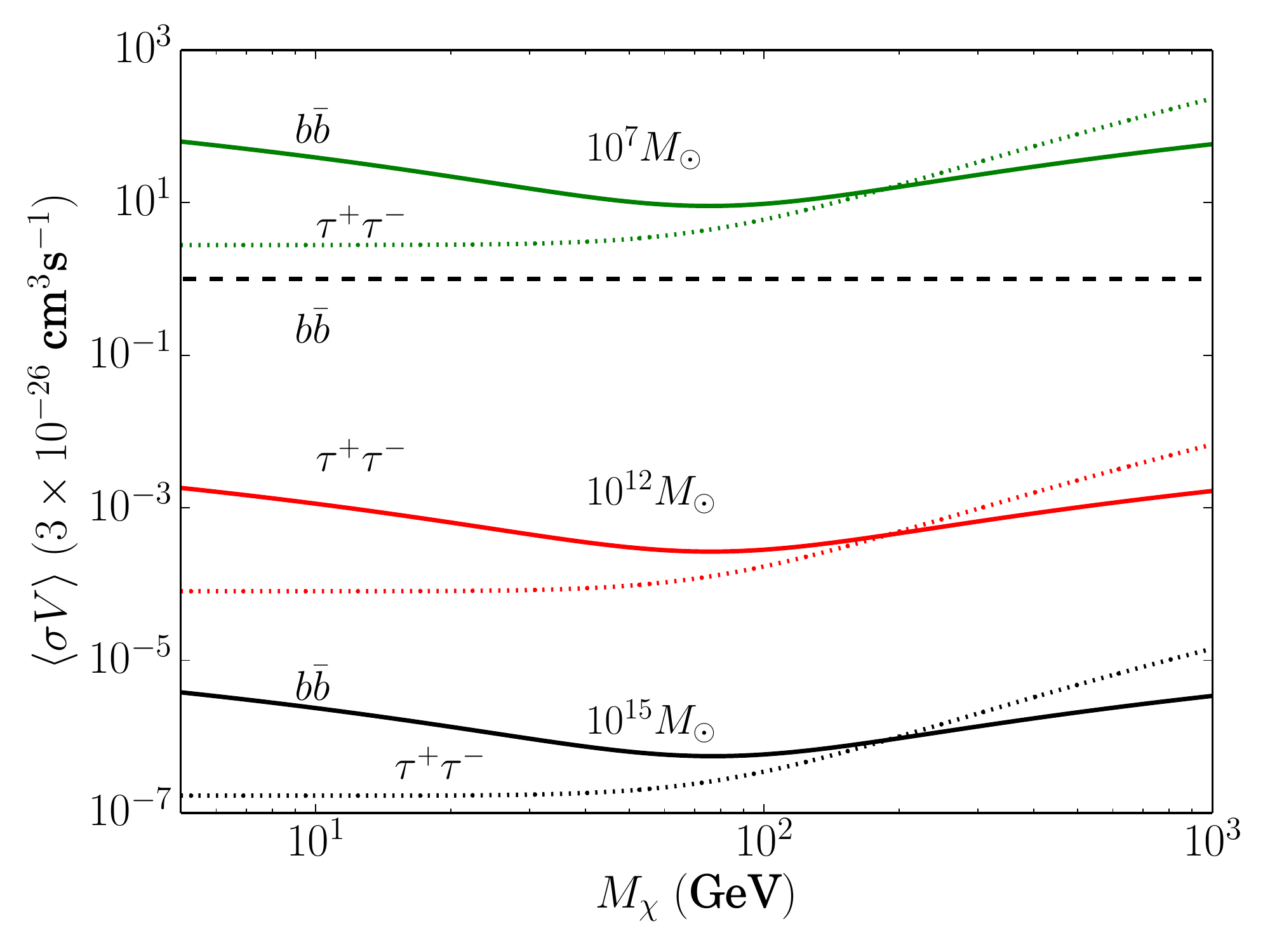}
\caption{Minimum detectable cross-section ($\langle \sigma V\rangle$) from 30 hour SKA integration time for $z = 0.01$ at 5 GHz as the neutralino mass $M_{\chi}$ is varied with annihilation channel $b\overline{b}$ in solid lines and $\tau^+\tau^-$ in dotted lines. Black lines correspond to halos with mass $10^{15}$ M$_{\odot}$, red lines to $10^{12}$ M$_{\odot}$ and green lines to $10^{7}$ M$_{\odot}$. The black dashed line is $\langle \sigma V\rangle = 3\times 10^{-26}$ cm$^3$ s$^{-1}$}
\label{fig:sigv_z0_5GHz}
\end{figure}

Figures~\ref{fig:sigv_z0}, \ref{fig:sigv_z1}, and \ref{fig:sigv_z0_5GHz} display the WIMP mass-dependent sensitivity cross-section $\langle \sigma V \rangle$ derived from the projected $1\sigma$ sensitivity of the SKA, with 30 hour integration time, at redshifts 0.01, 1.0 for frequencies 300 MHz, 1 GHz and 5 GHz. It is clear that the dwarf galaxies with $10^7$ M$_{\odot}$ provide non-detection constraints at best on par with the relic density bound ($\langle \sigma V \rangle \sim \mathcal{O}(10^{-26})$ cm$^3$ s$^{-1}$) for any considered redshift, mass and frequency combination. However, both galactic and cluster halos provide strong restrictions in the event of non-detection by the SKA. This remains true for the cluster halos out to $z \approx 1$. In the case of larger redshifts, none of the halos provide good non-detection constraints. Therefore, we see that the optimal short integration time search region for the SKA is redshifts below 1, concentrating observation on larger halos or considering small halos at much smaller redshifts. A comparison of these figures also demonstrates that the optimal observation window varies with neutralino mass, with lighter neutralinos, below $50$ GeV, favouring the lower frequency range while the heavier neutralino is more optimally detectable in frequencies between 1 and 5 GHz. Thus the SKA's frequency range is extremely well suited to studying neutralino-induced radio emissions across a broad range of neutralino masses.

\begin{figure}[htbp]
\centering
\includegraphics[scale=0.37]{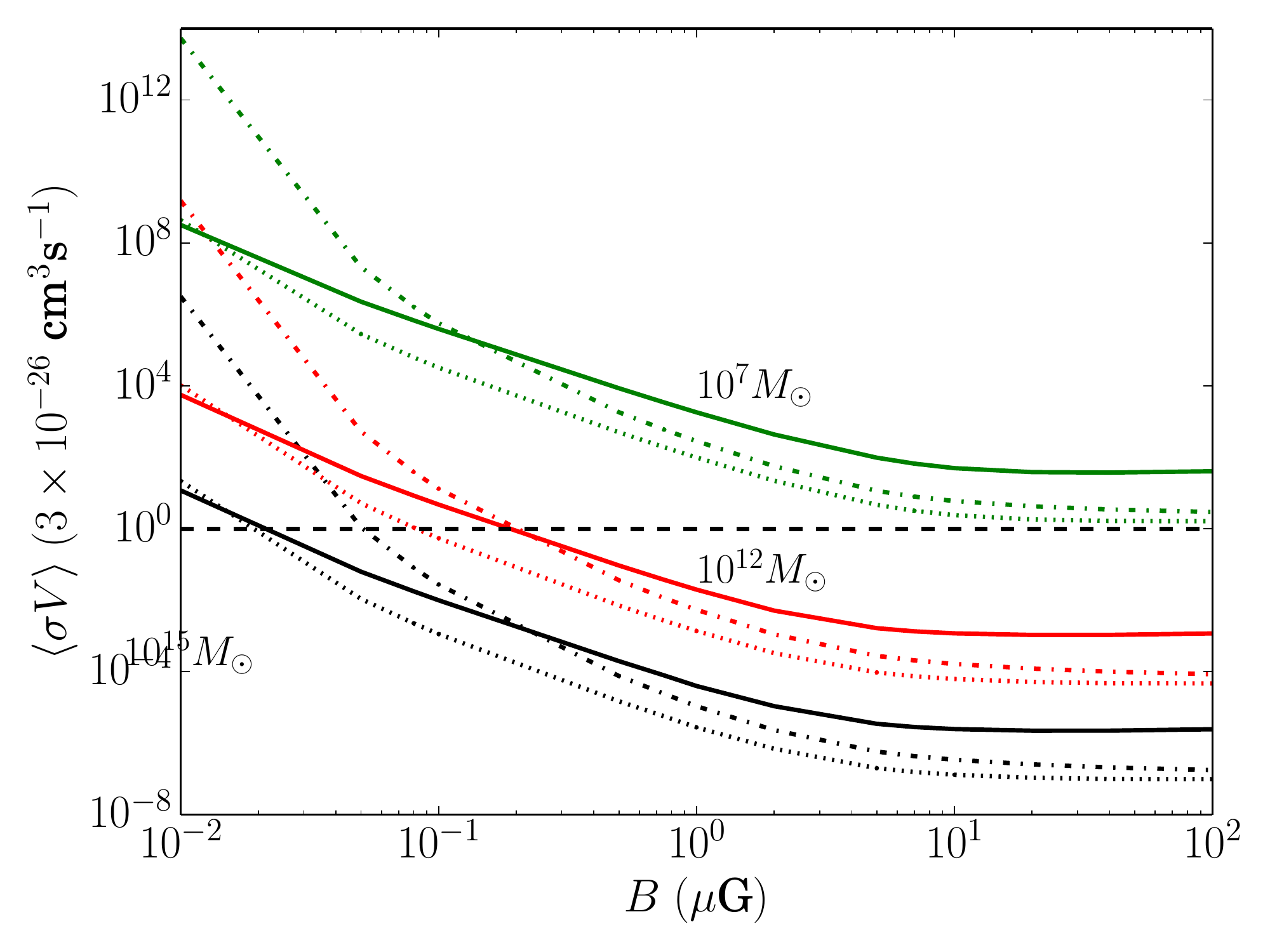}
\includegraphics[scale=0.37]{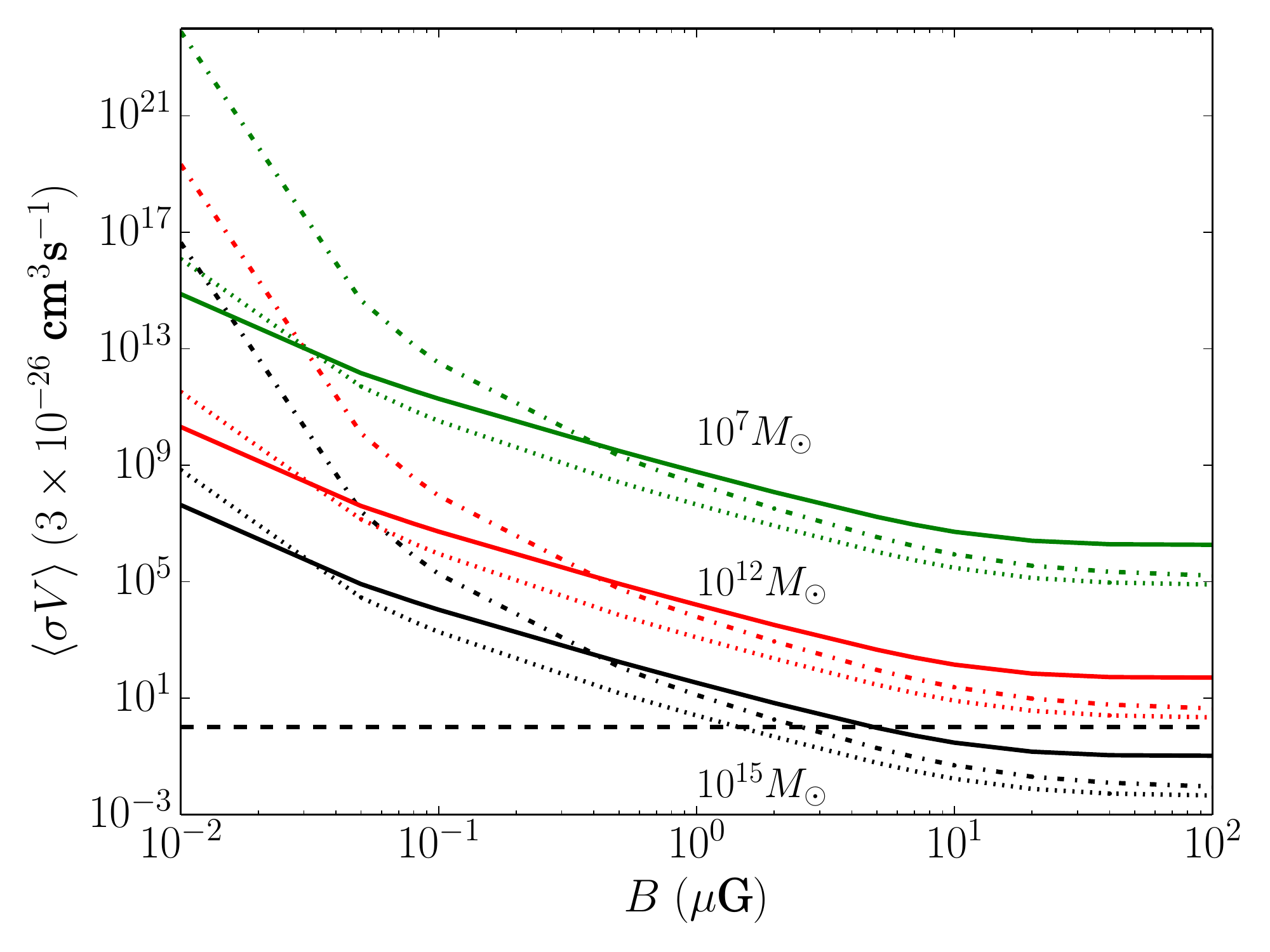}
\caption{Sensitivity cross-section ($\langle \sigma V\rangle$) from 30 hour SKA integration time as the magnetic field $B$ is varied for annihilation channel $b\overline{b}$ with $M_{\chi} = 60$ GeV. Solid lines reflect upper limits at 300 MHz, dotted lines are at 1 GHz while dash-dotted lines are at 5 GHz. Green lines represent dwarf galaxy halos, red lines are for galaxy halos and black lines for cluster halos. The black dashed line is $\langle \sigma V\rangle = 3\times 10^{-26}$ cm$^3$ s$^{-1}$. Left: $z = 0.01$, right: $z = 1.0$.}
\label{fig:sigv_b}
\end{figure}

Figure~\ref{fig:sigv_b} displays the effect of magnetic field strength on synchrotron emissions and how this impacts on the value of the minimum cross-section required for SKA observation at various frequencies. It is evident that, for the displayed 60 GeV $b\overline{b}$ neutralino, stronger magnetic fields, especially those associated with galaxy clusters, such those found in \cite{govoni2004}, from $\mathcal{O}(1)$ to $\mathcal{O}(10)$ $\mu$G provide an excellent test-bed for local observations ($z \le$ 1), with SKA non-detection producing restrictions well below the relic abundance value for both galaxies and clusters at all the displayed frequencies. However, increasingly strong fields offer no additional advantages as the synchrotron emission rapidly saturates. This demonstrates the importance of analysing the magnetic fields within target sources: here the SKA's capability of performing both Faraday Rotation and polarisation studies means that our single instrument can perform all the necessary observations to probe the nature of dark matter. Dwarf galaxies with mass $10^7$ M$_{\odot}$ can at best produce 1$\sigma$ non-detection constraints that are of the same order of magnitude of the reference value for the thermal relic abundance  limit ($\langle \sigma V\rangle = 3\times 10^{-26}$ cm$^3$ s$^{-1}$). Thus even the faintest sources detectable to SKA are capable of producing results competitive with current observations.

\section{Discussion and Conclusions}

\begin{figure}[htbp]
\centering
\includegraphics[scale=0.6]{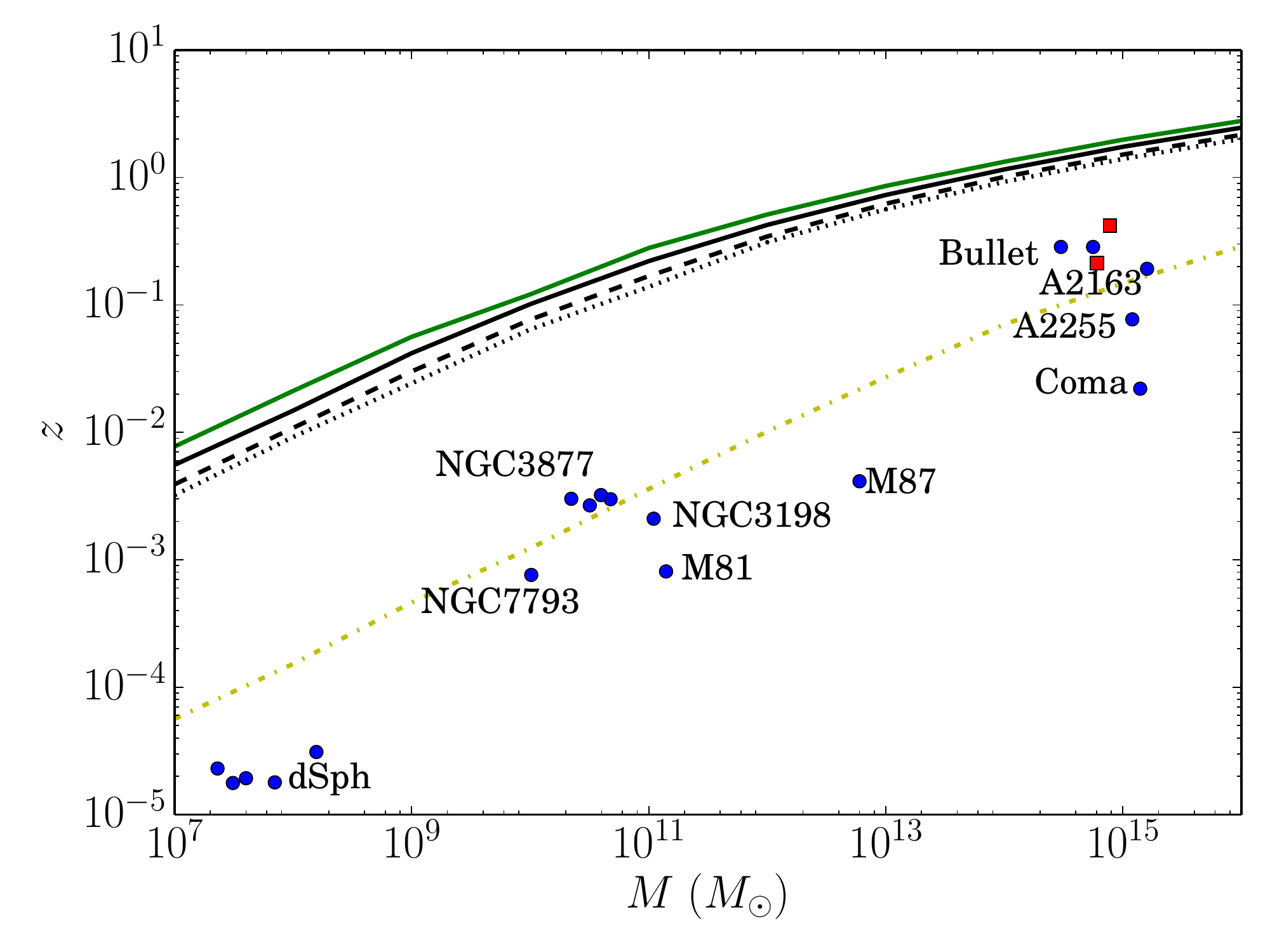}
\caption{Source visibility as a function of redshift and halo mass. Based on projected SKA sensitivity data for the relic abundance cross-section value with 30 hour integration time (black lines) and 1000 hour integration time (green lines). Solid lines are the $1\sigma$ sensitivity exclusion, dashed lines that of $2\sigma$ and dotted lines correspond to $3\sigma$. The yellow dash-dotted line corresponds to 30 hours of integration and 1$\sigma$ confidence with $\langle \sigma V \rangle = 3 \times 10^{-30}$ cm$^3$ s$^{-1}$. Annihilation channel $b\overline{b}$ is assumed with a neutralino mass of 60 GeV. The dSph group contains the dwarf spheroidal galaxies: Draco, Sculptor, Fornax, Carina and Sextans. Unlabelled Galaxies are: NGC3917, NGC3949 and NGC4010 (also note that the Bullet cluster components are shown separately). The data for the selection of sources included in this figure was taken from \cite{lokas2003-coma,lokas2005-draco,lokas2009-dsph,bradac2006-bullet,okabe2011-a2163,merritt1993-m87,kostov2006-ngc3198-m81,burns1995-a2255,jalocha2010-ngc7793,pietrzynski-ngc7793,gonzalez2010-ngcs,verheijen-ngcs}. For extremely local objects redshifts were estimated from the average distance. The square points represent the clusters RXJ2228.6+2037 and A1423~\cite{cassano2013}.}
\label{fig:loop}
\end{figure}

Figure~\ref{fig:loop} displays a source visibility bound as a function of a halo with mass $M$ and redshift $z$. This assumes annihilation channel $b\overline{b}$ (with 60 GeV mass) and a cross-section given by $\langle \sigma V \rangle = 3 \times 10^{-26}$ cm$^3$ s$^{-1}$ and shows the  exclusion curves for 30 hour and 1000 hour integration times at 1 GHz with 3 different confidence intervals (being $1\sigma$, $2\sigma$, and $3\sigma$ c.l.). This figure also contains a selection of sources, those with known radio halos are depicted with circles while the squares are those for which there is no strong evidence of a radio halo. This distinction is important as dark matter emissions should be visible from most structures, not just those with already discovered diffuse radio emissions and it is therefore necessary to probe both varieties of source. The region above the curve can only be observed by the SKA if the cross-section were larger than a given reference value (in this case the relic abundance cross-section $\langle \sigma V \rangle = 3 \times 10^{-26}$ cm$^3$ s$^{-1}$), while the region below the curve can be observed with cross-sections below the reference value. This demonstrates that with $\langle \sigma V \rangle = 3 \times 10^{-26}$ cm$^3$ s$^{-1}$ and longer integration times (1000 hours) dwarf galaxies of $\sim 10^8$ M$_{\odot}$ are detectable at $1\sigma$ out to $z = 0.01$ ($\sim 40$ Mpc). The dash-dotted curve shows off the fact that most of the displayed sources could provide non-detection constraints three or more orders of magnitude below the value $\langle \sigma V \rangle = 3 \times 10^{-26}$ cm$^3$ s$^{-1}$. 
The neutralino annihilation induced radio fluxes from nearby dwarf galaxies (at very small redshifts) can provide the strongest method of constraining the cross-section below the level of the reference cross-section value, as seen by the distance between the dSph group and the reference cross-section value curves in Fig.~\ref{fig:loop}. The extreme proximity of these dwarf galaxies makes their fluxes easily detectable and thus provides excellent detection prospects with the SKA as well as offering very strong constraints in the event of non-detection.\\
It is worth noting that constraints derived in this way may apply to a combination of the cross-section, sub-structure boost factor (as this is not a well known quantity), and the magnetic field amplitude in the halos. Further observational and simulation data are needed to remove this degeneracy by better constraining the magnitude of the sub-structure effect (we note that the substructure effect is also least apparent in dwarf galaxies) and by using the SKA to provide accurate Faraday rotation and polarisation measurements of the magnetic fields in cosmic structures. 
The polarized source counts at $\sim \mu$Jy levels expected to be detected with the SKA  are not known and are extrapolated from current radio surveys at $\sim 1$GHz. These estimates are currently made available for the SKA (see, e.g., {\cite{Govonietal2014}). This last study shows that extrapolating the counts from the deep survey in the GOODS-N field, at a sensitivity of $\approx 15$ $\mu$Jy and at a resolution of $1.6"$, one finds $\approx 315$ polarized sources per square degree  at a flux $\approx 1$ $\mu$Jy \cite{RudnickOwen2014}, a value in  agreement with the
extrapolation derived in \cite{Halesetal2014}, on the basis of the second data release of the Australia Telescope Large Area Survey (ATLAS DR2), at a sensitivity of $\approx 25$ $\mu$Jy/beam and at a resolution of $10"$.
Note that at a resolution of 1 arcmin, from NVSS polarization stacking, the expected number of polarized sources
at a flux level of 1 $\mu$Jy is of the order of $\approx 1300$ sources per square degree as calculated in \cite{Stiletal2014} with an overall uncertainty at the level of  $\approx 50\%$ (see discussion in \cite{Govonietal2014}). 
It is worth noting that the stacking result in \cite{Stiletal2014} is derived from hundreds of thousands of NVSS sources over $\approx 80\%$ of the sky while the expected counts in \cite{RudnickOwen2014} consider a very deep survey over a small field of view.
These expectations ensure that the magnetic field structure in the cosmic structures we consider in our work (which have dimensions of order of half to a deg$^2$) can be proven with the SKA1 using a substantial (of order of at least $\sim 10^2$ FR measures per pointing) number of polarized sources that ensure a good analysis of the amplitude and spatial distribution of magnetic fields along the line of sight to our DM halos.
We must finally note that the predictions of the number of faint polarized radio sources that can be detected with SKA1 depend on the polarization properties of radio sources with a flux density below $\approx 1$ mJy. Total intensity source counts suggest a transition in the dominant population from
AGN to star-forming galaxies around this flux density and the properties of brighter radio sources may not be representative for this fainter population. The evolution of the luminosity functions, in total intensity and in polarization, will be one of the headline science results of future SKA1 surveys (see, e.g., discussion in \cite{Govonietal2014}).

Given that the SKA will be able to independently measure the magnetic fields in our DM halos, we note that the constraints based on heavier halos can be obtained, for our reference value of the magnetic field, out to a maximum redshift of $z \sim 2$, with $10^{15}$ M$_{\odot}$ halos remaining detectable at $1\sigma$ for a value $\langle \sigma V \rangle = 3 \times 10^{-26}$ cm$^3$ s$^{-1}$ out to this redshift with 1000 hour integration or to $z \sim 1.7$ with 30 hour integration. In the case of higher $\sigma$ detections the limiting redshift for the reference cross-section is around $z \sim 1.4$.
This argument demonstrates the potential for SKA non-detection, in large halos, to constrain the annihilation cross-section both from deeper field observations (requiring more integration time), which allow constraint below the relic density bound, and local observations which provide constraints at least 3 orders of magnitude below the relic abundance density bound even with short integration times as seen in Figs.~\ref{fig:sigv_z0} and \ref{fig:sigv_b}.

The constraints derived here are highly competitive in the multi-frequency strategy for indirect detection of dark matter signals. However, a detailed discussion of the foreground emission from standard astrophysical processes has to be considered in the specific study of the telescope sensitivity to dark matter signals (see, e.g., \cite{Regisetal2014} for a discussion). 
There are two main caveats in the forecasts for dark matter detection in the radio frequency band. The
first stems from the fact that, for an extended radio emission, the confusion issue becomes stronger
and stronger as one tries to probe fainter and fainter fluxes. Thus, source subtraction procedures
become crucial and this can affect the estimated sensitivities. The impact of this effect on
the actual sensitivity is hardly predictable at the present time, especially for the SKA, since it will
depend on the properties of the detected sources, the efficiency of deconvolution algorithms, and
the accuracy of the telescope beam shape.
The second caveat is that by bringing down the observational threshold, one can possibly start to
probe the very low levels of possible non-thermal emission associated to the tiny rate of star formation
in dwarf galaxies, or in galaxies and galaxy clusters. The dark matter contribution should be then disentangled
from such astrophysical background. The superior angular resolution of the SKA will allow for the
precise mapping of emissions, putatively either dark matter or baryonically induced, and will enable their
correlation with the stellar or dark matter profiles (obtained via optical and/or kinematic measurements).
It is evident, however, that SKA non-detection holds the potential to be far more stringent than existing bounds on the thermally averaged annihilation cross-section, such as those in\cite{ackerman2010} from the Fermi collaboration, \cite{hooper2011} using Fermi-LAT data, and \cite{bottino2008} from the WMAP haze data.} Specifically, the Fermi collaboration results place the most stringent restrictions, of $\mathcal{O}(10^{-27})$ cm$^3$ s$^{-1}$, on only neutralinos below 10 GeV, with neutralinos around 50 GeV being bounded by $\langle \sigma V \rangle < 10^{-25}$ cm$^3$ s$^{-1}$. 

Additionally, the realistic analysis of the prospects of the CTA~\cite{silverwood2014} show that it provides weak constraints, around $\langle \sigma V \rangle <  10^{-25}$ cm$^3$ s$^{-1}$ for neutralinos below 100 GeV, but are maximally constraining for 1000 GeV at $\langle \sigma V \rangle < 10^{-26}$ cm$^3$ s$^{-1}$. The comparison of the Fermi results and the CTA prospects with the predictions presented here shows that the SKA is sufficiently sensitive to probe a far larger region of the cross-section parameter space than has been previously accessed astrophysically, making the SKA a strong contender as a ``dark matter machine" in future indirect searches for unveiling the signatures of neutralino annihilation. This is particularly reinforced by the fact that the frequency range of the SKA encompasses the regions of greatest importance to the neutralino spectrum, i.e. lower frequencies around 1 GHz and below, which allow for the possibility to distinguish between annihilation channels and for the optimal detection constraints, and higher frequencies around 5 GHz and upwards, which are important in constraining the neutralino mass through the natural cut-off behaviour.

Following to the results displayed for magnetic field variation, the detection of neutralino-induced radio emission is maximised within halos that host stronger magnetic fields, of order $\mathcal{O}$($10 \; \mu$G), conditions which could hold for the central regions of galaxies and galaxy clusters and that can be easily verified with the SKA observations. 

It must be noted, however, that the SKA phase-1 will have still limited ability to detect radio emission from the halos of small structures that are not at small redshifts, making concentration on nearby or very large structures necessary for optimal detection sensitivities with the technique here discussed.\\ 
However, it is evident that highly local dwarf spheroidal galaxies offer some of the best environments for either detection of neutralino annihilation or constraining the cross-section, as they are both rich in dark matter and close enough to provide highly detectable fluxes. Moreover, the ability of the SKA to probe structures at large redshifts allows for the observation of more primeval galaxies and galaxy clusters that may contain fewer foreground sources of baryon-induced radio emission. Additionally the radio emission spectra display qualitative features that distinguish the strength of the magnetic field and the mass of the neutralino, as these are both influential in determining the high-frequency cut-off of the spectrum. SKA-observable cut-off frequencies corresponding to either light neutralinos or weak magnetic fields, in either case the necessary frequencies for the scenarios displayed here will be best observed with SKA phase-2. The characteristic annihilation channel is also evident in the low-frequency slope of the spectrum, with $b\overline{b}$ producing steeper spectra than $\tau^+\tau^-$ at 60 GeV with a cross-over around 1 GHz, while at 500 GeV $b\overline{b}$ is also marginally steeper than $W^+W^-$ with a higher frequency cross-over. These properties are of great importance to the detection of non-thermal radio emission resulting from neutralino annihilation, as they provide tell-tale markers distinguishing the synchrotron emission of electrons produced in this manner from other sources of such radiation within the halo. It is of great significance that such radio spectra might fall within the observation window of the SKA, in particular this would aid in the unambiguous detection of neutralino annihilation products or place additional limits upon the annihilation cross-section, making the SKA a key player in piecing together the dark matter puzzle. This argument is enhanced by the fact that the SKA's frequency range covers the optimal detection regions for neutralinos with masses ranging from around 10 GeV to around 1000 GeV, while its sensitivity allows to probe far smaller values of the neutralino annihilation cross-section than has previously been achieved, thus allowing the SKA to survey a large range of the neutralino parameter space. We therefore conclude that radio observation, with future radio interferometers like the SKA, must be considered as a leading technique in the search for dark matter in the coming decades.

\section*{Acknowledgments}

S.C. acknowledges support by the South African Research Chairs Initiative of the Department of Science and Technology and National Research Foundation and by the Square Kilometre Array (SKA). P.M. and G.B. acknowledge support from the DST/NRF SKA post-graduate bursary initiative.

\end{document}